\documentstyle[epsfig]{elsart}

\newcommand \be{\begin{equation}}
\newcommand \ba{\begin{eqnarray}}
\newcommand \ee{\end{equation}}
\newcommand \ea{\end{eqnarray}}

\begin{document}
\runauthor{Zhou and Sornette}
\begin{frontmatter}
\title{Evidence of a Worldwide Stock Market Log-Periodic Anti-Bubble
Since Mid-2000}
\author[iggp]{\small{Wei-Xing Zhou}},
\author[iggp,ess,nice]{\small{Didier Sornette}\thanksref{EM}}
\address[iggp]{Institute of Geophysics and Planetary Physics, University of
California, Los Angeles, CA 90095-1567}
\address[ess]{Department of Earth and Space Sciences, University of
California, Los Angeles, CA 90095-1567}
\address[nice]{Laboratoire de Physique de la Mati\`ere Condens\'ee,
CNRS UMR 6622 and Universit\'e de Nice-Sophia Antipolis, 06108
Nice Cedex 2, France}
\thanks[EM]{Corresponding author. Department of Earth and Space
Sciences and Institute of Geophysics and Planetary Physics,
University of California, Los Angeles, CA 90095-1567, USA. Tel:
+1-310-825-2863; Fax: +1-310-206-3051. {\it E-mail address:}\/
sornette@moho.ess.ucla.edu (D. Sornette)\\
http://www.ess.ucla.edu/faculty/sornette/}

\begin{abstract}
Following our investigation of the USA Standard and Poor index
anti-bubble that started in August 2000 [Quantitative Finance 2,
468-481 (2002)], we analyze thirty eight world stock market
indices and identify 21 ``bearish anti-bubbles'' and 6 ``bullish
anti-bubbles.'' An ``anti-bubble'' is defined as a
self-reinforcing price trajectory with self-similar expanding
log-periodic oscillations. Mathematically, a ``bearish
antibubble'' is characterize by a power law decrease of the price
(or of the logarithm of the price) as a function of time and by
expanding log-periodic oscillations. We propose that bearish
anti-bubbles are created by positive price-to-price feedbacks
feeding overall pessimism and negative market sentiment further
strengthened by inter-personal interactions. Bullish anti-bubbles
are here identified for the first time. The most striking
discovery is that the majority of European and Western stock
market indices as well as other stock indices exhibit practically
the same log-periodic power law anti-bubble structure as found for
the USA S\&P500 index. These anti-bubbles are found to start
approximately at the same time, August 2000, in all these markets.
This shows a remarkable degree of synchronization worldwide. The
descent of the worldwide stock markets since 2000 is thus an
international event, suggesting the strengthening of
globalization.
\end{abstract}
\end{frontmatter}

%\tableofcontents{} \clearpage

\section{Introduction\label{s1:intro}}

Financial bubbles are loosely defined as phases of
over-valuations of stock market prices above the
fundamental prices. Such over-valuations may be
in accord with the theory of rational expectations,
leading to the concept of rational expectation bubbles
\cite{Blanchard,Blanchardwat,Luxsor,Malsorbub,sormalbubble}, may be due to
exogenous causes or sunspots (see for instance \cite{sunspots})
or may result from a variety of departures from
pure and perfect agent rationality
\cite{Shillervol,Shillerexu,Shefrin,Shleifer,bookcrash}.

A series of papers based on analogies with statistical physics
models have proposed that most financial crashes are the climax of
the so-called log-periodic power law signatures (LPPS) associated
with speculative bubbles resulting from imitation between
investors and their herding behavior
\cite{SJ1998,JS1999,JohSorLed99,CriCrash00,SorJoh01}. In addition,
a large body of empirical evidence supporting this proposition
have been presented
\cite{SJB,SJ1998,CriCrash00,JS2000,emergent,SorJoh01}. A
complementary line of research has established that, while the
vast majority of drawdowns occurring on the major financial
markets have a distribution which is well-described by an
exponential or a slightly fatter distribution in the class of
stretched exponentials, the largest drawdowns are occurring with a
significantly larger rate than predicted by extrapolating the bulk
of the distribution and should thus be considered as outliers
\cite{outl1,SorJoh01,outl2,crashcom}. A recent work
\cite{epsilondd} has merged these two lines of research in a
systematic way to offer a classification of crashes as either
events of an endogenous origin associated with preceding
speculative bubbles or as events of an exogenous origin associated
with the markets response to external shocks.  Two hallmarks of
criticality have been documented: (i) super-exponential power law
acceleration of the price towards a ``critical'' time $t_c$
corresponding to the end of the speculative bubble and (ii)
log-periodic modulations accelerating according to a geometric
series signaling a discrete hierarchy of time scales. Globally
over all the markets analyzed, Ref.~\cite{epsilondd} identified 49
outliers, of which 25 were classified as endogenous, 22 as
exogenous and 2 as associated with the Japanese ``anti-bubble''.
Restricting to the world market indices, Ref.~\cite{epsilondd}
found 31 outliers, of which 19 are endogenous, 10 are exogenous
and 2 are associated with the Japanese anti-bubble. The exogenous
crashes, not preceded by LPPS could be in each case associated
with an important piece of information impacting the market.

All these results taken together formulate a general hypothesis
according to which imitation between investors and their herding
behavior lead to speculative bubbles of financial markets with
accelerating overvaluation decorated by accelerating oscillatory
structures possibly followed by crashes or change of regimes. The
key concept is the existence of positive price-to-price feedbacks.
When speculative prices go up, creating wealth for some investors,
this may attract other investors by word-of-mouth interactions,
fuelling further price increases. This in turn promotes a
wide-spread interest in the media which promotes and amplifies the
self-fulfilling wishful thinking \cite{Roehnersor}, with seemingly
reasonable or rational theories advanced to justify the price
increases. These processes generate more investor demand, fuelling
further the expansion of the speculative bubble. The positive
price-to-price feedback mechanism has recently been formulated
mathematically in a nonlinear generalization of the Black-Scholes
stochastic differential equation \cite{Sorandersen} and in a
nonlinear model of stock market prices combining the positive
price-to-price feedback with nonlinear negative feedback due to
fundamental trading together with inertia \cite{Idesor1,Idesor2}.
It was there shown that the speculative bubble becomes unstable,
reflecting the fact that high prices are ultimately not
sustainable, since they are high only because of expectations of
further price increases. The bubble eventually bursts, and prices
come falling down. The feedback that fed the bubble carries the
seeds of its own destruction, and so the end of the bubble and the
crash are often unrelated to any really significant news on
fundamentals \cite{bookcrash}.

The same feedback mechanism may also produce a ``negative'' bubble
or ``bearish anti-bubble,'' that is, downward price movements propelling
further downward price movements, enhancing pessimism by
inter-personal interactions. Johansen and Sornette \cite{Nikkei99}
proposed indeed that such imitation and herding mechanism may also
lead to so-called ``anti-bubbles'' with decelerating market
devaluations following market peaks. The concept of
``anti-bubble'' was introduced to describe the long-term
depression of the Japanese index, the Nikkei, that has decreased
along a downward path marked by a succession of ups and downs
since its all-time high of 30 Dec. 1989
\cite{Nikkei99,evalNikkei}. The concept of anti-bubble restores a
certain degree of symmetry between the speculative behavior of the
``bull'' and ``bear'' market regimes. This degree of symmetry,
after the critical time $t_c$, corresponds to the existence of
``anti-bubbles,'' characterized by a power law decrease of the
price (or of the logarithm of the price) as a function of time
$t>t_c$, down from a maximum at $t_c$ (which is the beginning of
the anti-bubble) and by decelerating/expanding log-periodic
oscillations \cite{Nikkei99,evalNikkei}. Another anti-bubble was
found to describe the gold future prices after its all-time high
in 1980. The Russian market prior to and after its speculative
peak in 1997 also constitutes a remarkable example where both
bubble and anti-bubble structures appear simultaneously for the
same $t_c$. Several other examples have been described in emergent
markets \cite{emergent}.

In a recent paper \cite{SPpredict}, we have uncovered a remarkable
similarity in the behavior of the US S\&P500 index from 1996 to
August 2002 and of the Japanese Nikkei index from 1985 to 1992 (11
years shift), with particular emphasis on the structure of the
bearish phase which is qualified as an anti-bubble according to
the previous classification. Specifically, we found the existence
of a clear signature of herding in the decay of the S\&P500 index
since August 2000 with high statistical significance, in the form
of strong log-periodic components decorating a power law
relaxation.

Here, we show that the (bearish) anti-bubble that started around
August 2000 on the USA stock market is actually a world-wide phenomenon
with a high degree of correlation and synchronization between most
of the western markets.
To our knowledge, only during the crash of October 1987 and in its aftermath
did stock markets worldwide exhibited a similar or stronger correlation
\cite{Barro,Roll88}.

\section{Identification of anti-bubbles in world stock market
indexes\label{s1:ident}}

\subsection{Qualification of an anti-bubble}

Following the philosophy of Ref.~\cite{epsilondd} and references
therein (see also \cite{bookcrash} for a general review and
references therein), we qualify an anti-bubble by the existence of
a regime of stock market prices well-fitted by the expression
\begin{equation}
\ln p\left(t\right) \approx A + B \tau ^{\alpha} +C \tau^\alpha
\cos\left[ \omega \ln \left( \tau \right) +\phi \right]~,
\label{Eq:fit1}
\end{equation}
which embodies the log-periodic power law signature. Note that the
phase $\phi$ does nothing but provide a time scale $T$ since
$\omega \ln \left( \tau \right) +\phi = \omega \ln \left( \tau/T \right)$
with the definition $\phi = - \omega \ln T$. $\phi$ thus disappears
by the choice of $T$ as the time unit. This stresses the fact that, if
the phase is not a fundamental parameter of the fit
since it can be get rid of by a suitable gauge choice,
it contains nevertheless an important information on the existence of
a characteristic time scale.
expression (\ref{Eq:fit1}) obeys the symmetry of discrete scale
invariance \cite{SorDSI}, that has been proposed to be a hallmark
of cooperative behavior of interacting agents
\cite{JohSorLed99,CriCrash00,bookcrash}. The meaning of the
adjective ``well-fitted'' will be clarified below, first by
presenting visual evidences in figures and then more formally by
statistical tests. For a speculative bubble, we have
\begin{equation}
\tau = t_c-t~, \label{Eq:taub}
\end{equation}
which is the time to the end of the bubble occurring at $t_c$.
For an anti-bubble, we have
\begin{equation}
\tau = t-t_c~, \label{Eq:taua}
\end{equation}
which is the time since the beginning of the anti-bubbles at
$t_c$. The exponent $\alpha$ should be positive in order for the
price $p(t)$ to remain finite at $t=t_c$ \cite{CriCrash00}. In
general, speculative bubbles exhibiting the LPPS described by
(\ref{Eq:fit1}) with (\ref{Eq:taub}) are followed by crashes or
strong corrections that are ``outliers'' \cite{epsilondd}. It may
happen that some of these speculative bubbles transform rather
into anti-bubbles described by (\ref{Eq:fit1}) with
(\ref{Eq:taua}). As we said, this occurred for instance for the
Russian speculative bubble ending in 1997. In contrast,
anti-bubbles correspond in general to enduring corrections of
stock markets that follow a period of strong growth, as
exemplified by the trajectory of the Japanese Nikkei index
\cite{Nikkei99,SPpredict} which culminated in Dec. 1989 and then
has suffered a non-stop decay decorated by oscillations
\cite{Nikkei99,evalNikkei}.

Following our previous finding \cite{SPpredict} of a strong
influence of a log-periodic harmonic at the angular
log-frequency $2\omega$ for the S\&P500 index, we also present
fits including the effect of a harmonic at $2 \omega$. In this
goal, we postulate the formula
\be
\ln p\left( t\right) \approx A
+ B \tau ^{\alpha} +C \tau^\alpha \cos\left[ \omega \ln \left(
\tau \right) +\phi_1 \right] + D \tau^\alpha \cos\left[ 2\omega
\ln \left( \tau \right) +\phi_2 \right]~,
\label{Eq:fit4}
\ee
which differs from equation (\ref{Eq:fit1}) by the addition of the
last term  proportional to the amplitude $D$.
The two phases $\phi_1$ and $\phi_2$ now define two time scales
$T_1=e^{-\phi_1/\omega}$ and $T_2=e^{-\phi_2/\omega}$. It is thus no
longer possible to make them disappear simultaneously by a choice of
time unit.

\subsection{Methodology}

For each stock market described below,
we use equations (\ref{Eq:fit1}) and (\ref{Eq:fit4})
to fit the logarithm of the stock
market indices over an interval starting from a time
$t_{\rm start}$ and ending in September, 30, 2002.
If we knew the critical time $t_c$, then an obvious
choice would be $t_{\rm start} = t_c$. This choice would be
optimal since it allows us to use expression (\ref{Eq:fit1}) for
the longer possible time span compatible with the occurrence
of the anti-bubble. Not knowing $t_c$ precisely, in order
to be consistent with the meaning of expression (\ref{Eq:fit1})
with (\ref{Eq:taua}), we should ensure that $t_{\rm start} \geq t_c$.

Furthermore, in accordance with the intuitive meaning of an
anti-bubble, we would like to take $t_{\rm start}$ close to the
last strong maximum in 2000 and then carry out a sensitivity
analysis with respect to $t_{\rm start}$. Fortunately, we shall
show that the critical times $t_c$ estimated from the fit of the
data for different $t_{\rm{start}}$ do not disperse much. In other
words, the fits with different $t_{\rm{start}}$ are robust and
$t_c$ is not very sensitive to $t_{\rm{start}}$. We shall come
back to this point later in Sec.~\ref{s1:tc} which will be focused
on the best possible characterization of $t_c$.

As part of the sensitivity analysis
with respect to $t_{\rm start}$, we shall also use the following
trick, which ensures that the impact of $t_{\rm start}$ is minimized.
Since nothing informs us a priori about the time ordering of
$t_{\rm start}$ and $t_c$, following \cite{SPpredict},
we modify expression (\ref{Eq:taua}) into
\begin{equation}
\tau=|t-t_c|~. \label{Eq:absTau}
\end{equation}
This definition (\ref{Eq:absTau}) has the advantage of removing
the constraint of $t_c<t_{\rm{start}}$ in the
optimization. We see this as an advantage because this constraint
has not deep meaning and does not contain
any specific information on the data as it results solely from the
analyzing procedure and the arbitrary choices for
$t_{\rm{start}}$.
It thus enables us to test the robustness of fits by
scanning different $t_{\rm{start}}$ \cite{SPpredict}.
Not knowing $t_c$ precisely, in order
to be consistent with the meaning of expression (\ref{Eq:fit1})
with (\ref{Eq:taua}), we ensure that $t_{\rm start} \geq t_c$
by trial and error: a given chosen $t_{\rm start}$
is accepted only if the fit gives a critical time $t_c \leq t_{\rm start}$.

While the definition (\ref{Eq:taua}) together with the logarithmic
as well as power law singularities associated with formula
(\ref{Eq:fit1}) imposes that $t_c < t_{\rm start}$ for an
anti-bubble, the definition (\ref{Eq:absTau}) allows for the
critical time $t_c$ to lie anywhere within the time series. In
that case, the part of the time series for $t<t_c$ corresponds to
an accelerating ``bubble'' phase while the part $t>t_c$
corresponds to a decelerating ``anti-bubble'' phase. Definition
(\ref{Eq:absTau}) has thus the advantage of introducing a degree
of flexibility in the search space for $t_c$ without much
additional cost. In particular, it allows us to avoid a thorough
scanning of $t_{\rm start}$ since the value of $t_c$ obtained with
this procedure is automatically adjusted without constraint. See
Ref.~\cite{SPpredict} for a discussion of the advantages and
potential problems associated with this procedure using
(\ref{Eq:absTau}). Here, we use the definition (\ref{Eq:absTau})
because it has proved to provide significantly better and more
stable fits with little need to vary $t_{\rm start}$.

For a given $t_{\rm start}$, we estimate the parameters $t_c$,
$\alpha$, $\omega$, $\phi$, $A$, $B$ and $C$ of (\ref{Eq:fit1}) by
minimizing the sum of the squared residues between the fit
function (\ref{Eq:fit1}) and the logarithm of the real index data.
Following \cite{JohSorLed99,CriCrash00}, the three linear
parameters $A$, $B$ and $C$ are slaved to the other parameters by
solving analytically a system of three linear equations and we are
left with optimizing four free parameters. To obtain the global
optimization solution, we employ the taboo search \cite{Taboo} to
determine an ``elite list'' of solutions as the initial conditions
of the ensuing line search procedures in conjunction with a
quasi-Newton method. The best fit thus obtained is regarded to be
globally optimized. A similar procedure is used to fit the index
with formula (\ref{Eq:fit4}). This formula has two
additional parameters compared with (\ref{Eq:fit1}), the amplitude
$D$ of the harmonic and its phase $\phi_2$. We follow a fit
procedure which is an adaptation of the slaving method of
\cite{JohSorLed99,CriCrash00}. This allows
us to slave the four parameters $A$, $B$, $C$ and $D$ to the other
parameters in the search for the best fit. With this approach, we
find that the search of the optimal parameters is very stable and
provides fits of very good quality in spite of the remaining five
free parameters.

We apply this procedure to 38 stock indices all over the world
including:
\begin{itemize}
\item eight indices in Americas
(Argentina, Brazil, Chile, USA-Dow Jones, Mexico, USA-NASDAQ, Peru, and
Veneruela),
\item fourteen indices in Asia/Pacific (Australia, China,
Hong Kong, India, Indonesia, Japan, Malaysia, New Zealand,
Pakistan, Philippines, South Korea, Sri Lanka, Taiwan and
Thailand),
\item fourteen indices in Europe (Austria, Belgium, Czech,
Denmark, France, Germany, Netherlands, Norway, Russia, Slovakia,
Spain, Switzerland, Turkey and United Kingdom) and
\item two indices in
Africa/Middle East (Egypt and Israel).
\end{itemize}
By the obvious criterion to obtain at least a solution in the
fitting procedure, we find no evidence of an anti-bubble in the
following eleven indices: Austria, Chile, China, Egypt, Malaysia,
New Zealand, Pakistan, Philippines, Slovakia, Sri Lanka and
Venezuela. Interestingly, except for Austria and New Zealand, they
are emergent markets in developing countries.

The rest of the paper is thus devoted to the study of the remaining 27
indexes out of our initial list of 38.

\subsection{Bearish anti-bubbles}

\begin{figure}
\begin{center}
\epsfig{file=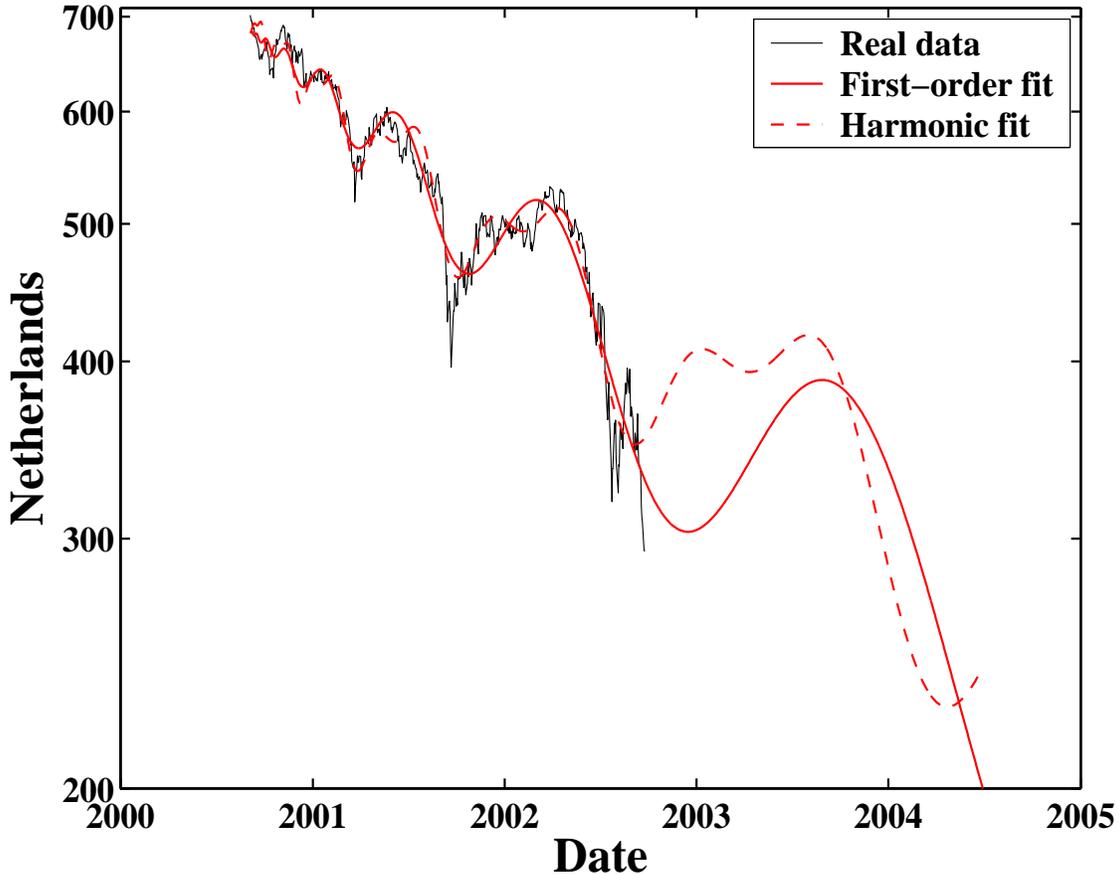,width=15cm, height=12cm}
\end{center}
\caption{The stock market index of Netherlands (fine noisy
line) and its fits with the simple log-periodic formula (\ref{Eq:fit1})
(thick line) and with formula (\ref{Eq:fit4}) incorporating
an harmonics at $2\omega$ (dashed
line). The starting date for the fits is $t_{\rm{start}}=$
04-Sep-2000 and the ending date of the fit is September, 30, 2002.
The parameter values of the fit with (\ref{Eq:fit1}) are
$t_c=$ 28-Aug-2000, $\alpha=1.05$, $\omega=9.16$, $\phi=1.63$,
$A=6.53$, $B=-0.00055$, and $C=-0.00018$. The r.m.s. of the fit
errors is 0.043. The parameter values of the fit with (\ref{Eq:fit4}) are
$t_c=$ 20-Jul-2000, $\alpha=0.80$, $\omega=11.07$, $\phi_1=4.75$,
$\phi_2=0.91$, $A=6.60$, $B=-0.00282$, $C= 0.00064$, and $D=
0.00032$. The r.m.s. of the fit errors is 0.0372. The
formula including the harmonics reduces the r.m.s. of fit
errors by 13.6\%.}
\label{Fig:IDNetherlands}
\end{figure}

\begin{figure}
\begin{center}
\epsfig{file=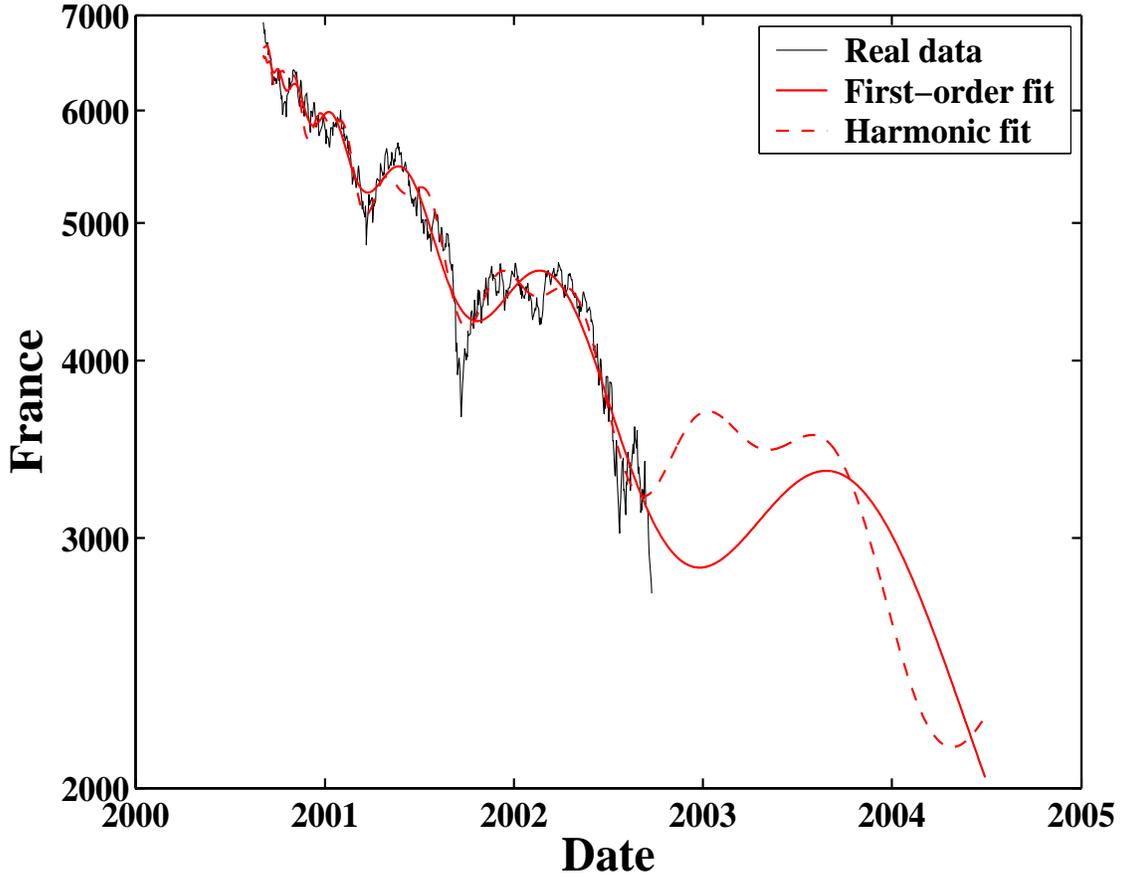,width=15cm, height=12cm}
\end{center}
\caption{The stock market index of France (fine noisy line)
and its fits with the simple log-periodic formula (\ref{Eq:fit1}) (thick line)
and with formula (\ref{Eq:fit4}) incorporating an harmonics
(dashed line). The
starting date for the fits is $t_{\rm{start}}=$ 04-Sep-2000. The
parameter values of the fit with (\ref{Eq:fit1}) are $t_c=$ 30-Aug-2000,
$\alpha=0.92$, $\omega=8.88$, $\phi=3.54$, $A=8.79$, $B=-0.00139$,
and $C=-0.00033$. The r.m.s. of the fit errors is 0.0399. The
parameter values of the fit with (\ref{Eq:fit4}) are $t_c=$ 04-Jul-2000,
$\alpha=0.71$, $\omega=11.24$, $\phi_1=0.28$, $\phi_2=4.25$,
$A=8.90$, $B=-0.00576$, $C=-0.00091$, and $D= 0.00048$. The r.m.s.
of the fit errors is 0.0347.
The formula including the harmonics  reduces the
r.m.s. of fit errors by 12.9\%.} \label{Fig:IDFrance}
\end{figure}

\begin{figure}
\begin{center}
\epsfig{file=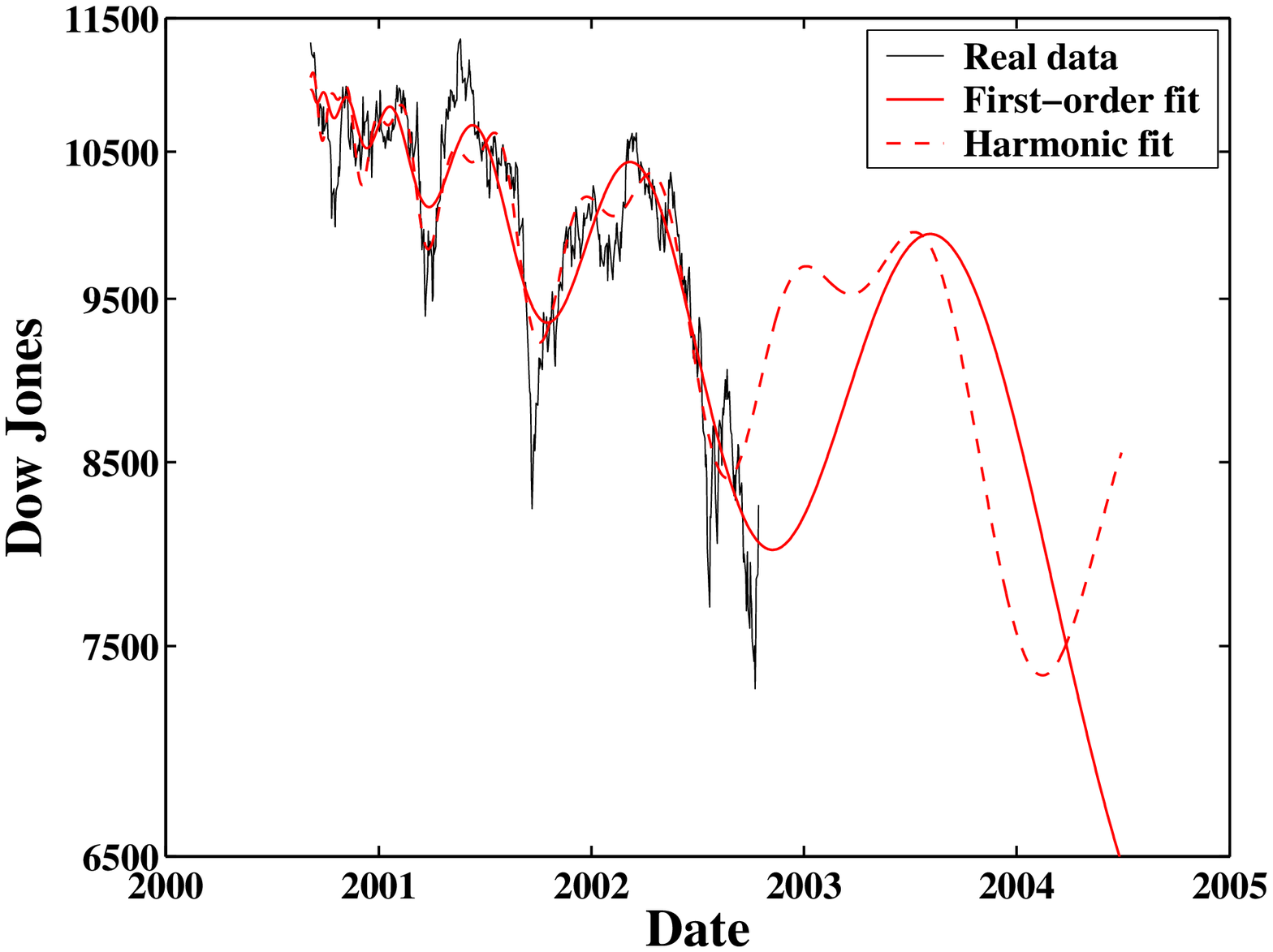,width=15cm, height=12cm}
\end{center}
\caption{The stock market DowJones index of USA (fine noisy line)
and its fits with the simple log-periodic formula (\ref{Eq:fit1}) (thick line)
and with formula (\ref{Eq:fit4}) incorporating an harmonics
(dashed line). The
starting date for the fits is $t_{\rm{start}}=$ 06-Sep-2000. The
parameter values of the fit with (\ref{Eq:fit1}) are $t_c=$ 15-Aug-2000,
$\alpha=1.05$, $\omega=9.76$, $\phi=4.01$, $A=9.30$, $B=-0.00017$,
and $C=-0.00011$. The r.m.s. of the fit errors is 0.0337. The
parameter values of the fit with (\ref{Eq:fit4}) are $t_c=$ 18-Jun-2000,
$\alpha=0.72$, $\omega=12.09$, $\phi_1=0.83$, $\phi_2=2.16$,
$A=9.33$, $B=-0.00145$, $C=-0.00065$, and $D=-0.00034$. The r.m.s.
of the fit errors is 0.03. The
formula including the harmonics reduces the r.m.s.
of fit errors by 11.1\%.} \label{Fig:IDDowJones}
\end{figure}

\begin{figure}
\begin{center}
\epsfig{file=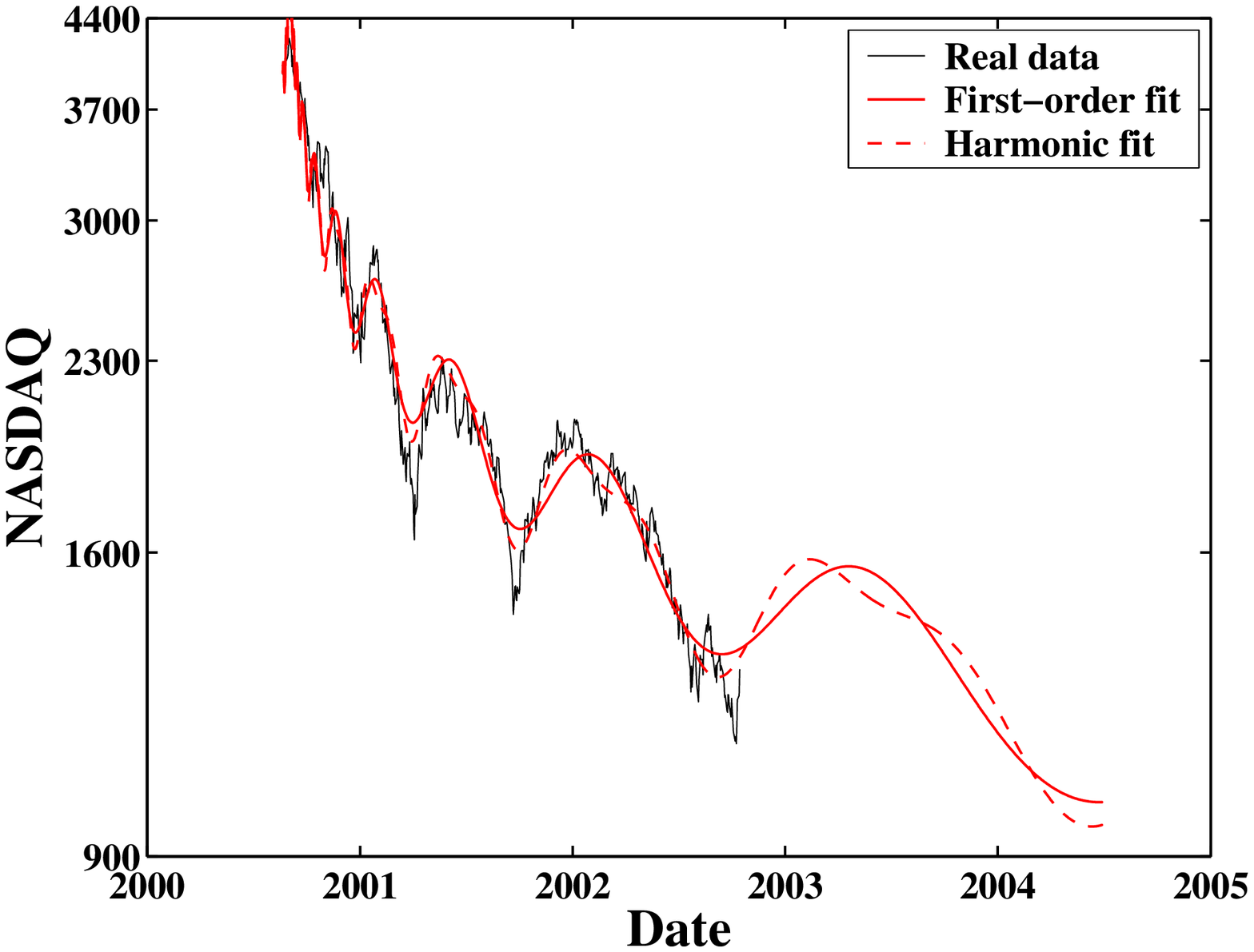,width=15cm, height=12cm}
\end{center}
\caption{The stock market NASDAQ index of USA (fine noisy line)
and its fits with the simple log-periodic formula (\ref{Eq:fit1}) (thick line)
and with formula (\ref{Eq:fit4}) incorporating an harmonics
(dashed line). The
starting date for the fits is $t_{\rm{start}}=$ 20-Aug-2000. The
parameter values of the fit with (\ref{Eq:fit1}) are $t_c=$ 02-Sep-2000,
$\alpha=0.26$, $\omega=10.00$, $\phi=3.37$, $A=8.74$,
$B=-0.25128$, and $C=-0.02360$. The r.m.s. of the fit errors is
0.0636. The parameter values of the fit with (\ref{Eq:fit4}) are $t_c=$
02-Sep-2000, $\alpha=0.26$, $\omega=10.00$, $\phi_1=3.37$,
$\phi_2=0.00$, $A=8.77$, $B=-0.26922$, $C=-0.02473$, and
$D=-0.00918$. The r.m.s. of the fit errors is 0.057.
The formula including the harmonics  reduces the r.m.s. of fit
errors by 10.3\%.}
\label{Fig:IDNASDAQ}
\end{figure}

\begin{figure}
\begin{center}
\epsfig{file=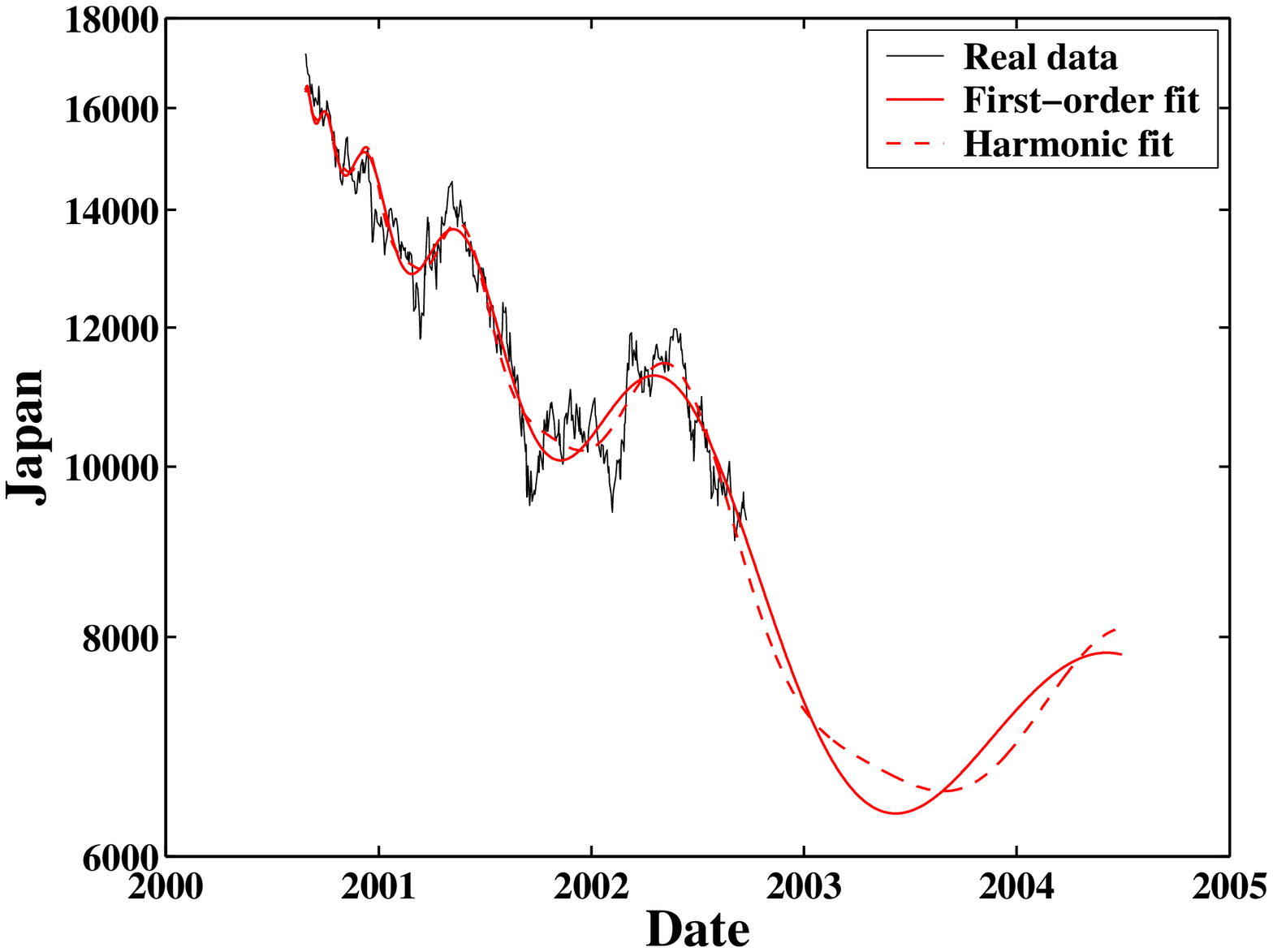,width=15cm, height=12cm}
\end{center}
\caption{The stock market index of Japan (fine noisy line)
and its fits with the simple log-periodic formula (\ref{Eq:fit1}) (thick line)
and with formula (\ref{Eq:fit4}) incorporating an harmonics
(dashed line). The
starting date for the fits is $t_{\rm{start}}=$ 28-Aug-2000. The
parameter values of the fit with (\ref{Eq:fit1}) are $t_c=$ 06-Aug-2000,
$\alpha=0.79$, $\omega=7.74$, $\phi=3.34$, $A=9.74$, $B=-0.00340$,
and $C=-0.00086$. The r.m.s. of the fit errors is 0.0396. The
parameter values of the fit with (\ref{Eq:fit4}) are $t_c=$ 06-Aug-2000,
$\alpha=0.77$, $\omega=7.74$, $\phi_1=3.35$, $\phi_2=0.00$,
$A=9.74$, $B=-0.00372$, $C=-0.00089$, and $D= 0.00021$. The r.m.s.
of the fit errors is 0.0369. The
formula including the harmonics reduces the
r.m.s. of fit errors by  6.9\%.} \label{Fig:IDJapan}
\end{figure}

\begin{figure}
\begin{center}
\epsfig{file=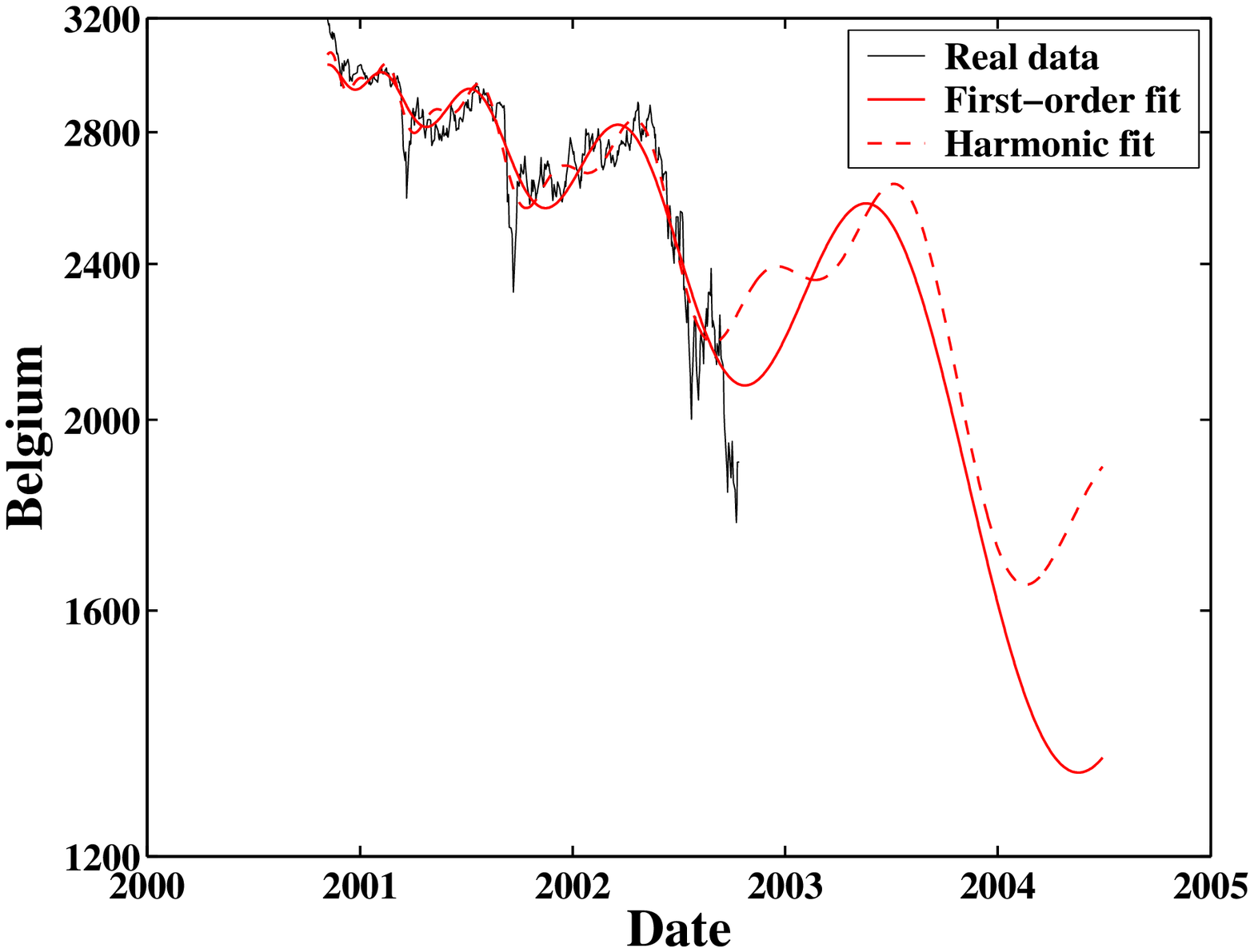,width=15cm, height=12cm}
\end{center}
\caption{The stock market index of Belgium (fine noisy line)
and its fits with the simple log-periodic formula (\ref{Eq:fit1}) (thick line)
and with formula (\ref{Eq:fit4}) incorporating an harmonics
(dashed line). The
starting date for the fits is $t_{\rm{start}}=$ 06-Nov-2000. The
parameter values of the fit with (\ref{Eq:fit1}) are $t_c=$ 25-Jun-2000,
$\alpha=1.52$, $\omega=12.20$, $\phi=2.85$, $A=8.02$,
$B=-0.00001$, and $C= 0.00000$. The r.m.s. of the fit errors is
0.0336. The parameter values of the fit with (\ref{Eq:fit4}) are $t_c=$
21-Jun-2000, $\alpha=1.14$, $\omega=12.24$, $\phi_1=2.67$,
$\phi_2=0.49$, $A=8.05$, $B=-0.00012$, $C= 0.00005$, and
$D=-0.00002$. The r.m.s. of the fit errors is 0.0283. The
formula including the harmonics reduces the r.m.s. of fit
errors by 15.9\%.}
\label{Fig:IDBelgium}
\end{figure}

\begin{figure}
\begin{center}
\epsfig{file=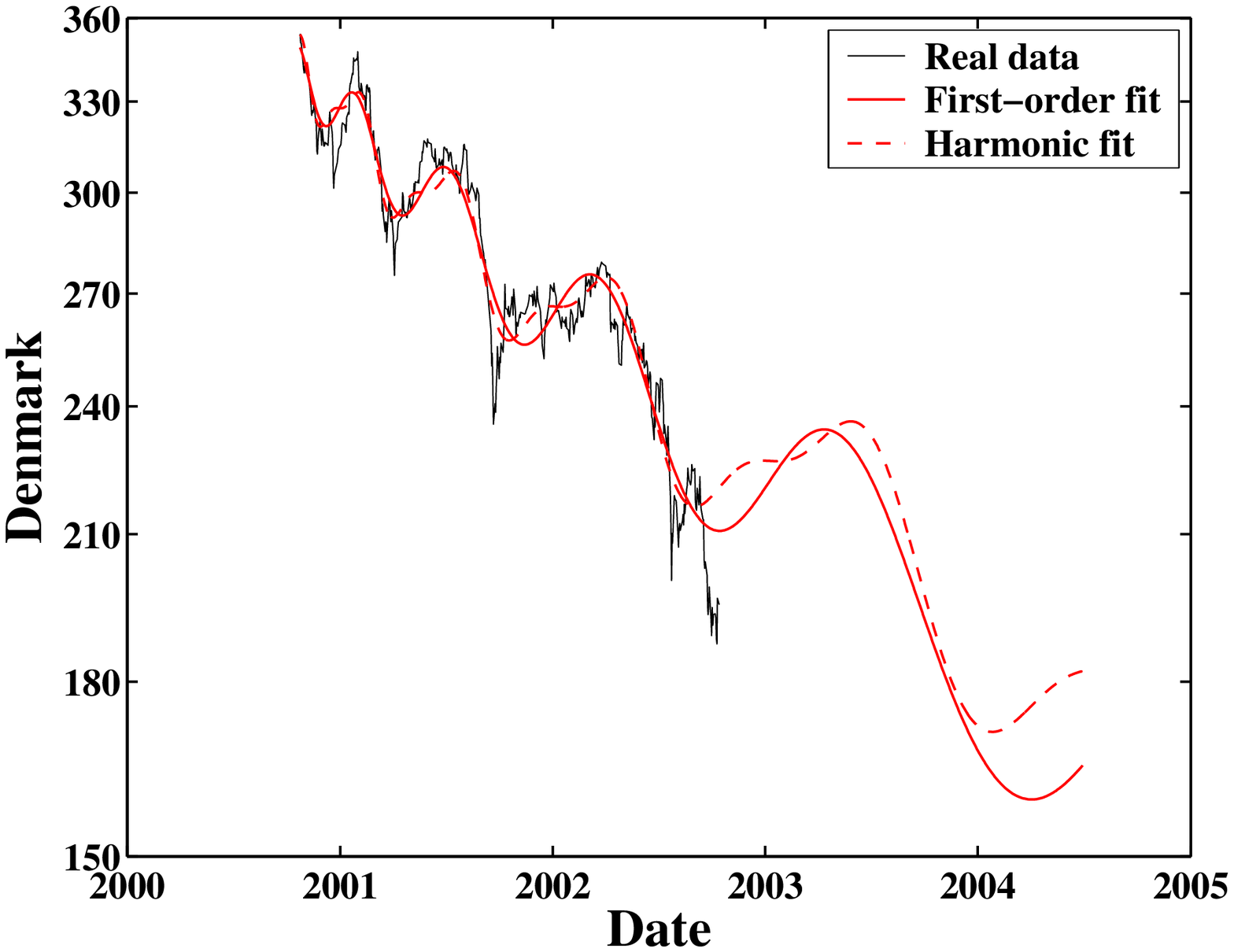,width=15cm, height=12cm}
\end{center}
\caption{The stock market index of Denmark (fine noisy line)
and its fits with the simple log-periodic formula (\ref{Eq:fit1}) (thick line)
and with formula (\ref{Eq:fit4}) incorporating an harmonics
(dashed line). The
starting date for the fits is $t_{\rm{start}}=$ 24-Oct-2000. The
parameter values of the fit with (\ref{Eq:fit1}) are $t_c=$ 03-May-2000,
$\alpha=0.78$, $\omega=13.37$, $\phi=0.62$, $A=5.98$,
$B=-0.00273$, and $C= 0.00047$. The r.m.s. of the fit errors is
0.0316. The parameter values of the fit with (\ref{Eq:fit4}) are $t_c=$
04-May-2000, $\alpha=0.64$, $\omega=13.35$, $\phi_1=0.87$,
$\phi_2=0.00$, $A=6.04$, $B=-0.00751$, $C= 0.00105$, and $D=
0.00036$. The r.m.s. of the fit errors is 0.0286. The
formula including the harmonics reduces the r.m.s. of fit
errors by  9.6\%.}
\label{Fig:IDDenmark}
\end{figure}

\begin{figure}
\begin{center}
\epsfig{file=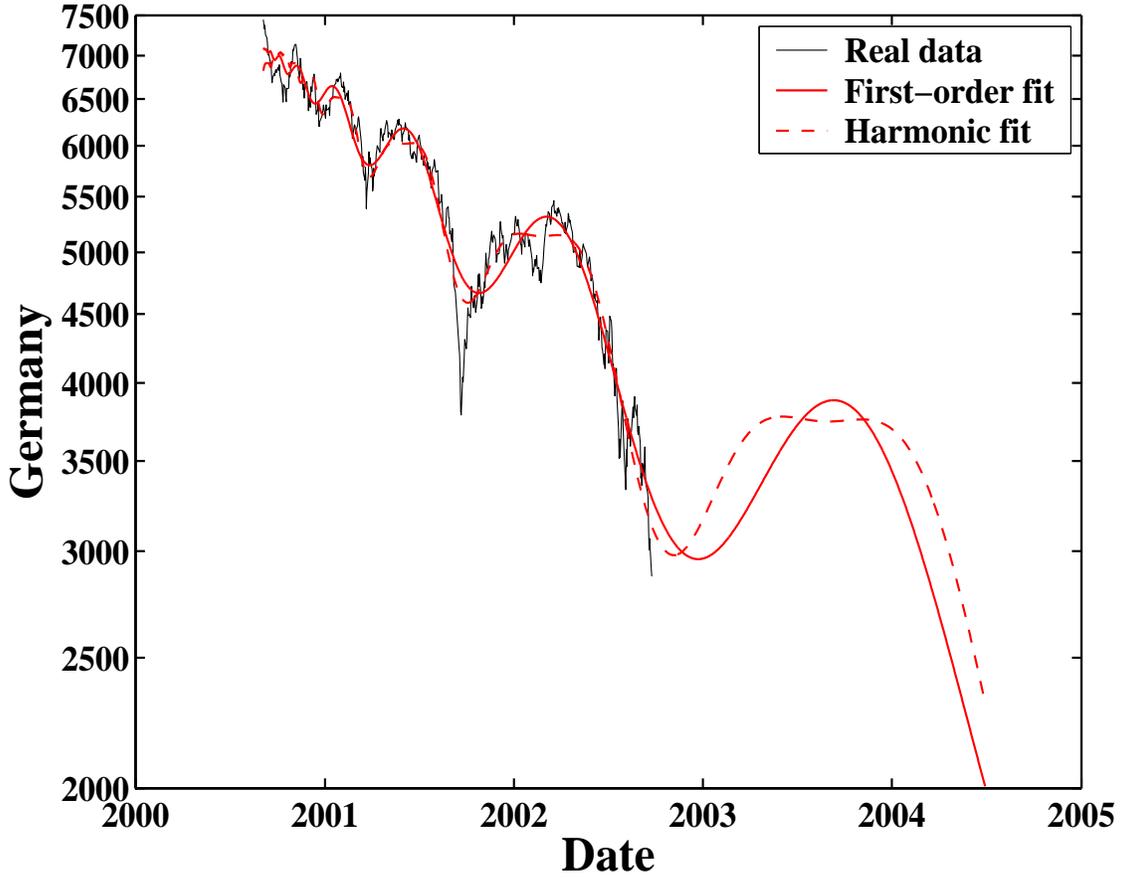,width=15cm, height=12cm}
\end{center}
\caption{The stock market index of Germany (fine line)
and its fits with the simple log-periodic formula (\ref{Eq:fit1}) (thick line)
and with formula (\ref{Eq:fit4}) incorporating an harmonics
(dashed line). The
starting date for the fits is $t_{\rm{start}}=$ 04-Sep-2000. The
parameter values of the fit with (\ref{Eq:fit1}) are $t_c=$ 31-Aug-2000,
$\alpha=1.05$, $\omega=9.02$, $\phi=5.66$, $A=8.87$, $B=-0.00057$,
and $C= 0.00019$. The r.m.s. of the fit errors is 0.0439. The
parameter values of the fit with (\ref{Eq:fit4}) are $t_c=$ 06-Oct-2000,
$\alpha=0.94$, $\omega=8.47$, $\phi_1=3.61$, $\phi_2=4.58$,
$A=8.86$, $B=-0.00120$, $C= 0.00041$, and $D= 0.00012$. The r.m.s.
of the fit errors is 0.0396. The
formula including the harmonics reduces the
r.m.s. of fit errors by 10.0\%.} \label{Fig:IDGermany}
\end{figure}

\begin{figure}
\begin{center}
\epsfig{file=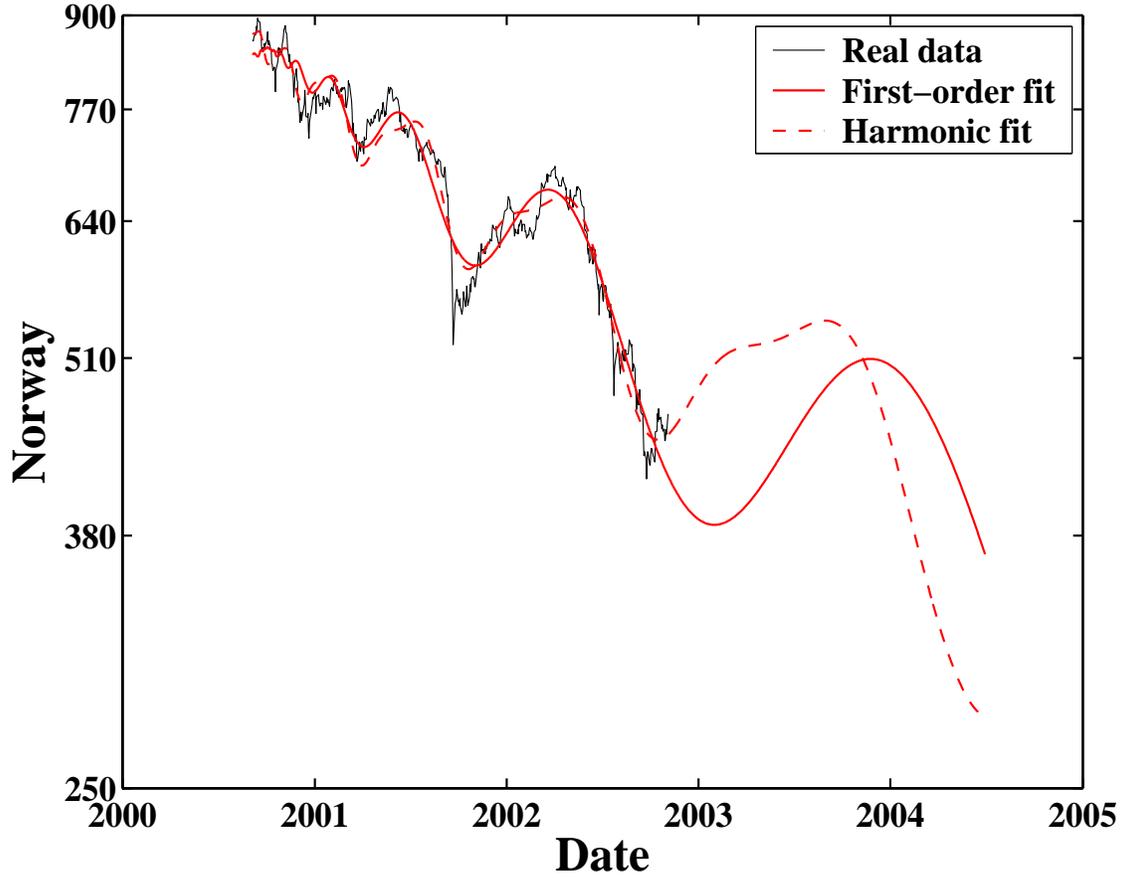,width=15cm, height=12cm}
\end{center}
\caption{The stock market index of Norway (fine line) and
its fits with the simple log-periodic formula (\ref{Eq:fit1}) (thick line)
and with formula (\ref{Eq:fit4}) incorporating an harmonics (dashed line). The
starting date for the fits is $t_{\rm{start}}=$ 05-Sep-2000. The
parameter values of the fit with (\ref{Eq:fit1}) are $t_c=$ 02-Oct-2000,
$\alpha=1.02$, $\omega=8.21$, $\phi=4.77$, $A=6.75$, $B=-0.00060$,
and $C= 0.00022$. The r.m.s. of the fit errors is 0.0396. The
parameter values of the fit with (\ref{Eq:fit4}) are $t_c=$ 13-Jul-2000,
$\alpha=0.87$, $\omega=10.92$, $\phi_1=2.03$, $\phi_2=1.78$,
$A=6.82$, $B=-0.00161$, $C=-0.00045$, and $D= 0.00013$. The r.m.s.
of the fit errors is 0.0358. The formula including the harmonics reduces the
r.m.s. of fit errors by  9.7\%.} \label{Fig:IDNorway}
\end{figure}

\begin{figure}
\begin{center}
\epsfig{file=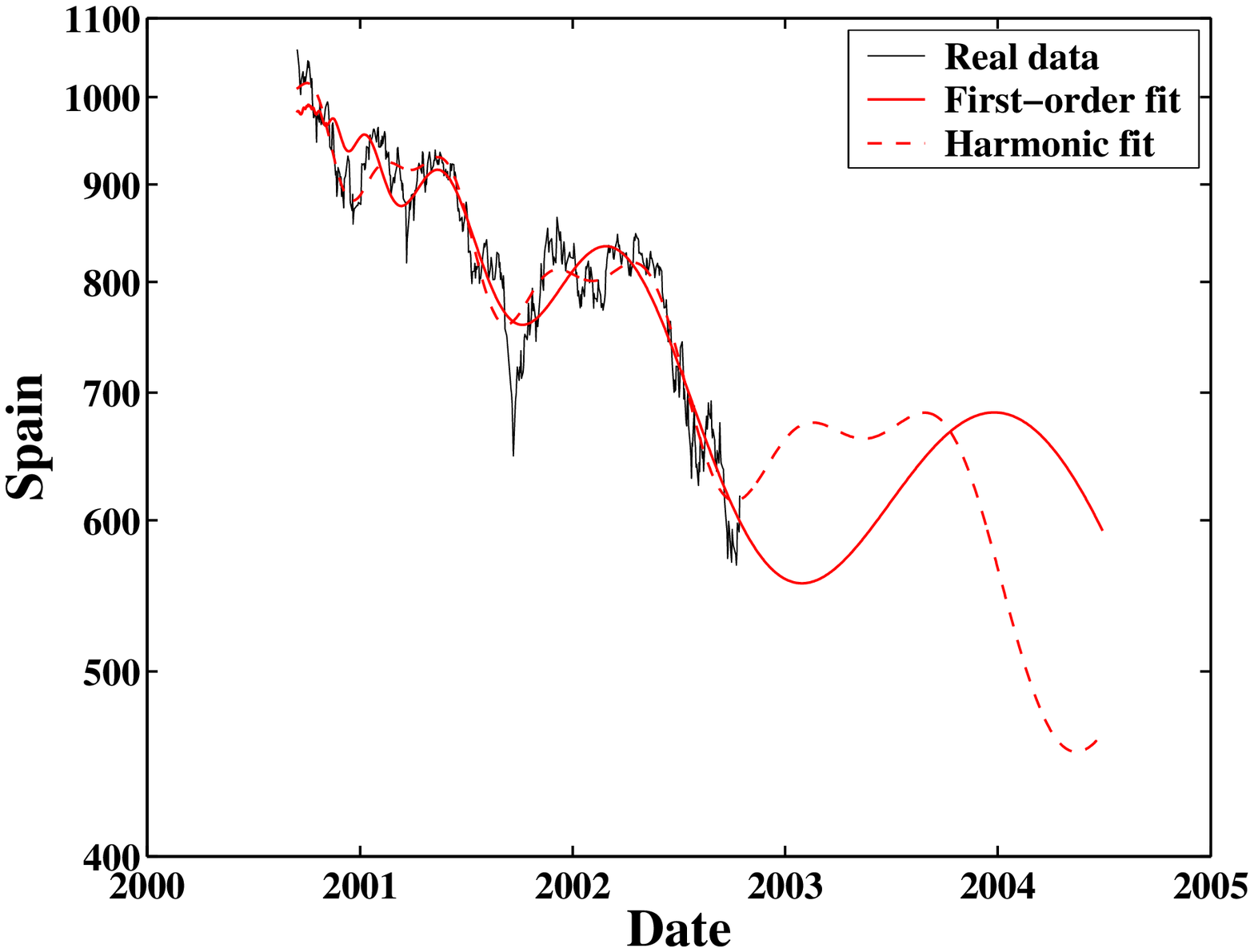,width=15cm, height=12cm}
\end{center}
\caption{The stock market index of Spain (fine noisy line)
and its fits with the simple log-periodic formula (\ref{Eq:fit1}) (thick line)
and with formula (\ref{Eq:fit4}) incorporating an harmonics
(dashed line). The
starting date for the fits is $t_{\rm{start}}=$ 14-Sep-2000. The
parameter values of the fit with (\ref{Eq:fit1}) are $t_c=$ 04-Oct-2000,
$\alpha=0.93$, $\omega=7.52$, $\phi=3.15$, $A=6.90$, $B=-0.00082$,
and $C= 0.00031$. The r.m.s. of the fit errors is 0.0398. The
parameter values of the fit with (\ref{Eq:fit4}) are $t_c=$ 03-Jul-1999,
$\alpha=0.88$, $\omega=15.70$, $\phi_1=1.75$, $\phi_2=1.28$,
$A=7.14$, $B=-0.00114$, $C=-0.00018$, and $D= 0.00008$. The r.m.s.
of the fit errors is 0.0346. The harmonic formula reduces the
r.m.s. of fit errors by 12.9\%.} \label{Fig:IDSpain}
\end{figure}

\begin{figure}
\begin{center}
\epsfig{file=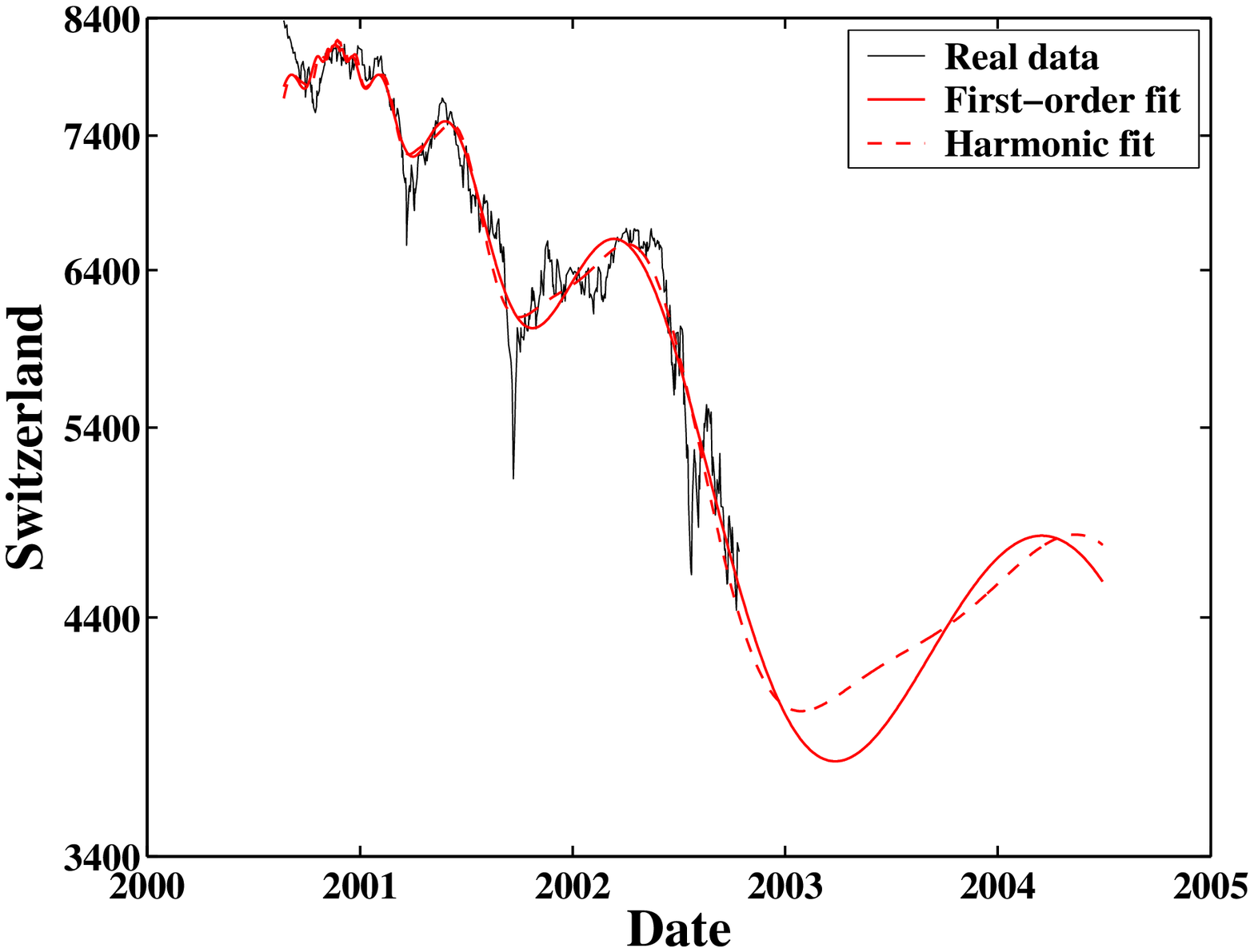,width=15cm, height=12cm}
\end{center}
\caption{The stock market index of Switzerland (fine noisy
line) and its fits with the simple log-periodic formula
(\ref{Eq:fit1}) (thick line)
and with formula (\ref{Eq:fit4}) incorporating an harmonics
(dashed line).
   The starting date for the fits is $t_{\rm{start}}=$
23-Aug-2000. The parameter values of the fit with (\ref{Eq:fit1}) are
$t_c=$ 18-Nov-2000, $\alpha=1.00$, $\omega=6.76$, $\phi=5.16$,
$A=9.01$, $B=-0.00070$, and $C=-0.00026$. The r.m.s. of the fit
errors is 0.0367. The parameter values of the fit with (\ref{Eq:fit4}) are
$t_c=$ 23-Nov-2000, $\alpha=0.95$, $\omega=6.72$, $\phi_1=5.54$,
$\phi_2=0.00$, $A=9.01$, $B=-0.00098$, $C=-0.00035$, and
$D=-0.00007$. The r.m.s. of the fit errors is 0.0352. The
formula including the harmonics reduces the r.m.s. of fit
errors by  4.0\%.}
\label{Fig:IDSwitzerland}
\end{figure}

\begin{figure}
\begin{center}
\epsfig{file=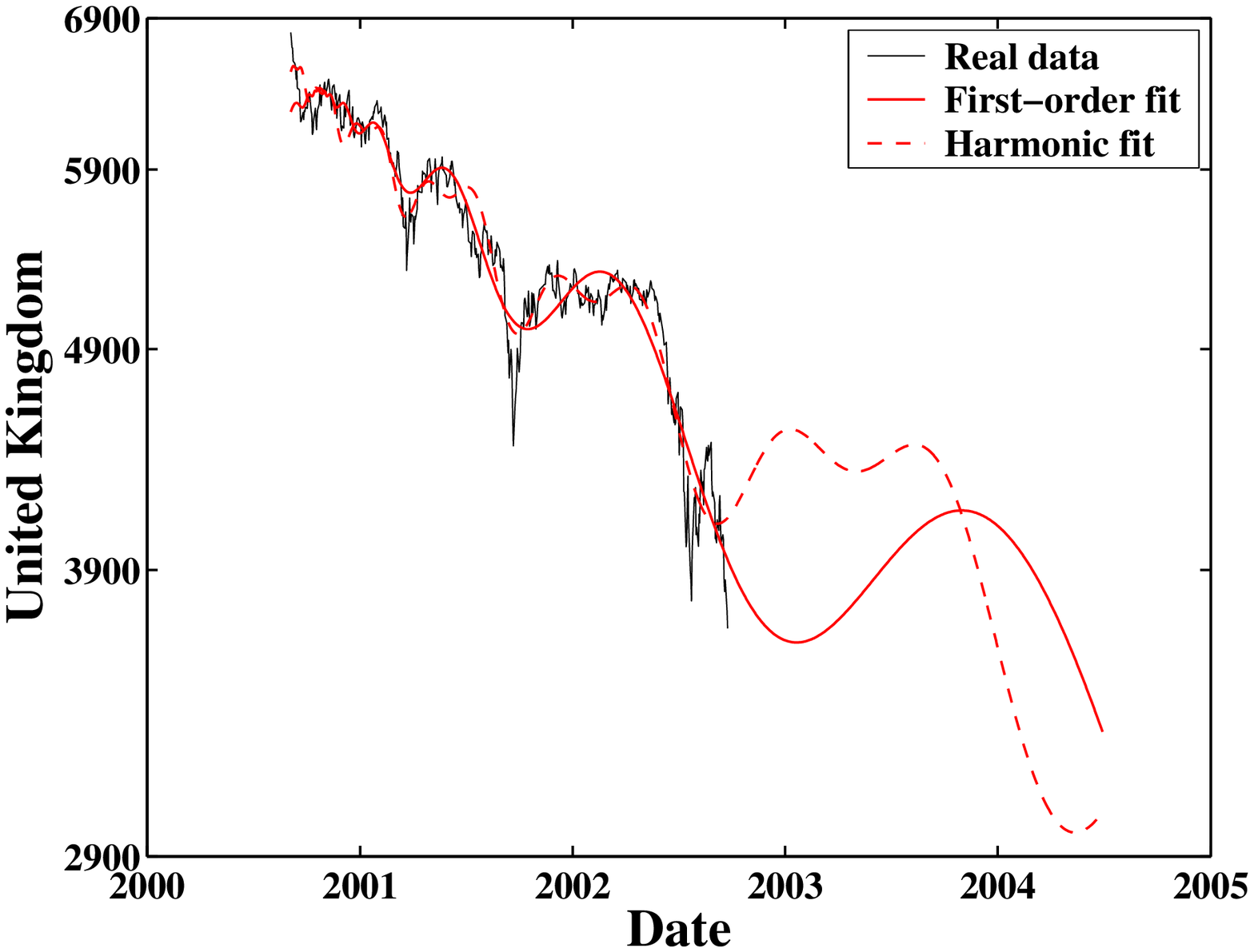,width=15cm, height=12cm}
\end{center}
\caption{The stock market index of UK (fine noisy line)
and its fits with the simple log-periodic formula (\ref{Eq:fit1}) (thick line)
and with formula (\ref{Eq:fit4}) incorporating an harmonics
(dashed line).
The starting
date for the fits is $t_{\rm{start}}=$ 04-Sep-2000. The parameter
values of the fit with (\ref{Eq:fit1}) are $t_c=$ 23-Oct-2000,
$\alpha=1.00$, $\omega=7.58$, $\phi=0.00$, $A=8.77$, $B=-0.00055$,
and $C=-0.00017$. The r.m.s. of the fit errors is 0.0317. The
parameter values of the fit with (\ref{Eq:fit4}) are $t_c=$ 21-Jul-2000,
$\alpha=0.84$, $\omega=10.78$, $\phi_1=3.56$, $\phi_2=1.52$,
$A=8.83$, $B=-0.00153$, $C=-0.00030$, and $D=-0.00016$. The r.m.s.
of the fit errors is 0.0282. The
formula including the harmonics reduces the
r.m.s. of fit errors by 11.2\%.} \label{Fig:IDUK}
\end{figure}

\begin{figure}
\begin{center}
\epsfig{file=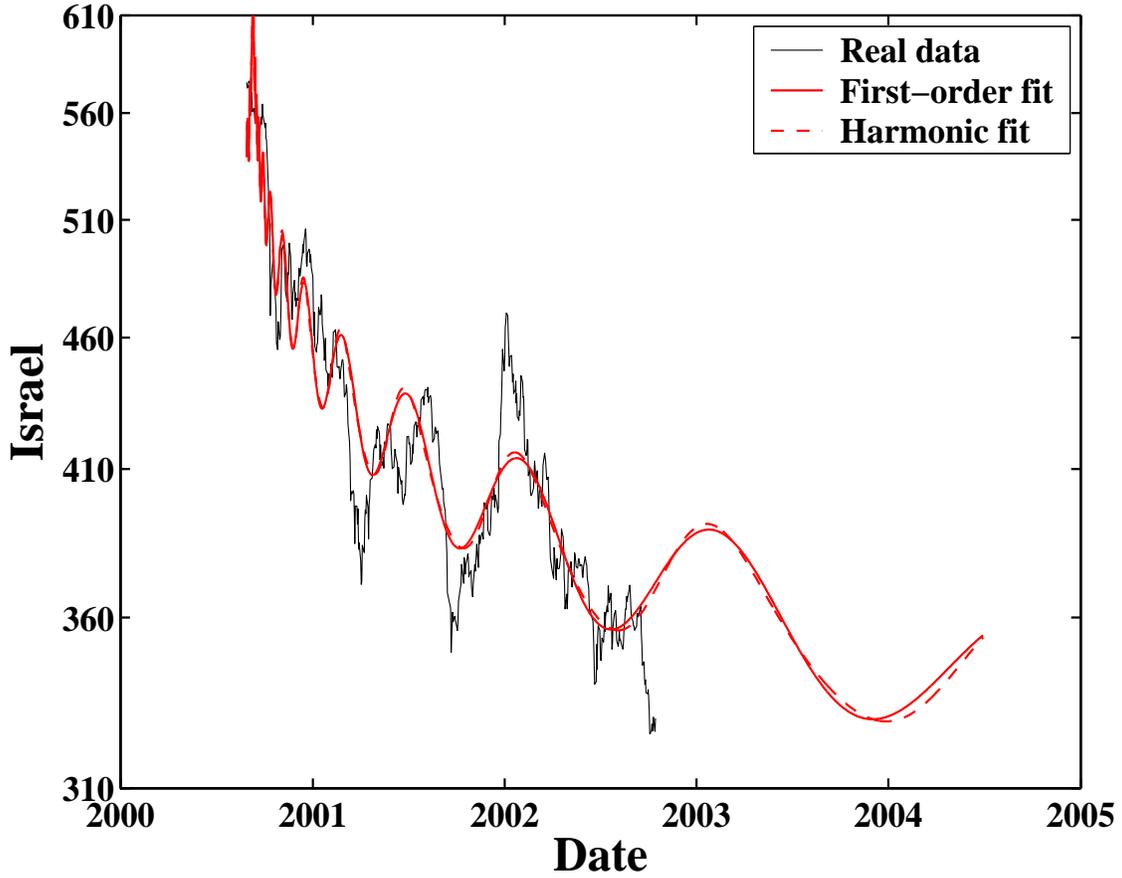,width=15cm, height=12cm}
\end{center}
\caption{The stock market index of Israel (fine noisy line)
and its fits with the simple log-periodic formula (\ref{Eq:fit1}) (thick line)
and with formula (\ref{Eq:fit4}) incorporating an harmonics
(dashed line). The
starting date for the fits is $t_{\rm{start}}=$ 28-Aug-2000. The
parameter values of the fit with (\ref{Eq:fit1}) are $t_c=$ 09-Sep-2000,
$\alpha=0.18$, $\omega=11.45$, $\phi=4.02$, $A=6.61$,
$B=-0.20470$, and $C= 0.01772$. The r.m.s. of the fit errors is
0.0451. The parameter values of the fit with (\ref{Eq:fit4}) are $t_c=$
09-Sep-2000, $\alpha=0.19$, $\omega=11.41$, $\phi_1=4.26$,
$\phi_2=0.00$, $A=6.60$, $B=-0.19459$, $C= 0.01728$, and
$D=-0.00173$. The r.m.s. of the fit errors is 0.0449.
The formula including the harmonics reduces the r.m.s. of fit
errors by  0.4\%.}
\label{Fig:IDIsrael}
\end{figure}

\begin{figure}
\begin{center}
\epsfig{file=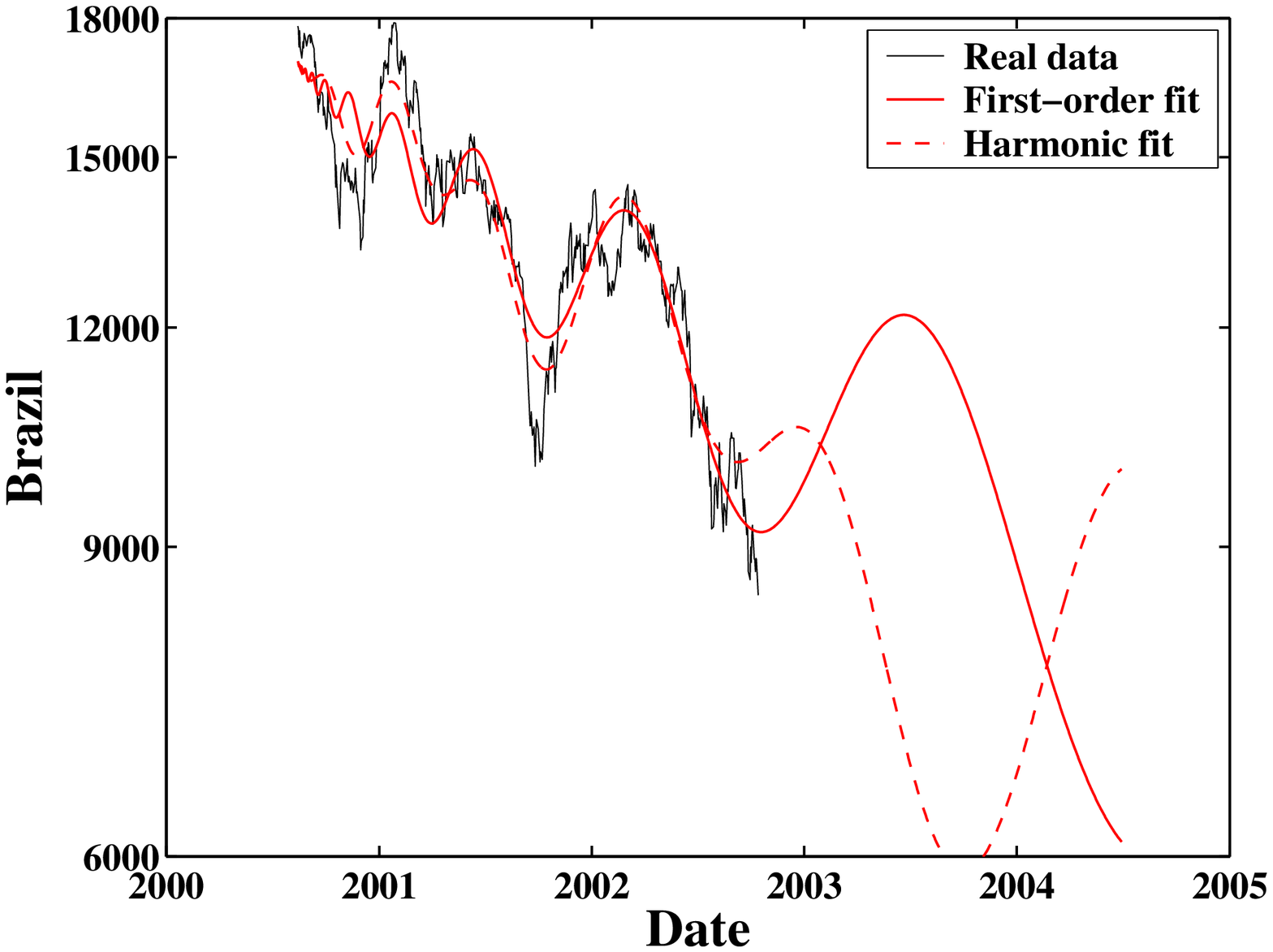,width=15cm, height=12cm}
\end{center}
\caption{The stock market index of Brazil (fine noisy line)
and its fits with the simple log-periodic formula (\ref{Eq:fit1}) (thick line)
and with formula (\ref{Eq:fit4}) incorporating an harmonics
(dashed line). The
starting date for the fits is $t_{\rm{start}}=$ 14-Aug-2000. The
parameter values of the fit with (\ref{Eq:fit1}) are $t_c=$ 12-Aug-2000,
$\alpha=0.87$, $\omega=10.15$, $\phi=4.74$, $A=9.74$,
$B=-0.00136$, and $C= 0.00056$. The r.m.s. of the fit errors is
0.062. The parameter values of the fit with (\ref{Eq:fit4}) are $t_c=$
23-Feb-2000, $\alpha=1.08$, $\omega=7.99$, $\phi_1=6.24$,
$\phi_2=4.69$, $A=9.83$, $B=-0.00034$, $C=-0.00007$, and
$D=-0.00008$. The r.m.s. of the fit errors is 0.0541. The
formula including the harmonics reduces the r.m.s. of fit
errors by 12.8\%.}
\label{Fig:IDBrazil}
\end{figure}

\begin{figure}
\begin{center}
\epsfig{file=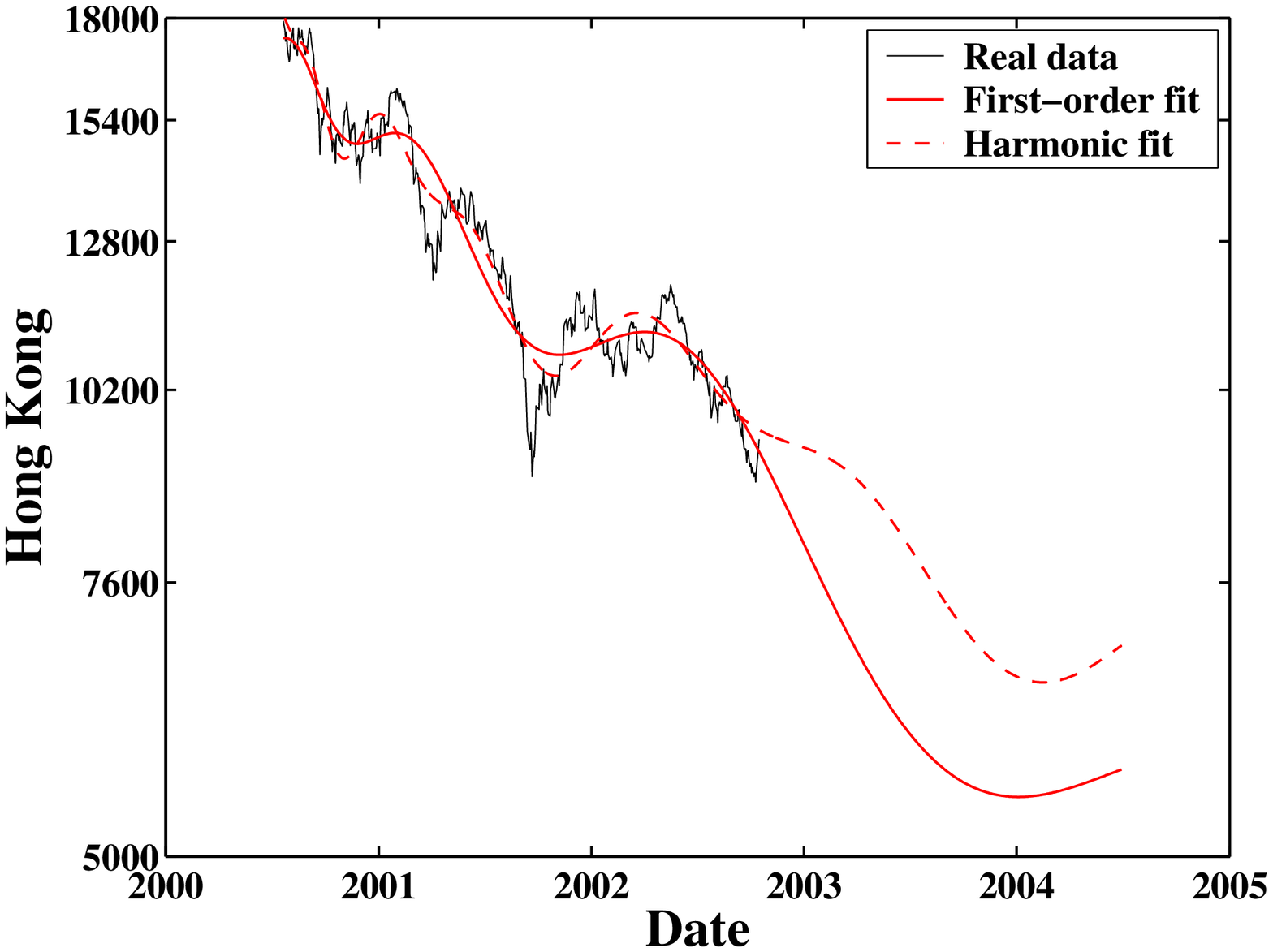,width=15cm, height=12cm}
\end{center}
\caption{The stock market index of Hong-Kong (fine noisy line)
and its fits with the simple log-periodic formula (\ref{Eq:fit1}) (thick line)
and with formula (\ref{Eq:fit4}) incorporating an harmonics
(dashed line). The
starting date for the fits is $t_{\rm{start}}=$ 21-Jul-2000. The
parameter values of the fit with (\ref{Eq:fit1}) are $t_c=$ 26-Feb-2000,
$\alpha=0.90$, $\omega=7.65$, $\phi=1.95$, $A=9.90$, $B=-0.00169$,
and $C=-0.00029$. The r.m.s. of the fit errors is 0.0527. The
parameter values of the fit with (\ref{Eq:fit4}) are $t_c=$ 30-Jan-2000,
$\alpha=0.41$, $\omega=7.52$, $\phi_1=2.21$, $\phi_2=0.41$,
$A=10.39$, $B=-0.07168$, $C=-0.00488$, and $D= 0.00303$. The
r.m.s. of the fit errors is 0.0463. The
formula including the harmonics reduces
the r.m.s. of fit errors by 12.2\%.} \label{Fig:IDHongKong}
\end{figure}

\begin{figure}
\begin{center}
\epsfig{file=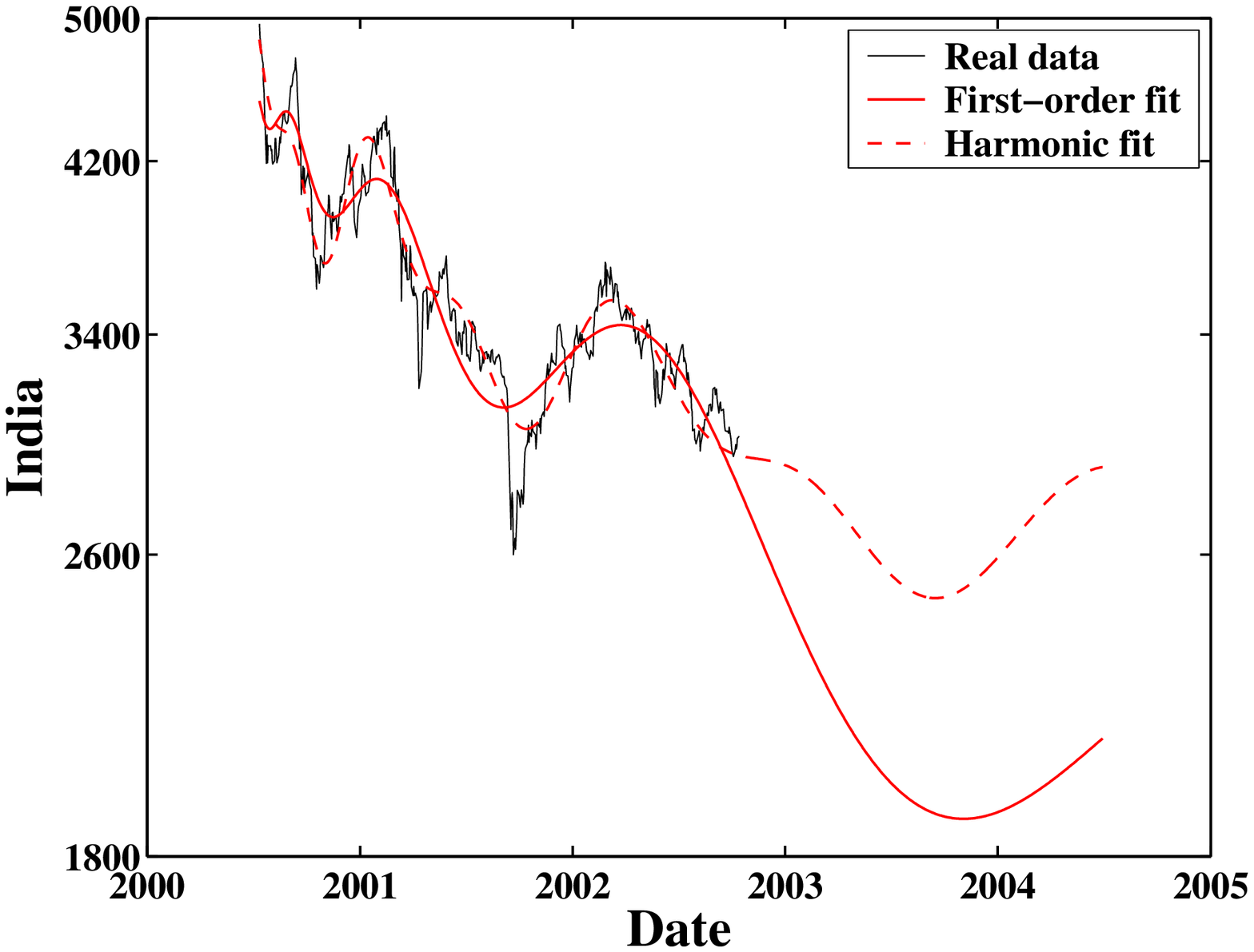,width=15cm, height=12cm}
\end{center}
\caption{The stock market index of India (fine noisy line)
and its fits with the simple log-periodic formula (\ref{Eq:fit1}) (thick line)
and with formula (\ref{Eq:fit4}) incorporating an harmonics
(dashed line). The
starting date for the fits is $t_{\rm{start}}=$ 12-Jul-2000. The
parameter values of the fit with (\ref{Eq:fit1}) are $t_c=$ 26-May-2000,
$\alpha=0.78$, $\omega=6.34$, $\phi=5.61$, $A=8.47$, $B=-0.00292$,
and $C=-0.00085$. The r.m.s. of the fit errors is 0.0489. The
parameter values of the fit with (\ref{Eq:fit4}) are $t_c=$ 04-Dec-1999,
$\alpha=0.03$, $\omega=8.84$, $\phi_1=5.96$, $\phi_2=0.73$,
$A=17.34$, $B=-7.54442$, $C=-0.06297$, and $D= 0.03948$. The
r.m.s. of the fit errors is 0.0412. The
formula including the harmonics reduces
the r.m.s. of fit errors by 15.7\%.} \label{Fig:IDIndia}
\end{figure}

\begin{figure}
\begin{center}
\epsfig{file=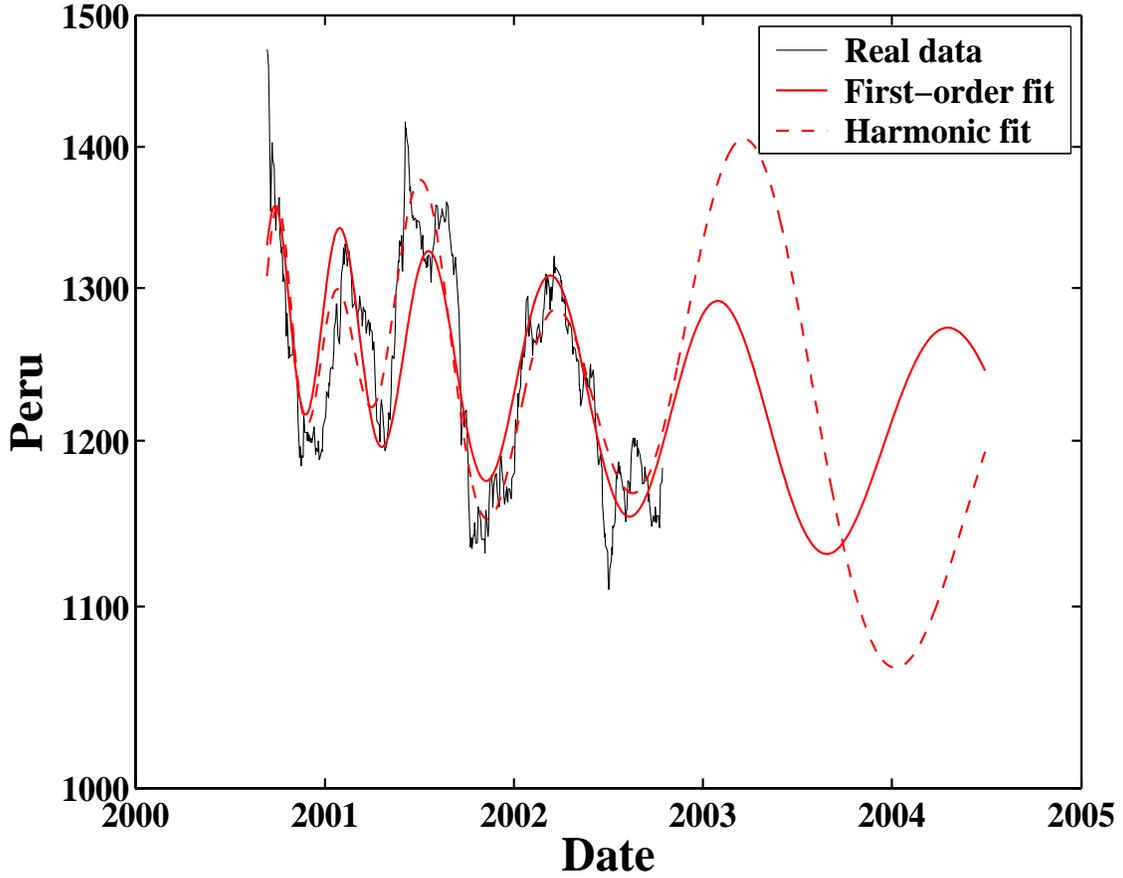,width=15cm, height=12cm}
\end{center}
\caption{The stock market index of Peru (fine noisy line)
and its fits with the simple log-periodic formula (\ref{Eq:fit1}) (thick line)
and with formula (\ref{Eq:fit4}) incorporating an harmonics
(dashed line). The
starting date for the fits is $t_{\rm{start}}=$ 11-Sep-2000. The
parameter values of the fit with (\ref{Eq:fit1}) are $t_c=$ 28-Oct-1999,
$\alpha=0.15$, $\omega=19.80$, $\phi=1.16$, $A=7.45$,
$B=-0.11761$, and $C=-0.02090$. The r.m.s. of the fit errors is
0.0322. The parameter values of the fit with (\ref{Eq:fit4}) are $t_c=$
26-Feb-2000, $\alpha=0.54$, $\omega=7.69$, $\phi_1=2.98$,
$\phi_2=4.99$, $A=7.19$, $B=-0.00167$, $C= 0.00120$, and $D=
0.00192$. The r.m.s. of the fit errors is 0.0276. The
formula including the harmonics reduces the r.m.s. of fit
errors by 14.4\%.}
\label{Fig:IDPeru}
\end{figure}

\begin{figure}
\begin{center}
\epsfig{file=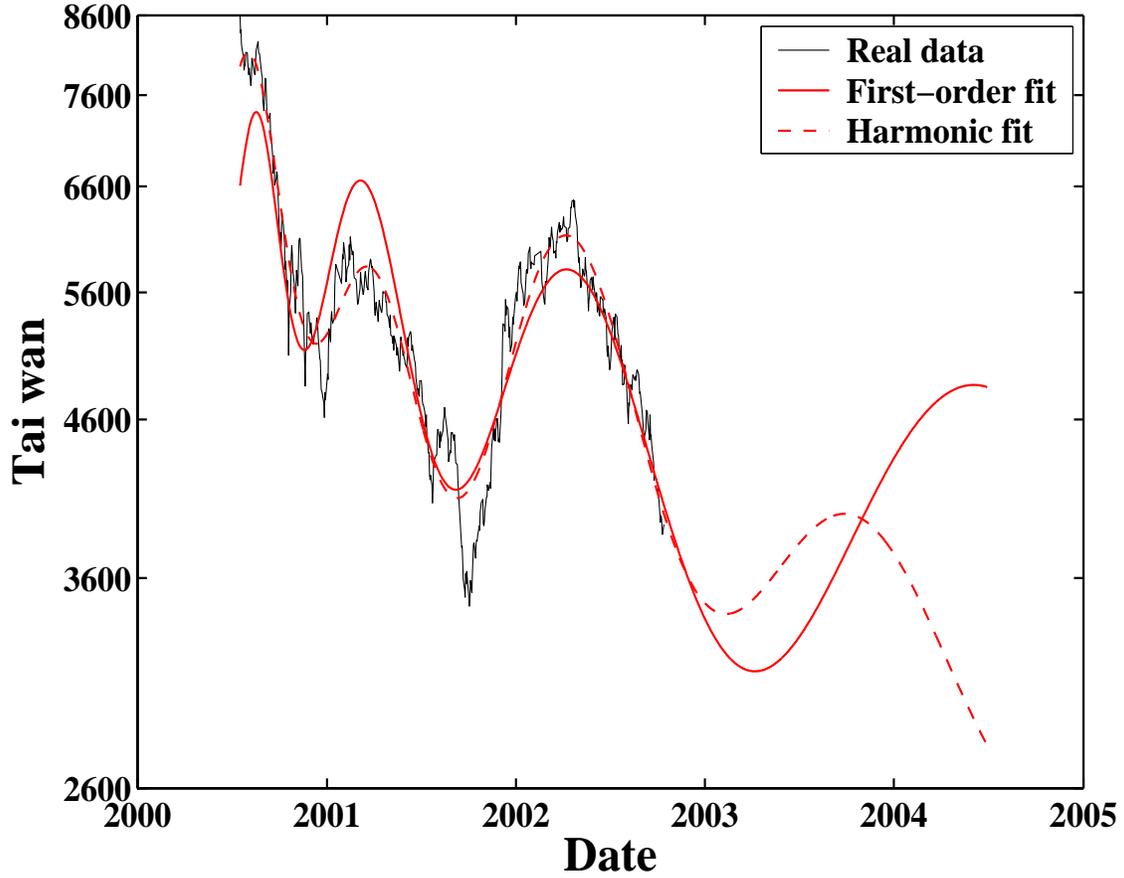,width=15cm, height=12cm}
\end{center}
\caption{The stock market index of Taiwan (fine noisy line)
and its fits with the simple log-periodic formula (\ref{Eq:fit1}) (thick line)
and with formula (\ref{Eq:fit4}) incorporating an harmonics
(dashed line). The
starting date for the fits is $t_{\rm{start}}=$ 17-Jul-2000. The
parameter values of the fit with (\ref{Eq:fit1}) are $t_c=$ 24-Jan-2000,
$\alpha=0.38$, $\omega=9.22$, $\phi=1.01$, $A=9.26$, $B=-0.06302$,
and $C= 0.01769$. The r.m.s. of the fit errors is 0.085. The
parameter values of the fit with (\ref{Eq:fit4}) are $t_c=$ 07-May-1999,
$\alpha=0.31$, $\omega=7.34$, $\phi_1=1.93$, $\phi_2=4.44$,
$A=9.95$, $B=-0.18106$, $C=-0.01451$, and $D= 0.02150$. The r.m.s.
of the fit errors is 0.0541. The
formula including the harmonics reduces the
r.m.s. of fit errors by 36.3\%.} \label{Fig:IDTaiwan}
\end{figure}

\begin{figure}
\begin{center}
\epsfig{file=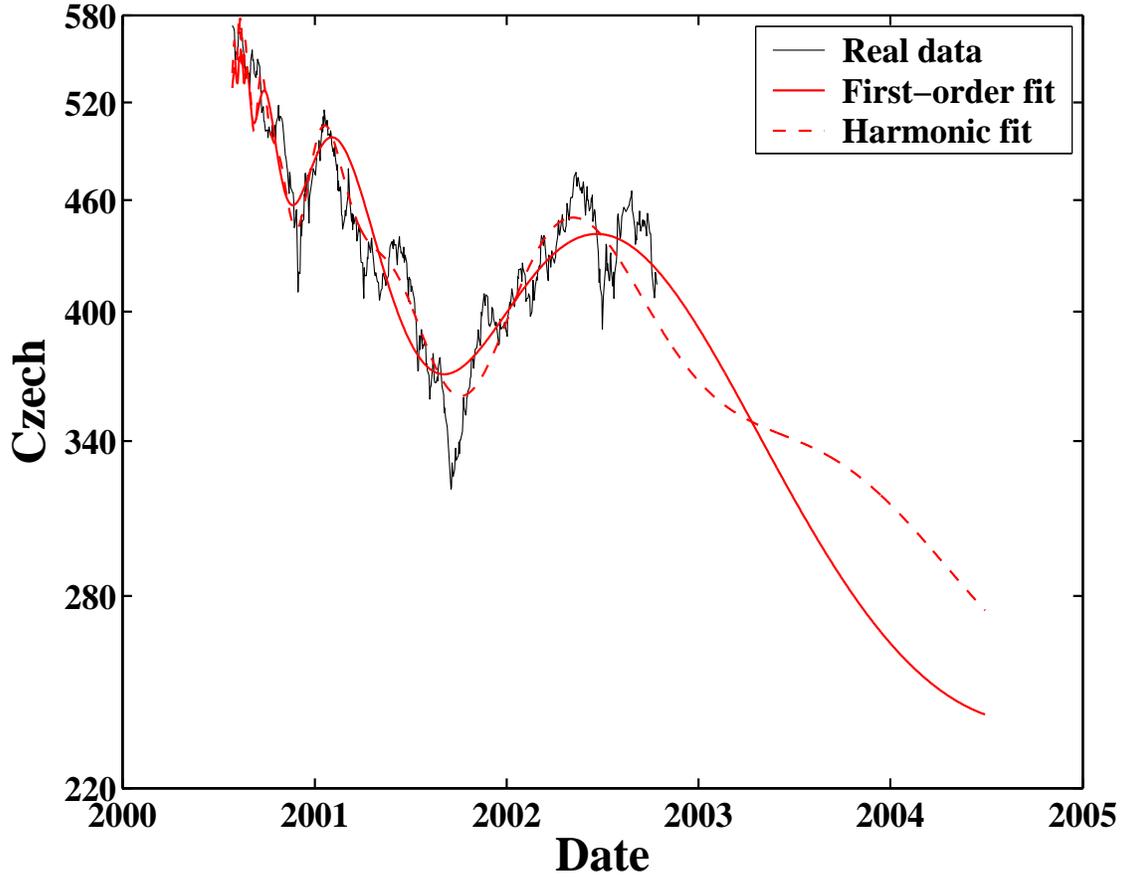,width=15cm, height=12cm}
\end{center}
\caption{The stock market index of Czech republic (fine noisy line)
and its fits with the simple log-periodic formula (\ref{Eq:fit1}) (thick line)
and with formula (\ref{Eq:fit4}) incorporating an harmonics
(dashed line). The
starting date for the fits is $t_{\rm{start}}=$ 28-Jul-2000. The
parameter values of the fit with (\ref{Eq:fit1}) are $t_c=$ 13-Aug-2000,
$\alpha=0.53$, $\omega=4.61$, $\phi=1.19$, $A=6.32$, $B=-0.01246$,
and $C= 0.00520$. The r.m.s. of the fit errors is 0.048. The
parameter values of the fit with (\ref{Eq:fit4}) are $t_c=$ 11-Aug-2000,
$\alpha=0.40$, $\omega=4.59$, $\phi_1=1.27$, $\phi_2=0.77$,
$A=6.39$, $B=-0.03390$, $C= 0.00969$, and $D=-0.00446$. The r.m.s.
of the fit errors is 0.0394. The
formula including the harmonics reduces the
r.m.s. of fit errors by 17.8\%.} \label{Fig:IDCzech}
\end{figure}

\begin{figure}
\begin{center}
\epsfig{file=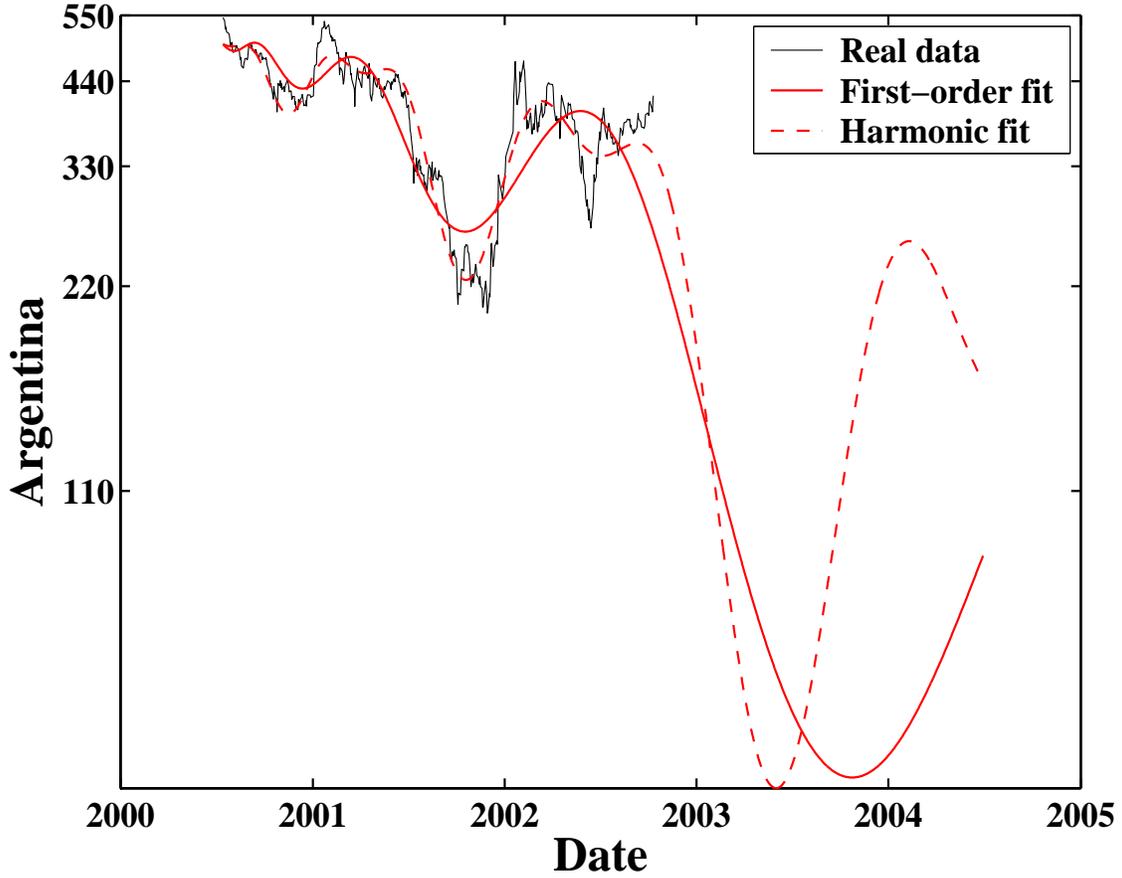,width=15cm, height=12cm}
\end{center}
\caption{The stock market index of Argentina (fine noisy line)
and its fits with the simple log-periodic formula (\ref{Eq:fit1}) (thick line)
and with formula (\ref{Eq:fit4}) incorporating an harmonics
(dashed line). The
starting date for the fits is $t_{\rm{start}}=$ 14-Jul-2000. The
parameter values of the fit with (\ref{Eq:fit1}) are $t_c=$ 02-May-2000,
$\alpha=1.54$, $\omega=7.25$, $\phi=5.28$, $A=6.23$, $B=-0.00003$,
and $C=-0.00002$. The r.m.s. of the fit errors is 0.1223. The
parameter values of the fit with (\ref{Eq:fit4}) are $t_c=$ 23-Aug-1999,
$\alpha=2.00$, $\omega=11.23$, $\phi_1=0.67$, $\phi_2=4.40$,
$A=6.25$, $B=-0.00000$, $C=-0.00000$, and $D= 0.00000$. The r.m.s.
of the fit errors is 0.082. The
formula including the harmonics reduces the
r.m.s. of fit errors by 32.9\%.} \label{Fig:IDArgentina}
\end{figure}

\begin{figure}
\begin{center}
\epsfig{file=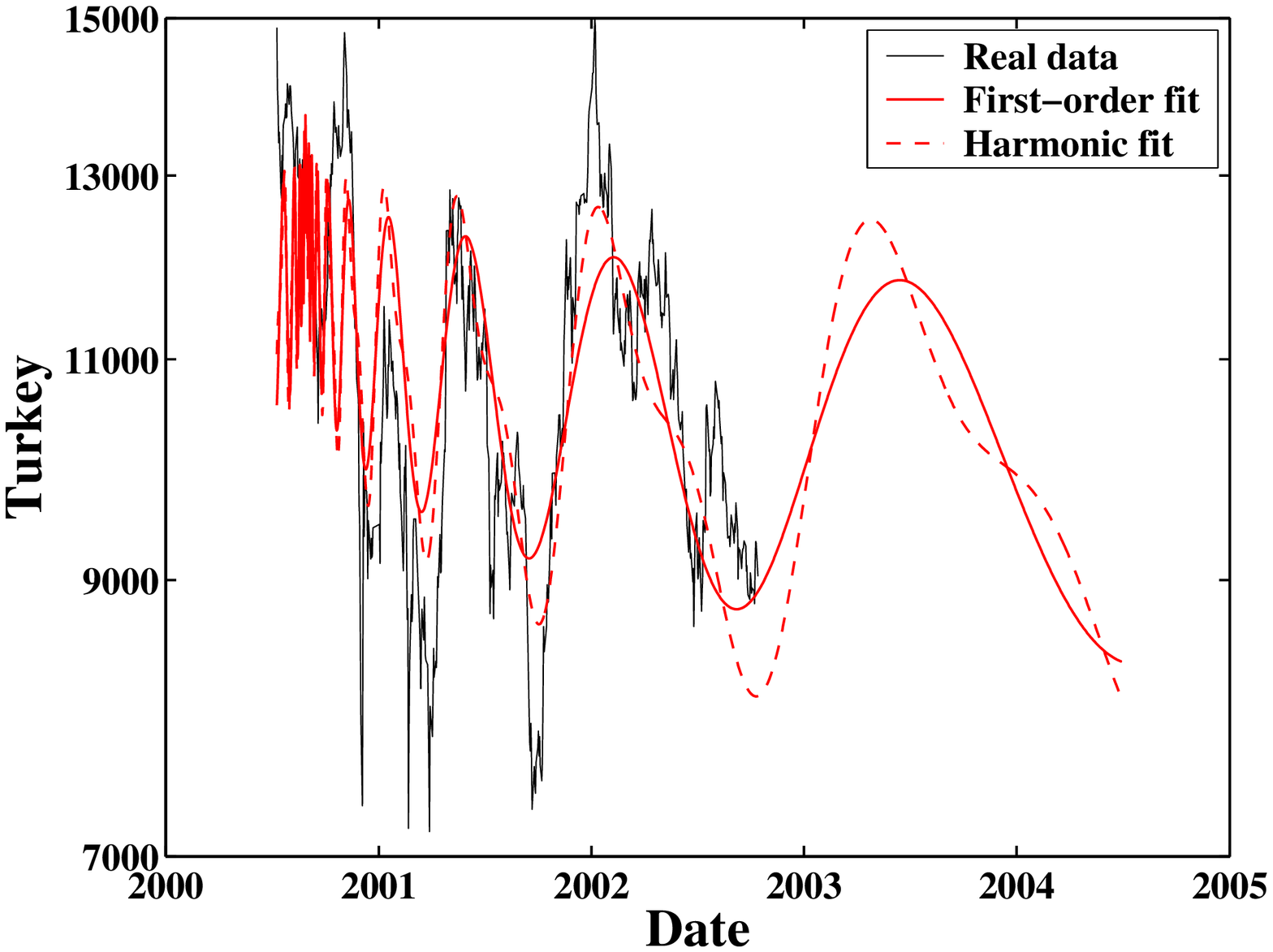,width=15cm, height=12cm}
\end{center}
\caption{The stock market index of Turkey (fine noisy line)
and its fits with the simple log-periodic formula (\ref{Eq:fit1}) (thick line)
and with formula (\ref{Eq:fit4}) incorporating an harmonics
(dashed line). The
starting date for the fits is $t_{\rm{start}}=$ 10-Jul-2000. The
parameter values of the fit with (\ref{Eq:fit1}) are $t_c=$ 28-Aug-2000,
$\alpha=0.13$, $\omega=9.56$, $\phi=2.88$, $A=9.62$, $B=-0.16041$,
and $C= 0.06365$. The r.m.s. of the fit errors is 0.1176. The
parameter values of the fit with (\ref{Eq:fit4}) are $t_c=$ 28-Aug-2000,
$\alpha=0.20$, $\omega=9.55$, $\phi_1=2.94$, $\phi_2=1.32$,
$A=9.52$, $B=-0.07816$, $C= 0.04712$, and $D= 0.01847$. The r.m.s.
of the fit errors is 0.1104. The
formula including the harmonics reduces the
r.m.s. of fit errors by  6.1\%.} \label{Fig:IDTurkey}
\end{figure}

Anti-bubbles are identified in 21 stock market
indices:
Netherlands, France, USA Dow Jones, USA NASDAQ, Japan, Belgium,
Denmark, Germany, Norway, Spain, Switzerland, United Kingdom,
Israel, Brazil, Hong Kong, India, Peru, Taiwan, Czech, Argentina
and Turkey. We refer to these anti-bubble by the term ``bearish'' to stress
that they are fitted by an overall decreasing power law (since
$B<0$ and $\alpha > 0$).
Figures \ref{Fig:IDNetherlands}-\ref{Fig:IDTurkey}
present the fits of these 21 bearish anti-bubbles by expression
(\ref{Eq:fit1}) and (\ref{Eq:fit4}). The
corresponding parameters are listed in Tables \ref{TbBearAB} and
\ref{TbBearAB4}.
Similar bearish anti-bubbles were observed
before in Latin-American, Asian and Western Stock markets
\cite{Nikkei99,emergent}.
We also show the extrapolation of these fits by formulas
(\ref{Eq:fit1}) and (\ref{Eq:fit4}) until mid-2004, in the
spirit of the analysis presented for the USA S\&P500 index \cite{SPpredict}.

These figures show that the log-periodic structures are
very prominent. For
instance, four to five log-periodic oscillations can be identified for
most of the cases presented. However, for some indices, the
log-periodic oscillations close to
$t_c$ are strongly affected by noise for some indices and are less clear-cut.

The majority of predicted critical times $t_c$ for the launch of the
anti-bubbles fall between August and November, 2000: eight in August, two
in September, three in October and one in November.
This is in agreement with the determination $t_c = $ Aug-09-2000
for the S\&P500 index \cite{SPpredict}, suggesting a worldwide
synchronization of the start of a bearish anti-bubble phase.

This analysis, paralleling that presented for the USA S\&P500
index \cite{SPpredict}, suggests that many of the stock markets
shown here are in a phase of recovery that started close to the
last date, September 30, 2002, used to perform the fit. This
recovery is predicted by the extrapolations of (\ref{Eq:fit1}) and
(\ref{Eq:fit4}) to extend until some time in 2003 depending upon
the markets (see the figures) before a recession resumes for a
while. However, we do not claim that these extrapolations should
be valid beyond roughly the end of 2003. The statement applies to
the markets of The Netherlands, France, USA (Dow Jones and
NASDAQ), Belgium, Denmark, Germany, Norway, Spain, Israel, Peru,
and Turkey. We note also that there is sometimes a substantial
difference between the timing of the recovery and the following
recession predicted by expression (\ref{Eq:fit1}) compared with
formula (\ref{Eq:fit4}). This is the case for the stock markets of
the Netherlands, France, the USA Dow Jones, Norway, Spain, and the
United Kingdom. In these cases, one should be careful in
interpreting these extrapolations as reliable forecasts. In some
cases, such as for the United Kingdom and Brazil, the two
extrapolations are so inconsistent as being meaningless. Thus, our
message here is not so much the forecasts but instead the
remarkable consistency in the log-periodicity of these anti-bubble
phases, shown by their common starting dates and similar
structures quantified by the power law exponents and the angular
log-frequencies.

In contrast, the markets of Japan, Switzerland, Hong Kong, India,
Taiwan, Czech and Argentina are extrapolated to continue their
overall descent roughly till the middle of 2003 or beyond before a
recovery sets in. The clear log-periodic structure since August
2000 shown in figure \ref{Fig:IDJapan} for the Japanese Nikkei
index is especially interesting because this structure follows the
large scale anti-bubble log-periodic pattern that started in
January 1990 \cite{Nikkei99} and continued at least until the
beginning of 2000 \cite{evalNikkei}. A possible interpretation of
the novel structure identified in figure \ref{Fig:IDJapan} is that
it is a sub-structure within a hierarchy of log-periodic patterns
occurring at many different scales, as found for instance in
Weierstrass functions (see \cite{Weir} for an interpretation of
Weierstrass functions and their generalizations in terms of
log-periodicity at many different scales) and suggested in
\cite{Drozdz}. We can expect more generally that similar
multiscale log-periodicity should exist in other markets. However,
these fine structures especially at the smaller scales (three
months, monthly, weekly, intraday, etc.) are greatly effected or
even spoiled by the intrinsically noisy nature of stock market
prices, due to the fact that many more effects contribute
potentially at small scales to scramble possible signals. Only at
the large time scales studied here can the cooperative behavior of
investors be systematically observed.

Another important observation is that the log-periodic
oscillations and power law decays are distinctly different with a
smaller number of oscillations and much larger ``noise'' for
Brazil, Hong-Kong, India, Peru, Taiwan, Czech republic, Argentina
and Turkey compared with the others. This may be explained by the
presence of stronger idiosyncratic influences, such as local
crises in South America. For these markets, the two extrapolations
obtained from expressions (\ref{Eq:fit1}) and (\ref{Eq:fit4})
diverge rapidly away from each other, making them quite
unreliable.

As shown in Table \ref{TbBearAB}, the power law exponents $\alpha$ of the
indices of Belgium and Argentina are significantly larger than $1$,
while those of Netherland, USA Dow Jones, Germany, Norway, Switzerland
and United Kingdom are close to or slightly greater than $1$.
In absence of the log-periodic oscillations, this would mean that
the overall shape of these indices would be concave
(downward plunging) rather
than convex (upward curvature) as they would be for $0<\alpha<1$.
Large values of $\alpha>1$ implies a steep downward overall
acceleration of the index. But in all cases when this occurs, this
is compensated by a large amplitude of the log-periodic oscillations.
In contrast, for
$0<\alpha<1$, the index initially drops fast in the early times
of the anti-bubble and then decelerates and approaches
a constant level at long times.

\begin{figure}
\begin{center}
\epsfig{file=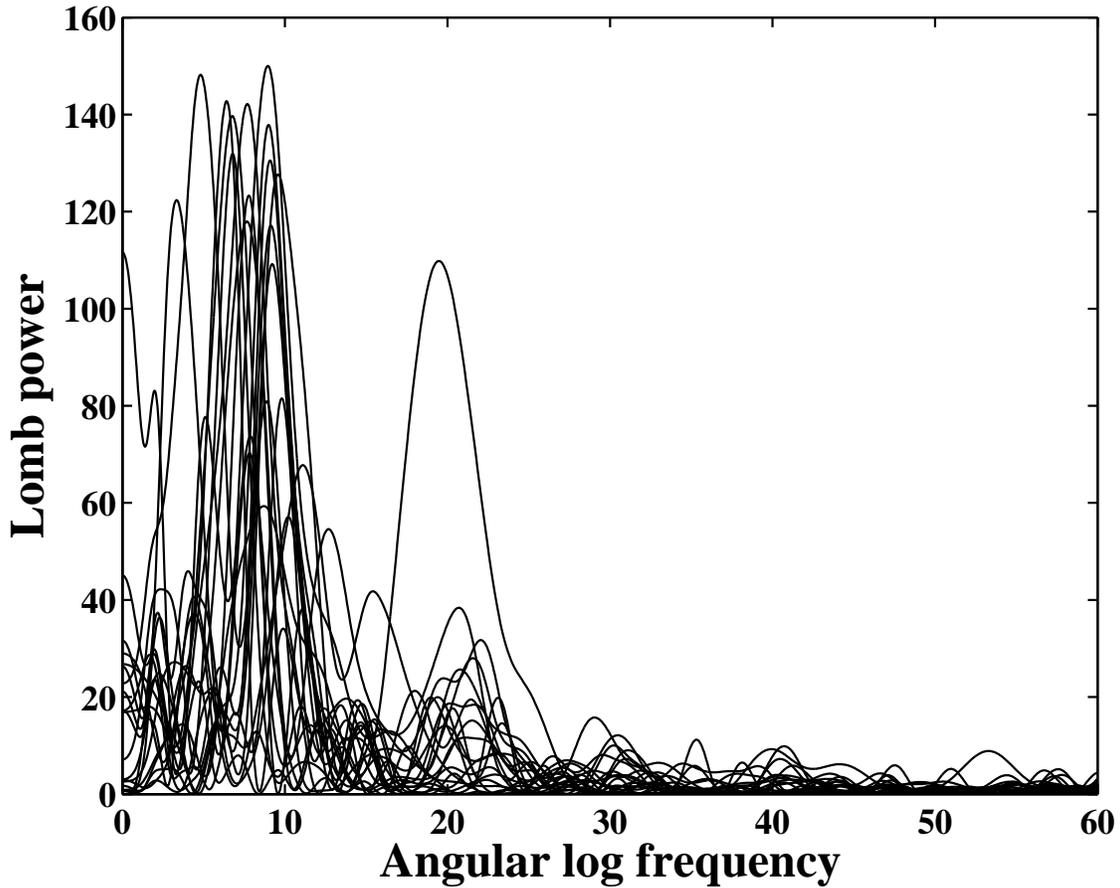,width=15cm, height=12cm}
\end{center}
\caption{Lomb periodograms for the 21 stock market indices shown
in previous figures as a function of the angular log-periodic
frequency $\omega$. These 21 stock market indices exhibit a
so-called bearish anti-bubble regime characterized by $B<0$.
The number of points in the Lomb analysis
is in the range 350-530, which implies that
the log-periodic signals are very significant
for most markets with high Lomb peaks. The relevance of the
log-periodicity is reflected in the quasi-universal value of
the angular log-periodic frequency found in the range $7-10$.
Notice also the presence of an harmonic at $2\omega$ in
many of the markets outlined by the cluster of secondary peaks
in the vicinity of $\omega=20$.}
\label{Fig:IDLomb}
\end{figure}

To quantify the significance level of the log-periodic
oscillations in these 21 anti-bubbles,
we adopt the Lomb analysis \cite{Press} on the residuals of the
logarithm of the indices by removing the power law
\cite{emergent}:
\begin{equation}
r(t) = {\ln p(t)- A - B\tau^\alpha \over C\tau^\alpha},
\label{Eq:res}
\end{equation}
where $\tau$ is defined in (\ref{Eq:taua}). If the log-periodic
formula is a correct representation of these indices, $r(t)$ should
be a pure cosine as a function of $\ln \tau$. Thus, a spectral
analysis of $r(t)$ as a function of the variable $\ln \tau$
should be a strong power peak.
Figure \ref{Fig:IDLomb}
presents the corresponding Lomb periodograms for
all 21 indices described in table \ref{TbBearAB} and shown
in figures \ref{Fig:IDNetherlands}-\ref{Fig:IDTurkey}.
Most of the Lomb spectral
peaks give a very significant signal of the existence of
log-periodic structures \cite{Zhoustatsi}.

In addition, an harmonic of a fundamental angular
log-frequency $\omega$ is visible at $2\omega$ in the Lomb
periodogram for many of the markets, as found previously for the
S\&P 500 index \cite{SPpredict}. This is the justification for
including an harmonic log-periodic oscillatory term
according to (\ref{Eq:fit4}). Table \ref{TbBearAB4} lists the
corresponding parameters of the fits of the 21 stock market
indices with expression (\ref{Eq:fit4}) and shows that using of
formula (\ref{Eq:fit4}) reduces the r.m.s. errors strongly for
most of the indexes. Only for Israel is the improvement of the fit
ambiguous. In section \ref{s2:wilk}, we shall come back to this
issue and provide rigorous and objective statistical tests on the
relevance of log-periodicity with a single angular log-frequency
$\omega$ and with the addition of its harmonics at
$2\omega$.

\subsection{Bullish anti-bubbles}

\begin{figure}
\begin{center}
\epsfig{file=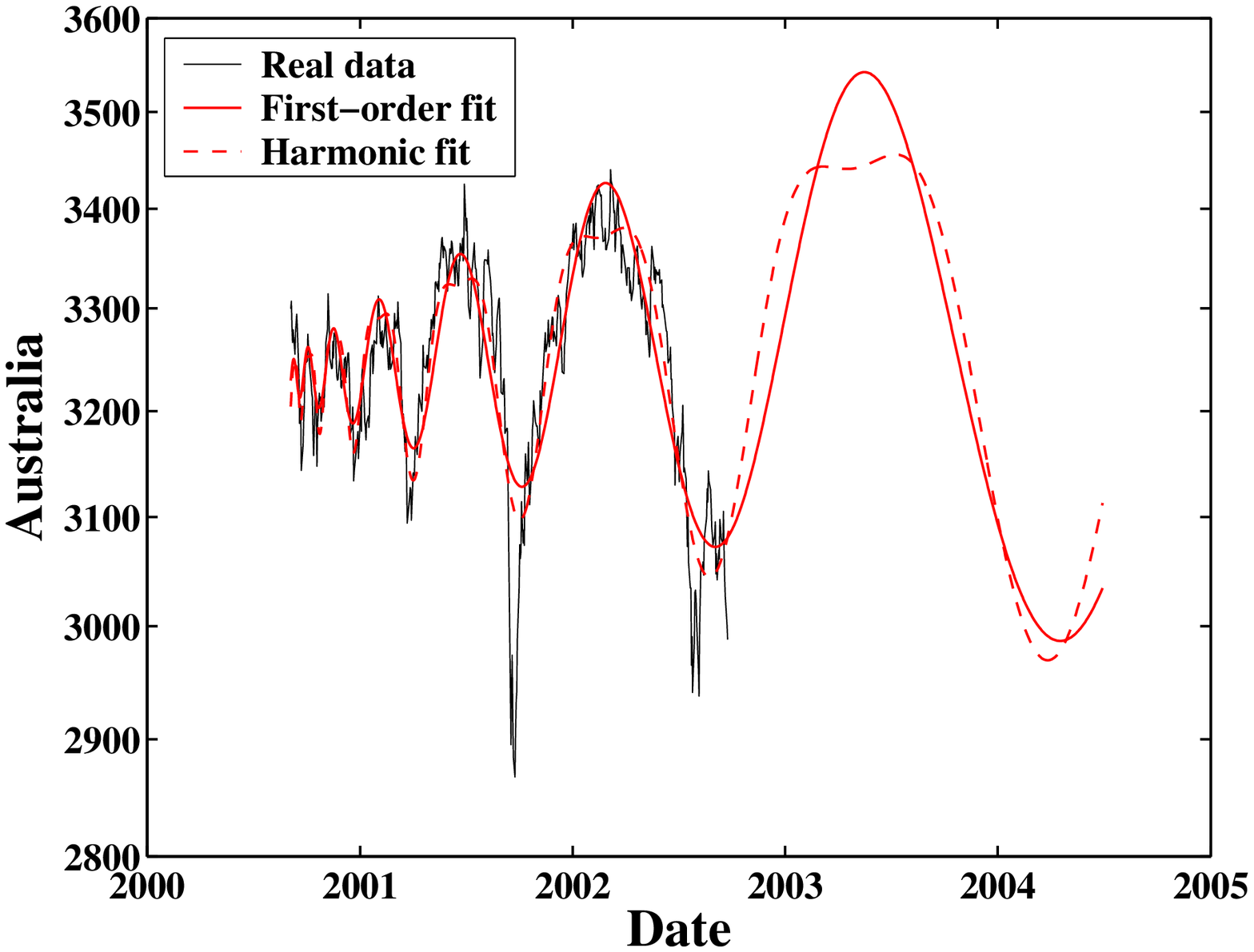,width=15cm, height=12cm}
\end{center}
\caption{The stock market index of Australia (fine line)
and its fits with the simple log-periodic formula (\ref{Eq:fit1}) (thick line)
and with formula (\ref{Eq:fit4}) incorporating an harmonics
(dashed line). The
starting date for the fits is $t_{\rm{start}}=$ 04-Sep-2000
and the ending date of the fit is September, 30, 2002. The
parameter values of the fit with (\ref{Eq:fit1}) are $t_c=$ 08-Aug-2000,
$\alpha=0.77$, $\omega=10.85$, $\phi=3.55$, $A=8.08$, $B=
0.00007$, and $C=-0.00037$. The r.m.s. of the fit errors is
0.0176. The parameter values of the fit with (\ref{Eq:fit4}) are $t_c=$
10-Aug-2000, $\alpha=0.64$, $\omega=10.89$, $\phi_1=3.42$,
$\phi_2=0.51$, $A=8.08$, $B= 0.00026$, $C=-0.00080$, and
$D=-0.00026$. The r.m.s. of the fit errors is 0.0153. The
formula including the harmonics reduces the r.m.s. of fit
errors by 13.2\%.}
\label{Fig:IDAustralia}
\end{figure}

\begin{figure}
\begin{center}
\epsfig{file=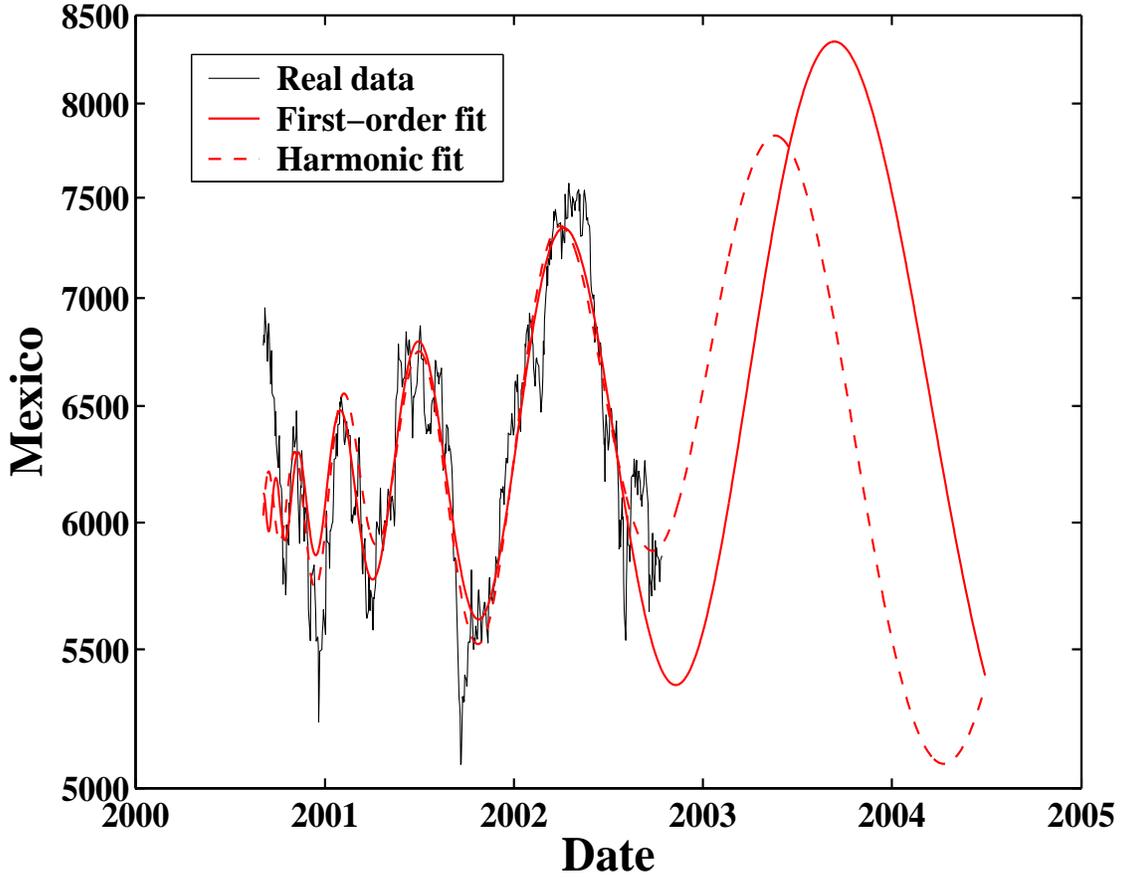,width=15cm, height=12cm}
\end{center}
\caption{The stock market index of Mexico (fine line)
and its fits with the simple log-periodic formula (\ref{Eq:fit1}) (thick line)
and with formula (\ref{Eq:fit4}) incorporating an harmonics
(dashed line). The
starting date for the fits is $t_{\rm{start}}=$ 04-Sep-2000. The
parameter values of the fit with (\ref{Eq:fit1}) are $t_c=$ 08-Aug-2000,
$\alpha=0.80$, $\omega=10.11$, $\phi=4.42$, $A=8.70$, $B=
0.00032$, and $C= 0.00086$. The r.m.s. of the fit errors is
0.0402. The parameter values of the fit with (\ref{Eq:fit4}) are $t_c=$
27-Jun-2000, $\alpha=0.73$, $\omega=5.97$, $\phi_1=4.23$,
$\phi_2=1.65$, $A=8.69$, $B= 0.00072$, $C= 0.00039$, and
$D=-0.00115$. The r.m.s. of the fit errors is 0.0391. The
formula including the harmonics reduces the r.m.s. of fit
errors by  2.6\%.}
\label{Fig:IDMexico}
\end{figure}

\begin{figure}
\begin{center}
\epsfig{file=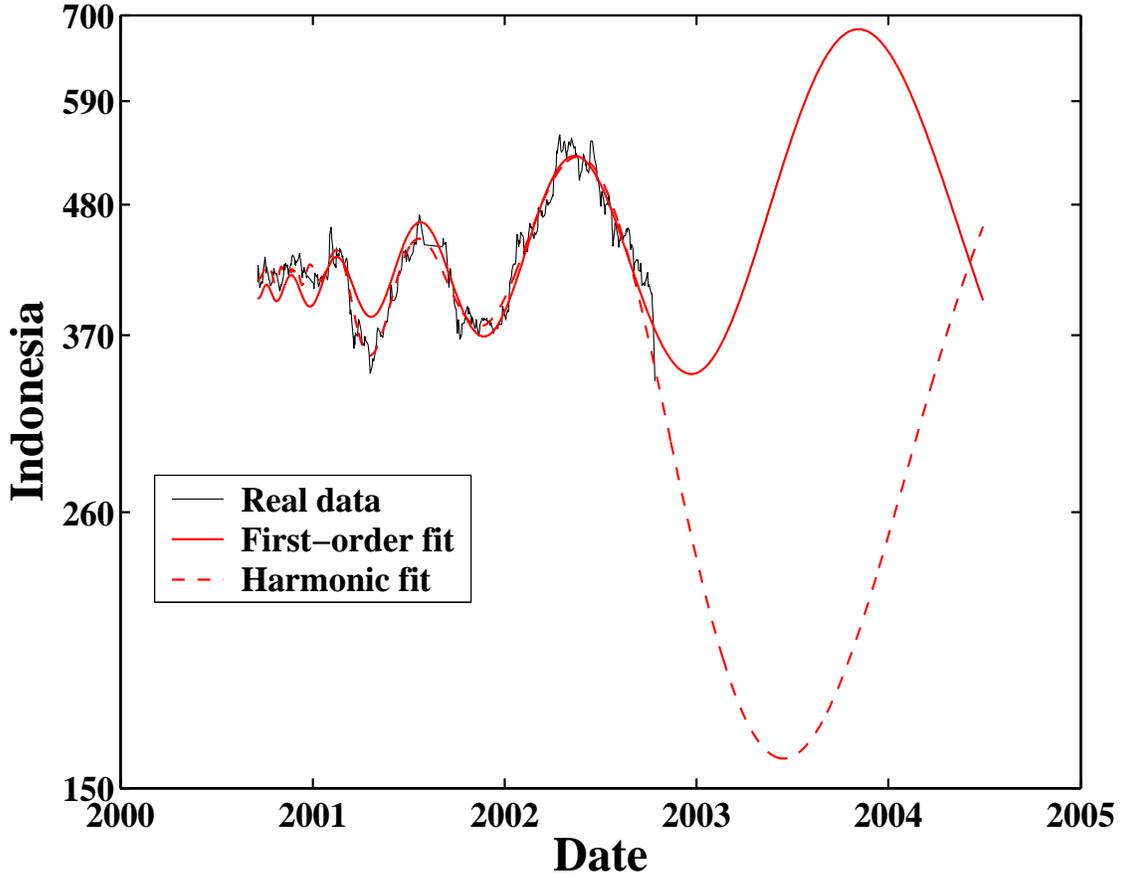,width=15cm, height=12cm}
\end{center}
\caption{The stock market index of Indonesia (fine line)
and its fits with the simple log-periodic formula (\ref{Eq:fit1}) (thick line)
and with formula (\ref{Eq:fit4}) incorporating an harmonics
(dashed line). The
starting date for the fits is $t_{\rm{start}}=$ 16-Sep-2000. The
parameter values of the fit with (\ref{Eq:fit1}) are $t_c=$ 09-Aug-2000,
$\alpha=1.06$, $\omega=10.30$, $\phi=5.83$, $A=5.99$, $B=
0.00009$, and $C=-0.00021$. The r.m.s. of the fit errors is 0.039.
The parameter values of the fit with (\ref{Eq:fit4}) are $t_c=$
16-Nov-2000, $\alpha=0.96$, $\omega=3.47$, $\phi_1=1.33$,
$\phi_2=3.10$, $A=6.04$, $B=-0.00025$, $C=-0.00050$, and
$D=-0.00062$. The r.m.s. of the fit errors is 0.0271. The
formula including the harmonics reduces the r.m.s. of fit
errors by 30.4\%.}
\label{Fig:IDIndonesia}
\end{figure}

\begin{figure}
\begin{center}
\epsfig{file=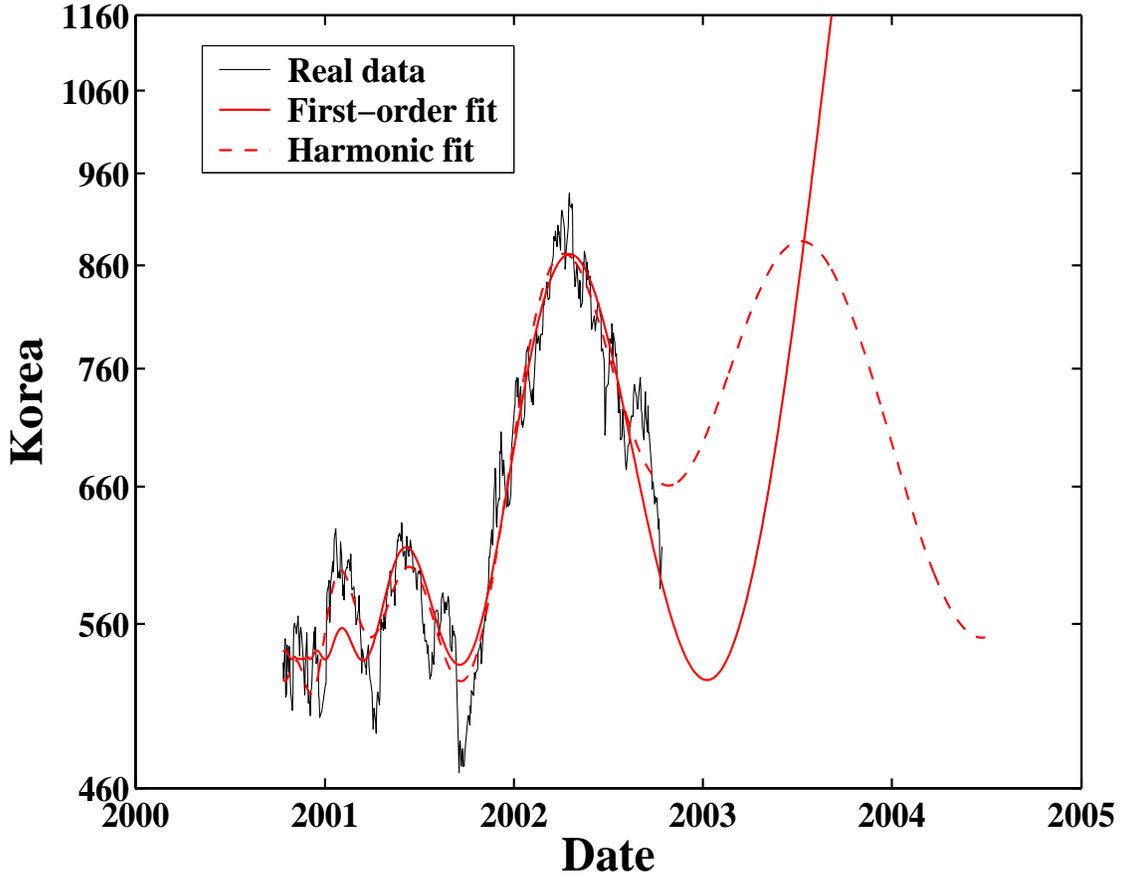,width=15cm, height=12cm}
\end{center}
\caption{The stock market index of Korea (fine line)
and its fits with the simple log-periodic formula (\ref{Eq:fit1}) (thick line)
and with formula (\ref{Eq:fit4}) incorporating an harmonics
(dashed line). The
starting date for the fits is $t_{\rm{start}}=$ 12-Oct-2000. The
parameter values of the fit with (\ref{Eq:fit1}) are $t_c=$ 15-Nov-2000,
$\alpha=1.37$, $\omega=6.69$, $\phi=5.67$, $A=6.29$, $B= 0.00005$,
and $C=-0.00005$. The r.m.s. of the fit errors is 0.05. The
parameter values of the fit with (\ref{Eq:fit4}) are $t_c=$ 10-Aug-2000,
$\alpha=1.00$, $\omega=5.04$, $\phi_1=1.90$, $\phi_2=1.90$,
$A=6.24$, $B= 0.00046$, $C=-0.00023$, and $D=-0.00025$. The r.m.s.
of the fit errors is 0.0428. The
formula including the harmonics reduces the
r.m.s. of fit errors by 14.4\%.} \label{Fig:IDKorea}
\end{figure}

\begin{figure}
\begin{center}
\epsfig{file=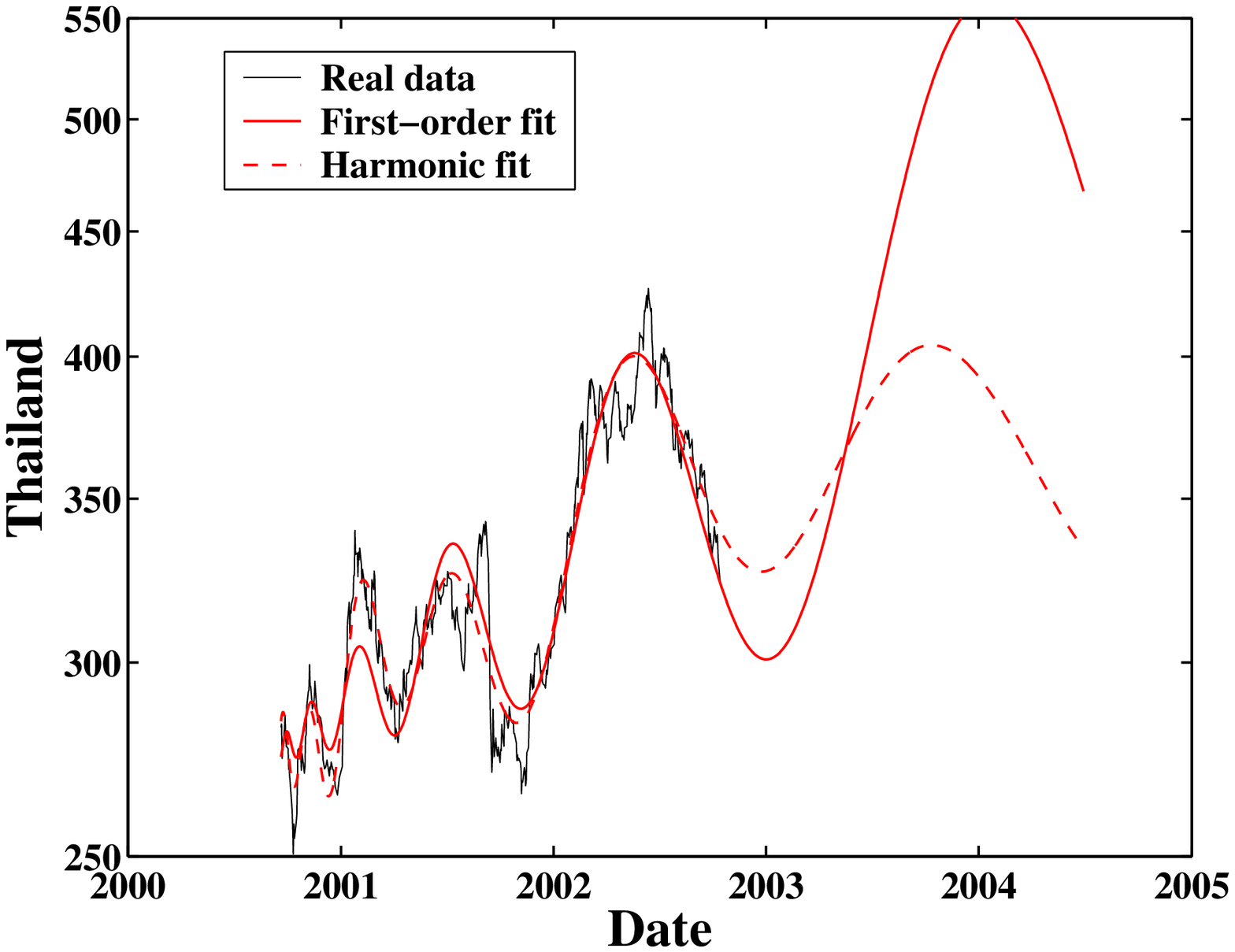,width=15cm, height=12cm}
\end{center}
\caption{The stock market index of Thailand (fine line)
and its fits with the simple log-periodic formula (\ref{Eq:fit1}) (thick line)
and with formula (\ref{Eq:fit4}) incorporating an harmonics
(dashed line). The
starting date for the fits is $t_{\rm{start}}=$ 20-Sep-2000. The
parameter values of the fit with (\ref{Eq:fit1}) are $t_c=$ 16-Aug-2000,
$\alpha=0.93$, $\omega=9.45$, $\phi=5.16$, $A=5.61$, $B= 0.00058$,
and $C=-0.00039$. The r.m.s. of the fit errors is 0.0437. The
parameter values of the fit with (\ref{Eq:fit4}) are $t_c=$ 27-Jul-2000,
$\alpha=0.43$, $\omega=5.18$, $\phi_1=3.62$, $\phi_2=2.22$,
$A=5.48$, $B= 0.02083$, $C= 0.00390$, and $D= 0.00662$. The r.m.s.
of the fit errors is 0.0354. The
formula including the harmonics reduces the
r.m.s. of fit errors by 19.0\%.} \label{Fig:IDThailand}
\end{figure}

\begin{figure}
\begin{center}
\epsfig{file=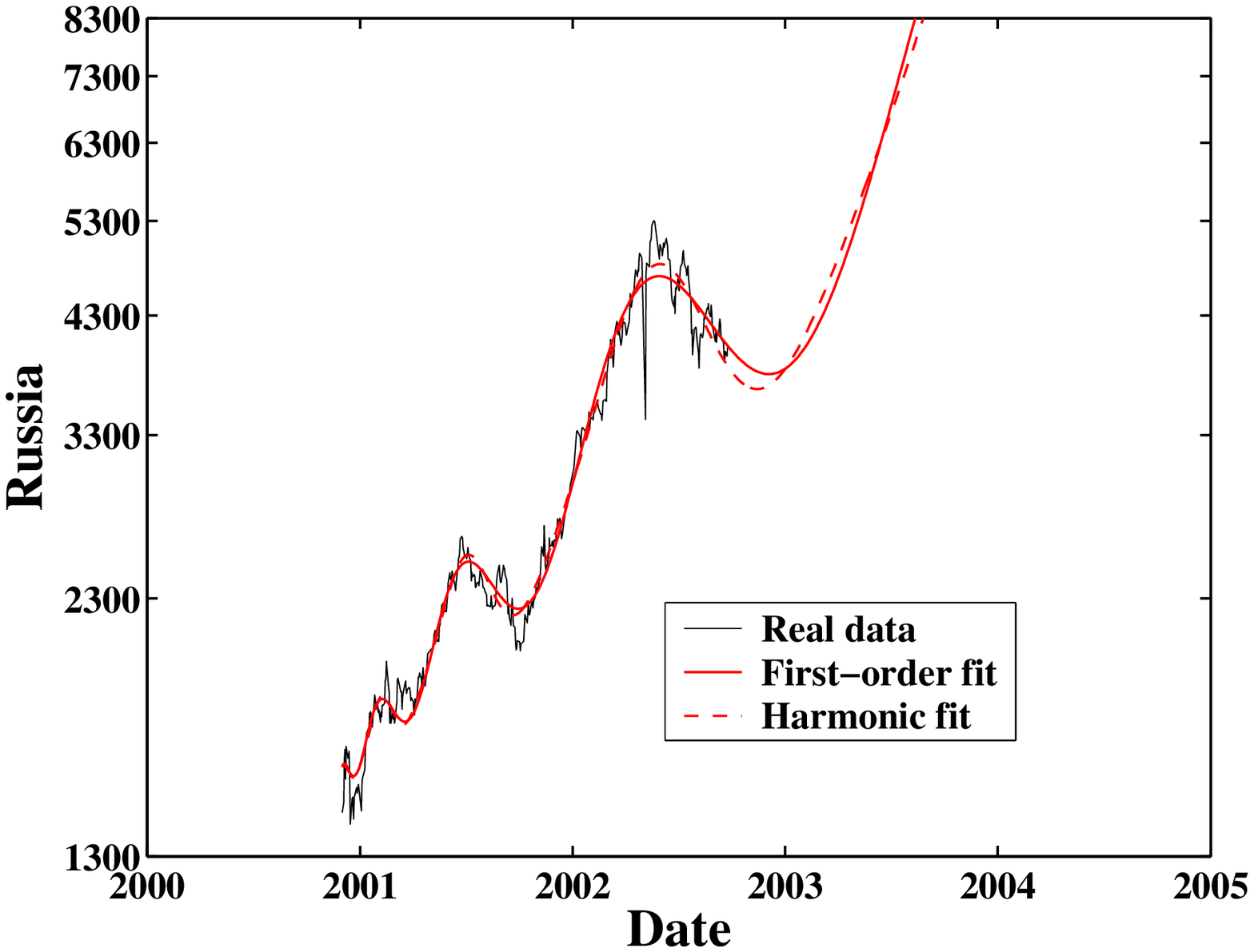,width=15cm, height=12cm}
\end{center}
\caption{The stock market index of Russia (fine line)
and its fits with the simple log-periodic formula (\ref{Eq:fit1}) (thick line)
and with formula (\ref{Eq:fit4}) incorporating an harmonics
(dashed line). The
starting date for the fits is $t_{\rm{start}}=$ 01-Dec-2000. The
parameter values of the fit with (\ref{Eq:fit1}) are $t_c=$ 08-Oct-2000,
$\alpha=0.92$, $\omega=7.93$, $\phi=3.33$, $A=7.24$, $B= 0.00276$,
and $C=-0.00068$. The r.m.s. of the fit errors is 0.0485. The
parameter values of the fit with (\ref{Eq:fit4}) are $t_c=$ 12-Oct-2000,
$\alpha=0.91$, $\omega=7.85$, $\phi_1=3.90$, $\phi_2=0.17$,
$A=7.23$, $B= 0.00307$, $C=-0.00075$, and $D= 0.00010$. The r.m.s.
of the fit errors is 0.0457. The
formula including the harmonics reduces the
r.m.s. of fit errors by  5.7\%.} \label{Fig:IDRussia}
\end{figure}

Figures \ref{Fig:IDAustralia}-\ref{Fig:IDRussia} present the fits
of six stock market indices (Australia, Mexico, Indonesia, South
Korea, Thailand and Russia) using formula (\ref{Eq:fit1}) and
(\ref{Eq:fit4}), with the corresponding parameters listed in Table
\ref{TbBullAB}. We have separated these 6 markets from the 21
previous ones because their fits with formulas (\ref{Eq:fit1}) and
(\ref{Eq:fit4}) give a positive coefficient $B$, corresponding to
an overall increasing market at large time scales. We thus call
them ``bullish'' to describe this overall increasing pattern. We
keep the terminology ``anti-bubble'' to refer to the fact that the
log-periodic oscillations are decelerating.  To the best of our
knowledge, the identification of such bullish anti-bubbles is
performed here for the first time. Notice that there are five
log-periodic oscillations for Australia, Indonesia and Thailand,
four for and Mexico and approximately three for South Korea and
Russia as can be seen in the figures
\ref{Fig:IDAustralia}-\ref{Fig:IDRussia}. This means that the
log-periodic structures in these stock market indices are quite
significant and convincing \cite{Zhoustatsi}. Table \ref{TbBullAB}
shows that the predicted critical times $t_c$ of the start of
these bullish anti-bubbles are again between August and November,
2000.

We have also performed a fit of these 6 stock market indices with
expression (\ref{Eq:fit4}) which accounts for the possible
presence of an harmonic log-periodic oscillatory term. The
fits are plotted in Figs.~\ref{Fig:IDAustralia}-\ref{Fig:IDRussia}
as dashed lines, whose parameters are presented in Table
\ref{TbBullAB4}. We find that the improvement of the fits using
expression (\ref{Eq:fit4}) compared with (\ref{Eq:fit1})
is very significant for Australia, Korea, Indonesia and Thailand.

\subsection{Statistical test of the log-periodic term\label{s2:wilk}}

Since expression (\ref{Eq:fit4}) contains formula (\ref{Eq:fit1})
as the special case $D=0$, we can use Wilk's theorem \cite{Rao}
and the statistical methodology of nested hypotheses to assess
whether the hypothesis that $D=0$ can be rejected. Similarly, we
can also test if $C=0$ can be rejected in (\ref{Eq:fit1}). We
consider the following three hypotheses.
\begin{enumerate}
\item{$H_0$: $C=0$, corresponding to use a simple and pure power law to fit the
stock market indices;}
\item{$H_1$: $D=0$, corresponding to the log-periodic function (\ref{Eq:fit1})
without any harmonics;}
\item{$H_2$: $D\ne0$,
corresponding to the log-periodic function (\ref{Eq:fit4})
which includes an harmonics at $2\omega$.}
\end{enumerate}
Our tests presented below show that $H_0$ can be rejected with
certainty in favor of $H_1$ for all the indexes which have been
analyzed. Moreover, we find that $H_1$ can be rejected in favor of
$H_2$ with high statistical significance for all except one index
(rejection level of $10^{-4}\%$). We stress that Wilk's
methodology of nested hypothesis testing automatically takes into
account the competition between (i) the improved fit obtained by
adding fitting parameters and the ``cost in parsimony'' of adding
these parameters.

The method proceeds as follows (see \cite{SorJoh01,SPpredict} for
recent implementations in similar contexts). Assuming a
Gaussian distribution of observational errors (residuals) at each
data point, the maximum likelihood estimation of the parameters
amounts exactly to the minimization of the sum of the square over
all data points (of number $n$) of the differences $\delta_j(i)$
between the mathematical formula and the data \cite{Press}. The
standard deviation $\sigma_j$ for hypothesis $H_j$ with $j=0,1,2$ of
the fits to the
data associated with (\ref{Eq:fit1}) and
(\ref{Eq:fit4}) is given by $1/n$ times the sum of the squares
over all data points of the differences $\delta_j^{(o)}(i)$
between the mathematical formula and the data, estimated for the
optimal parameters of the fit. The log-likelihoods corresponding
to the three hypotheses are thus given by
\be
L_j =
-n\ln\sqrt{2\pi}-n\ln\sigma_j - n/2~, \ee where the third term
results from the product of Gaussians in the likelihood, which is
of the form
$$
\propto \prod_{i=1}^n \exp[-(\delta_j^{(o)}(i))^2/2\sigma_j^2] =
\exp[-n/2]~,
$$
   from the
definition $\sigma_j^2 = (1/n) \sum_{i=1}^n
[\delta_j^{(o)}(i)]^2$. Then, according to Wilk's theorem of nested
hypotheses, the log-likelihood-ratio
\begin{equation}
T_{j,j+1} = -2 (L_j-L_{j+1}) = 2n(\ln\sigma_j - \ln\sigma_{j+1})~,
~~~j=0,1~~~~ \label{Eq:T}
\end{equation}
is a chi-square variable with $k$ degrees of freedom, where $k$ is
the number of restricted parameters \cite{Holden}. In the present
case, we have $k=1$.

The Wilk test thus amounts to calculating the probability
$P_{j,j+1}$ that the obtained value of $T_{j,j+1}$ can be
overpassed by chance alone. If this probability $P_{j,j+1}$ is
small, this means that chance is not a convincing explanation for
the large value of $T_{j,j+1}$ which becomes meaningful. This
implies a rejection of the hypothesis that $C=0$ (resp. $D=0$) is
sufficient to explain the data and favor the fit with $C \neq 0$
(resp. $D \neq 0$) as statistically significant. In other words,
if the observed value of the probability $1-P_{j,j+1}$ that
$T_{0,1}$ (respectively of $T_{1,2}$) does not exceed some
high-confidence level (say, the $99\%$ confidence level) of the
$\chi^{2}$, we then reject the hypothesis $H_1$ (respectively
$H_2$) in favor of the hypothesis $H_0$ (respectively $H_1$),
considering the additional term $C$ (respectively $D$) redundant.
Otherwise, we accept the hypothesis $H_1$ (respectively $H_2$,
considering the description with $H_0$ (respectively $H_1$)
insufficient.

For each stock index, we fit the corresponding time series
starting from $t_{\rm{start}}$ and ending on September 30, 2002 to
a simple power law, to the log-periodic function (\ref{Eq:fit1})
and to the formula (\ref{Eq:fit4}) respectively, and thus obtain
$\sigma_0$, $\sigma_1$ and $\sigma_2$. Then we can calculate
$T_{j,j+1}$ from (\ref{Eq:T}) and the corresponding probabilities
$P_{j,j+1}$. The results of the Wilk tests are presented in Table
\ref{TbWilk}. The values $T_{0,1}$ are extremely large for all
indices, which reject with extremely high statistical significance
the hypothesis that a pure power law is sufficient compared to a
log-periodic power law. For the test of $H_2$ against $H_1$,
$T_{1,2}$ is found very large ($> 40$) for most of the indices,
except for Israel ($T_{1,2}=2.7$) and Mexico ($T_{1,2}=25.9$).
Even in the case of Mexico, the improvement obtained by adding a
the harmonic term (hypothesis $H_2$) is nevertheless very
significant since $T_{1,2}=25.9$ corresponds to a probability of
rejection of $H_2$ $P_{1,2}$ less than $10^{-4}\%$. Thus, only for
Israel, we find that $D=0$ can not be rejected at the confidence
level of $95\%$. This reflect the fact that the reduction of the
r.m.s. errors when going from formula (\ref{Eq:fit1}) to
(\ref{Eq:fit4}) is less that $1\%$. The lack of significant
improvement can also be seen visually in Fig.~\ref{Fig:IDIsrael}.

Since the assumption of Gaussian noise is most probably an
under-estimation of the real distribution of noise amplitudes, the
very significant improvement in the quality of the fit brought by
the use of both formulas (\ref{Eq:fit1}) and (\ref{Eq:fit4})
quantified in Table \ref{TbWilk}
provides most probably a lower bound for the statistical
significance of the hypothesis that both $C$ and $D$ should be
chosen non-zero, above the $99.9999\%$ confidence level. Indeed, a
non-Gaussian noise with a fat-tailed distribution would be
expected to decrease the relevance of competing formulas, whose
performance could  be scrambled and be made fuzzy. The clear and
strong result of the Wilk tests with assumed Gaussian noises thus
confirm a very strong significance of both
formulas (\ref{Eq:fit1}) and (\ref{Eq:fit4}).

\subsection{Determination of $t_c$ \label{s1:tc}}

\begin{figure}
\begin{center}
\epsfig{file=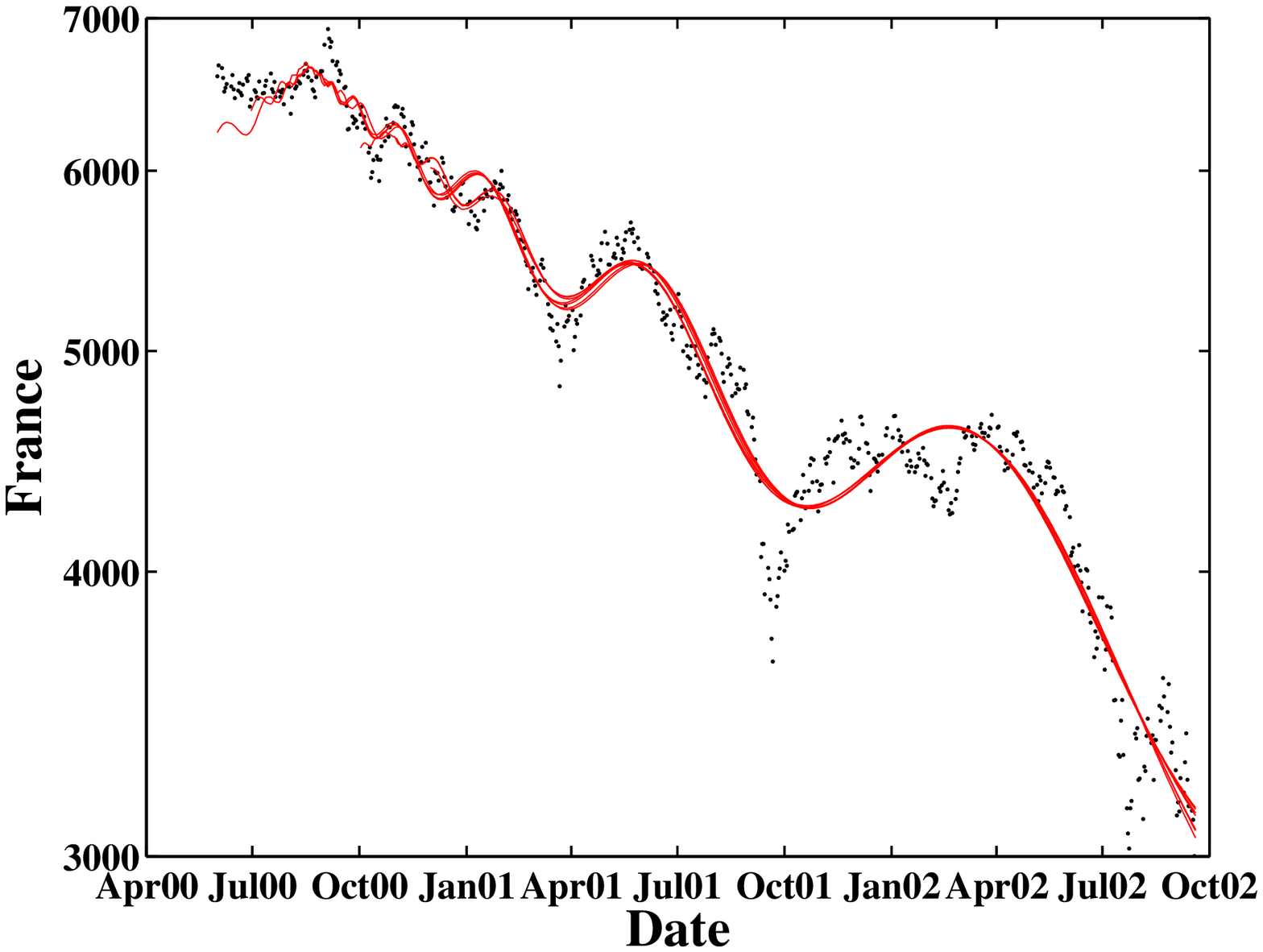,width=15cm, height=12cm}
\end{center}
\caption{The French stock index, a bearish anti-bubble, fitted
from $t_{\rm start}$ to September 2002 for different choices of
$t_{\rm start}$, spanning from Jun-01-2000 to Dec-01-2000. One see
that the fits are robust with respect to different starting date.}
\label{Fig:TCFrance}
\end{figure}

The critical time $t_c$ defines the real starting time of the
anti-bubbles and is an important parameter for quantifying the
synchronization between different stock markets. It is thus
important to investigate how robust is its determination by our
fitting procedure\footnote{We do not discuss here other approaches
for the estimation of $t_c$, such as using Shank's transformation,
the generalized $q$ analysis, the parametric fitting approach, and
so on (see Ref.~\cite{SPpredict} and reference therein).}. In this
section, we discuss two markets to illustrate the typical
situation, the French stock index as an example of a bearish
anti-bubble and the Australia stock index as an example of a
bullish anti-bubble. To test for the robustness of the
determination of $t_c$, we follow the analysis of \cite{SPpredict}
on the USA Standard and Poor index and take seven different values
for $t_{\rm{start}}$ from Jun-01-2000 to Dec-01-2000 for each
index. Figure \ref{Fig:TCFrance} shows the seven best fits, one
for each $t_{\rm{start}}$, for the French stock index. The
corresponding fitting parameters are listed in Table \ref{TbtcFr}.
Three fits in Fig.~\ref{Fig:TCFrance} are slightly different from
the rest especially in the early days of the anti-bubble. The
starting dates of these three fits are Oct-01-2000, Nov-01-2000
and Dec-01-2000. The slightly different nature of these three
cases is also reflected in discernable variations in the fitting
parameters listed in Table \ref{TbtcFr}. In particular, they
identify a critical $t_c$ at the end of October, 2000 rather than
mid-August, 2000. They also have slightly larger exponents
$\alpha$, lower log-frequencies $\omega$ and smaller $A$. Despite
these differences, all the fits are quite robust indicating a
critical time at or slightly after August, 2000.

\begin{figure}
\begin{center}
\epsfig{file=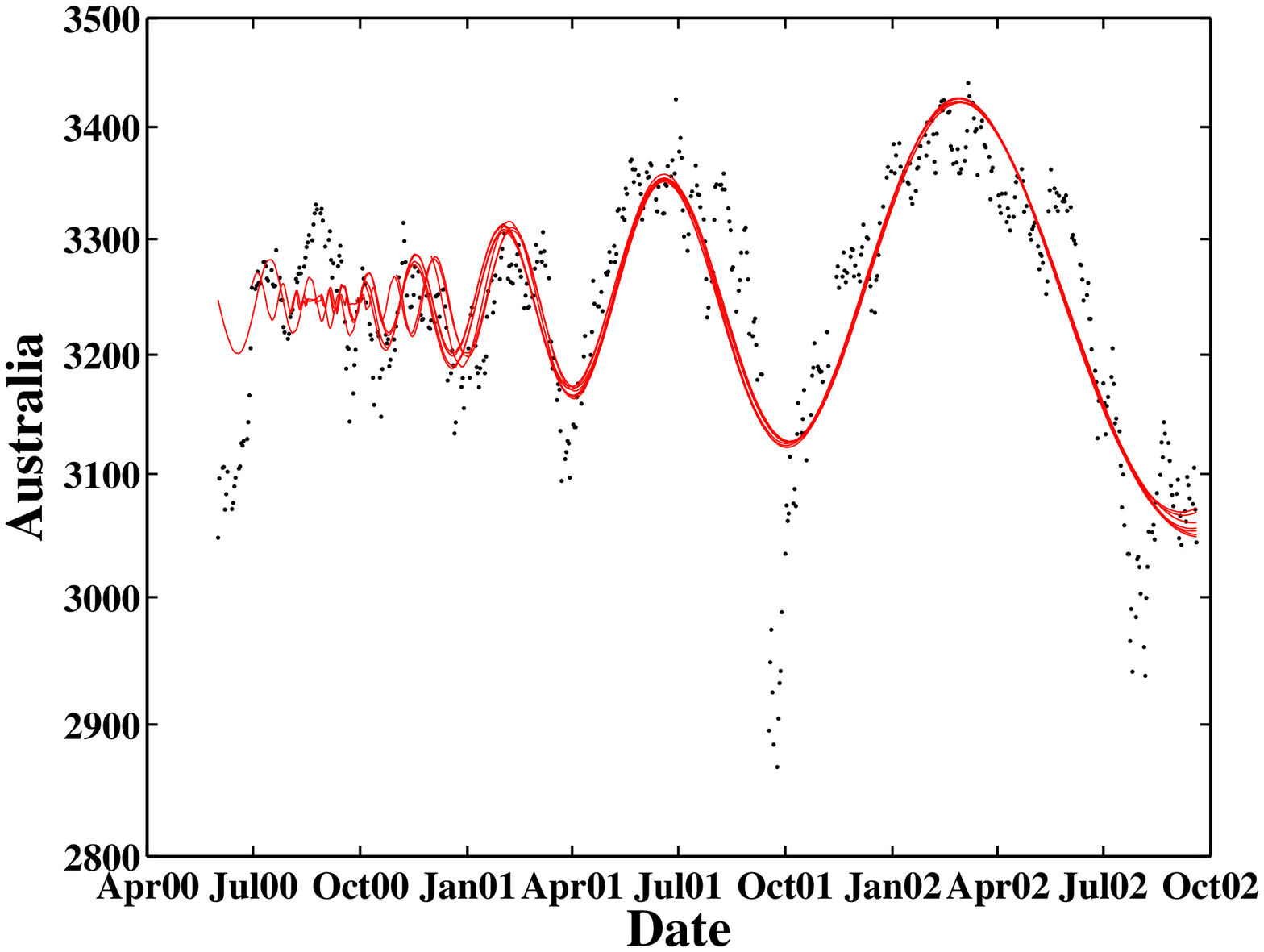,width=15cm, height=12cm}
\end{center}
\caption{The Australian stock index, a bearish anti-bubble, fitted
from $t_{\rm start}$ to September 2002 for different choices of
$t_{\rm start}$, spanning from Jun-01-2000 to Dec-01-2000. One see
that the fits are robust with respect to different starting date.}
\label{Fig:TCAustralia}
\end{figure}

Figure \ref{Fig:TCAustralia} shows the seven best fits,
one for each $t_{\rm{start}}$, for the
Australian stock index. The relevant parameters are
listed in Table \ref{TbtcAu}. We also observe three fits starting
on Jun-01-2000, Nov-01-2000 and Dec-01-2000 that have relatively
later predicted $t_c$, slightly larger exponents $\alpha$, lower
log-frequencies $\omega$. But the predicted index value at $t_c$
(i.e., $e^A$) are almost the same. The critical time $t_c$
of the Australian anti-bubble is also clustered around mid-August,
2000. This is consistent with the fact that both including extra
data earlier than $t_c$ ($t_{\rm{start}}\ll t_c$) and
truncating data after $t_c$ ($t_{\rm{start}}\gg t_c$) will reduce
the precision of the determination of $t_c$ and deteriorate the
quality of fits.

We nevertheless have to note that not all the indices give such
robust results. The log-periodic oscillations in the initial days
of some anti-bubbles are completely spoiled by noise, where
different effects overwhelm the herding behavior thought to be at
the origin of the log-periodic power law patterns. For instance,
the existence of an anti-bubble in the Peruvian stock index is
quite questionable in view of the particularities in its fitting
parameters.

\section{Correlation across different markets and synchronization
of the anti-bubbles\label{s2:xcorr}}

\begin{figure}
\begin{center}
\epsfig{file=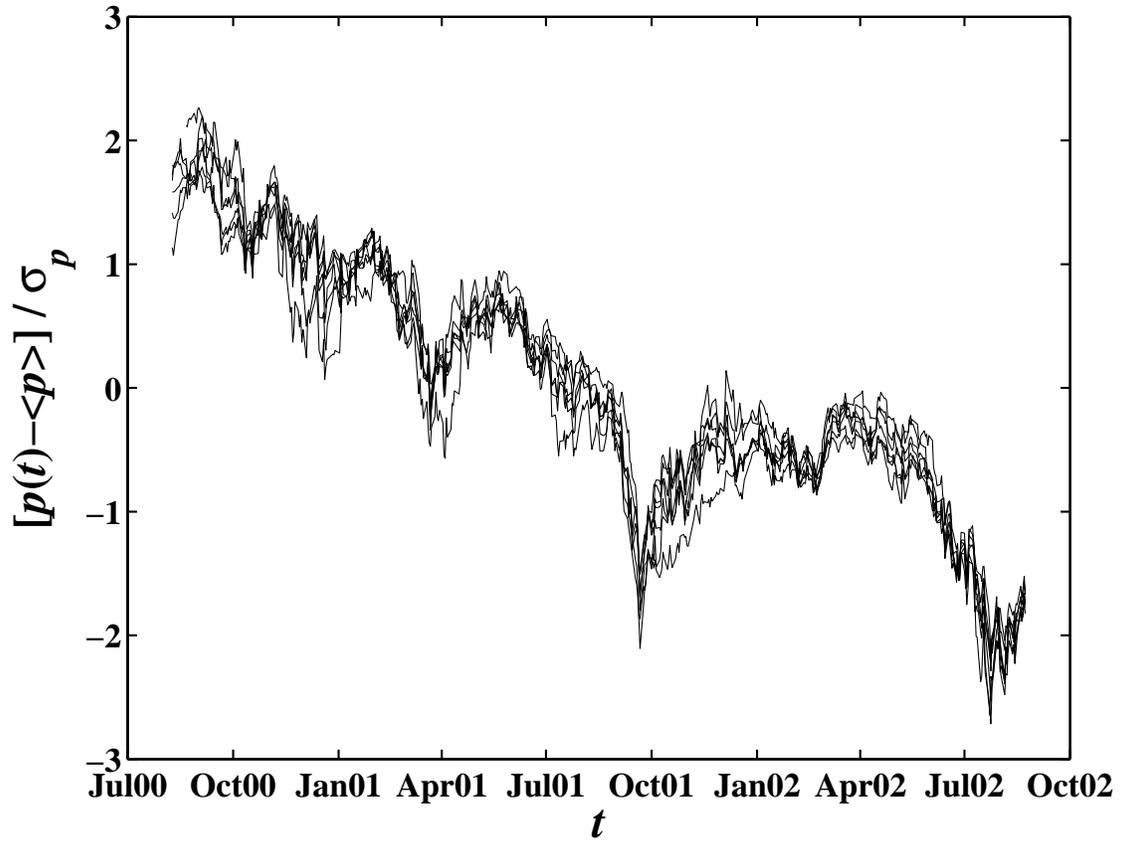,width=15cm}
\end{center}
\caption{The stock market indices of seven countries (US S\&P 500,
the Netherlands, France, Germany, Norway, UK, Spain) are
superimposed by plotting the normalized values $[p(t)-\langle p
\rangle]/\sigma_p$ of each index as a function of time, where
$\langle p \rangle$ is the mean whose substraction accounts for a
country-specific translation in price and $\sigma_p$ is the
standard deviation for each index which accounts for a
country-specific adjustment of scale. } \label{collapse}
\end{figure}

One of the most remarkable results obtained so far is that most of
the anti-bubbles started between August and November, 2000, with
very similar time evolutions as quantified by the formulas
(\ref{Eq:fit1}) and (\ref{Eq:fit4}) and by the Tables
\ref{TbBearAB} and \ref{TbBearAB4}. This suggests that the
triggering of almost simultaneously occurring anti-bubbles is an
international event. Figure~\ref{collapse} summarizes our main
message by superimposing the stock market indices of seven
countries (US S\&P 500, the Netherlands, France, Germany, Norway,
UK, Spain). The ordinate plots the normalized values
$[p(t)-\langle p \rangle]/ \sigma_p$ of each index as a function
of time, where $\langle p \rangle$ is the mean whose substraction
accounts for a country-specific translation in price and
$\sigma_p$ is the standard deviation for each index which accounts
for a country-specific adjustment of scale. This remarkable
collapse onto a single master curve does not rely on any
parametric fit. It demonstrates maybe more clearly than by any
other means the extraordinary strong synchronization of the
anti-bubble regime in the major western markets. Other markets
exhibit a higher variability and have not be represented on this
curve for clarity.

It is well-known that the October 1987 crash was an international event,
occurring within a few days in all major stock markets  \cite{Barro}.
It is also often been noted that smaller West-European stock markets  as well
as other markets around the world are influenced by dominating trends on
the USA market. In this spirit, in \cite{emergent},
a set of secondary stock markets were
shown to exhibit well-correlated ``anti-bubbles'' triggered by a rash of
crises on emerging markets in early 1994. In this case, the synchronization
occurred between West-European markets which were decoupled from the
USA markets.
This suggests that smaller stock
markets can weakly synchronize not only because of the over-arching
influence of the USA market, but also independent of the USA market
due to external
factors such as the Asian crisis of 1994.

Here, we have shown the occurrence of the synchronization of a large
majority of
markets with significant volumes
into a collective anti-bubble, that includes the USA markets, most of
the European markets as well as the developed Asian markets and a few
other markets worldwide (see the list given in table \ref{TbBearAB}).
Motivated by this result, we turn now to a series of
non-parametric tests exploring the nature and amplitude of this
worldwide synchronization, in order to attempt to cast addition light
on this remarkable event.

In the following, we investigate several measures of correlation,
or more generally of inter-dependence, between each index and the
USA S\&P500 index taken as a reference, in order to test whether
we could have otherwise detected the synchronization unravelled by
our log-periodic analysis. These measures of inter-dependence use
the cross-correlation of weekly returns, linear regressions of
indices and of their returns, a synchronization ratio of joint
occurrences of ups and downs and an event synchronization method
recently introduced \cite{EventSyn}. These different measures
confirm that the inter-dependence between the major western
markets has slightly increased as a function of time in the last
decade and especially since the Fall of 2000, confirming weakly
the qualitative message contained in our results of the occurrence
of a synchronized anti-bubble worldwide. However, these more
standard measures of dependence do not come near the log-periodic
analysis in the strength of the signal.

\subsection{Cross-correlation of weekly returns}

The formulas (\ref{Eq:fit1}) and (\ref{Eq:fit4}) have been
applied to the prices and the insight into the
existence of a synchronization
comes from a comparison between these fits on the index prices.
In order to study the cross-correlation between different indices,
we need to study the index returns which are approximately stationary
thus ensuring reasonable convergence properties of the correlation
estimators. We thus follow a procedure similar to that of
Ref.~\cite{Roll88} for the estimation of the cross-correlation
coefficients of monthly percentage changes in major stock market
indexes from June 1981 to September 1987.

In order to minimize noise, we smooth the price time series with
a causal Savitzky-Golay filter with eight points to the left
of each point (``present time''), zero
point to its right and a fourth order polynomial \cite{Press}. This
provides a smoothed price time series $\bar{p}(t)$. We
then construct the
return time series and then obtain the cross-correlation functions of
$\bar{p}(t)$.
We use weekly returns, as a compromise between daily and monthly returns
to minimize noise and maximize the data set size. The weekly returns
are defined on the smoothed price time series $\bar{p}(t)$ as
\begin{equation}
r(t) = \ln[\bar{p}(t)/\bar{p}(t-7)]~. \label{Eq:r}
\end{equation}
We calculate the
correlation coefficients $C(t)$ of the stock indices in
a moving window of 65 trading days (or about a quarter
in calendar days). We present our results obtained for the
cross-correlation between the USA S\&P500 index and the stock market indices of
the Netherlands, France, Japan, Germany, UK, Hong
Kong, Australia, Russia and China, which are typical.

\begin{figure}
\begin{center}
\epsfig{file=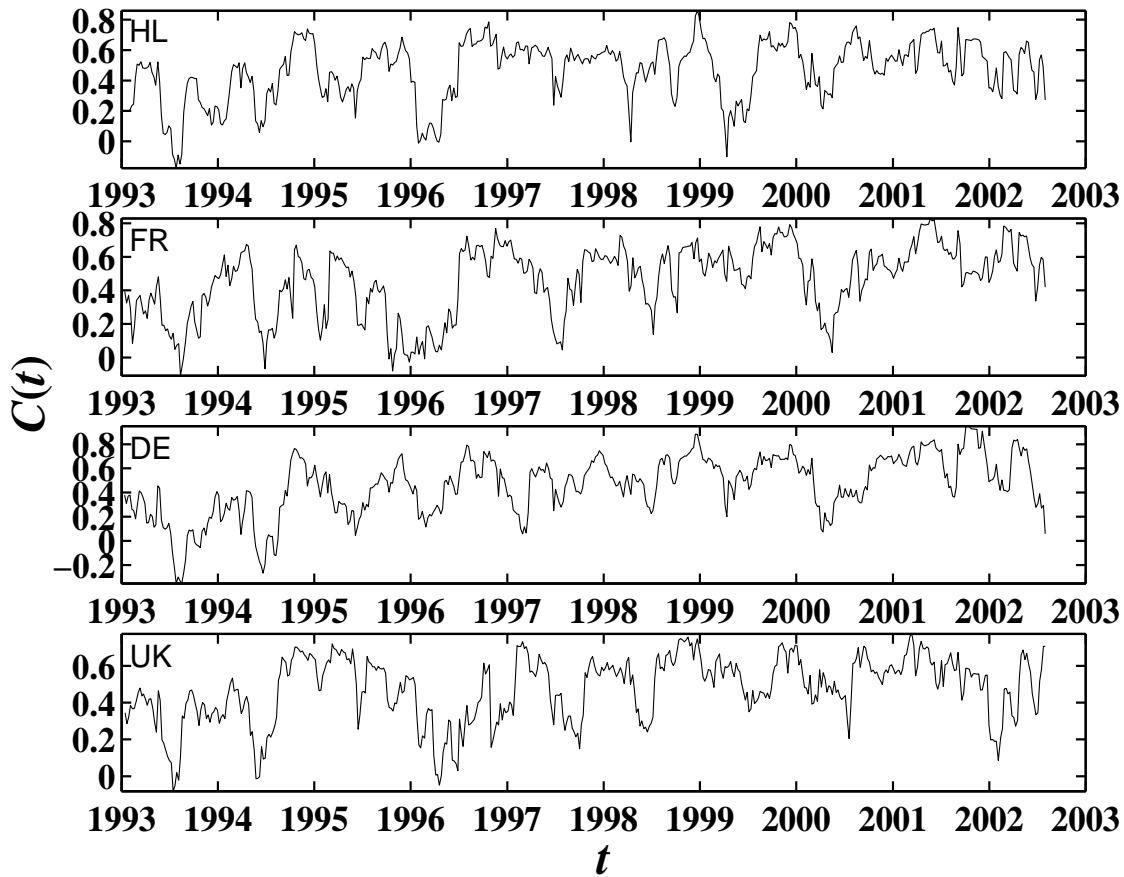,width=15cm, height=12cm}
\end{center}
\caption{Cross-correlation coefficients between the USA
   S\&P500 index and four European stock market indexes
(Netherlands, France, Germany and United Kingdom from top to
bottom) for weekly returns on filtered prices in a moving
three-month window. Surrogate tests obtained by reshuffling the
returns show that the large peaks and troughs are statistically
significant. One can observe a slow overall increase of the
correlation coefficient $C(t)$ over this decade. The variations
are estimated to be $0.12$ by bootstrapping simulations.}
\label{Fig:XCorr01}
\end{figure}

\begin{figure}
\begin{center}
\epsfig{file=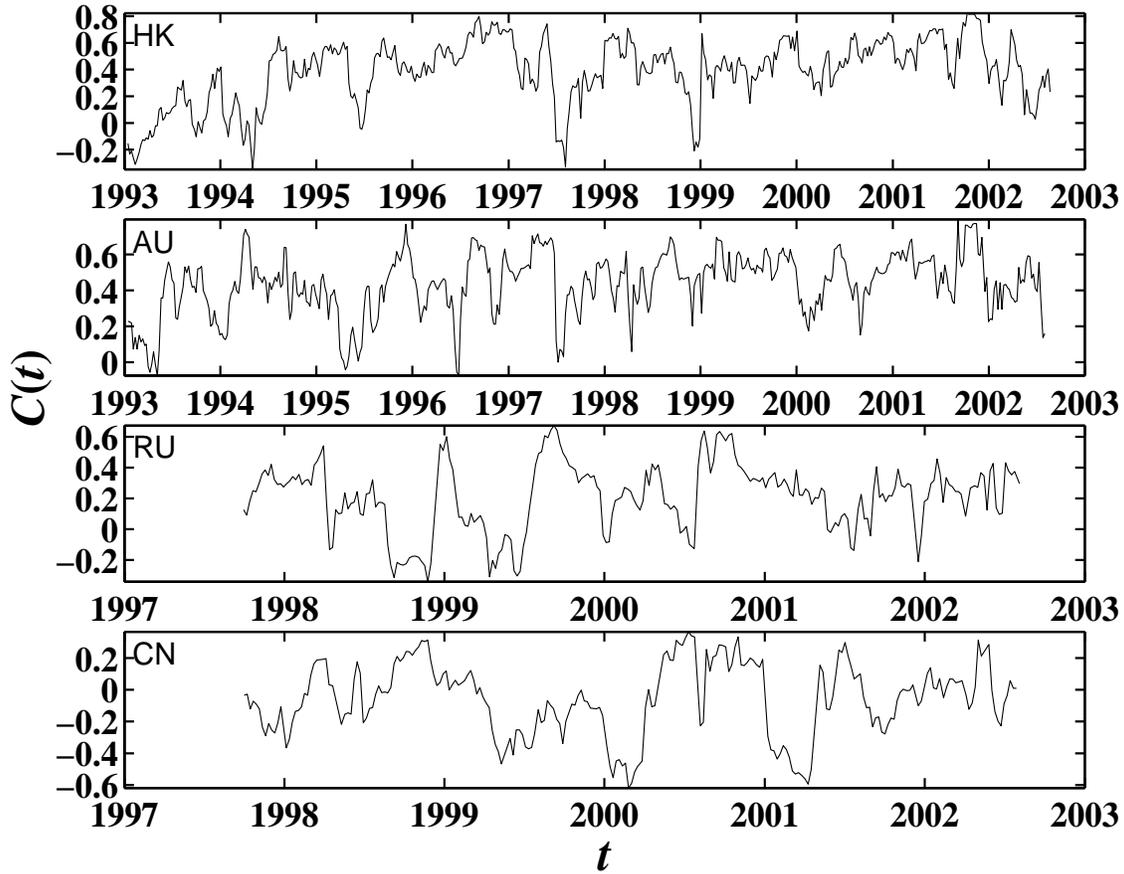,width=15cm, height=12cm}
\end{center}
\caption{Cross-correlation coefficients between the USA
   S\&P500 index and four non-European indexes (Hong
Kong, Australia, Russia and China from top to bottom) for weekly
returns on filtered prices in a moving three-month window. While
one can observe a slow overall increase of the correlation
coefficient $C(t)$ over this decade for Hong Kong and for
Australia, no such trend is seen for Russia and China which are
characterized by weaker correlations and more pronounced regime
switches. The variations are estimated to be $0.12$ by
bootstrapping simulations.} \label{Fig:XCorr02}
\end{figure}

As illustrated in Fig.~\ref{Fig:XCorr01}, the European markets
have rather strong correlations with the American market with an
average correlation coefficient of $0.44 \pm 0.05$. The
cross-correlation coefficients of the smoothed weekly returns for
Hong Kong, Australia, Russia and China are shown in
Fig.~\ref{Fig:XCorr02} as a function of time. Their average
cross-correlation coefficients are relatively weaker than those
for the European markets, with values respectively equal to
$0.30\pm 0.05$, $0.34\pm 0.05$, $0.35\pm 0.05$ and $0.21\pm 0.05$
for Japan, Hong Kong, Australia and Russia. The average
cross-correlation coefficient for China is slightly negative
($-0.06 \pm 0.05$), indicating that the Chinese stock market seems
practically uncorrelated from the western markets. The
uncertainties and fluctuations of the variables $C(t)$ are
determined by a bootstrap simulation of 1000 series of reshuffled
returns which gives a standard deviation $\sigma_C = 0.12$.

Interestingly, Fig.~\ref{Fig:XCorr01} shows
that the cross-correlation coefficients of the
European markets with the American market increases slowly with time.
This property is weaker for Japan and
Hong Kong and is completely absent for Russia and China. While
qualitatively compatible, the
evidence for a slow
increase of the cross-correlation is not sufficiently precise to
relate precisely to our previous finding of a strong synchronization of
an anti-bubble regime since the summer of 2000.

\begin{figure}
\begin{center}
\epsfig{file=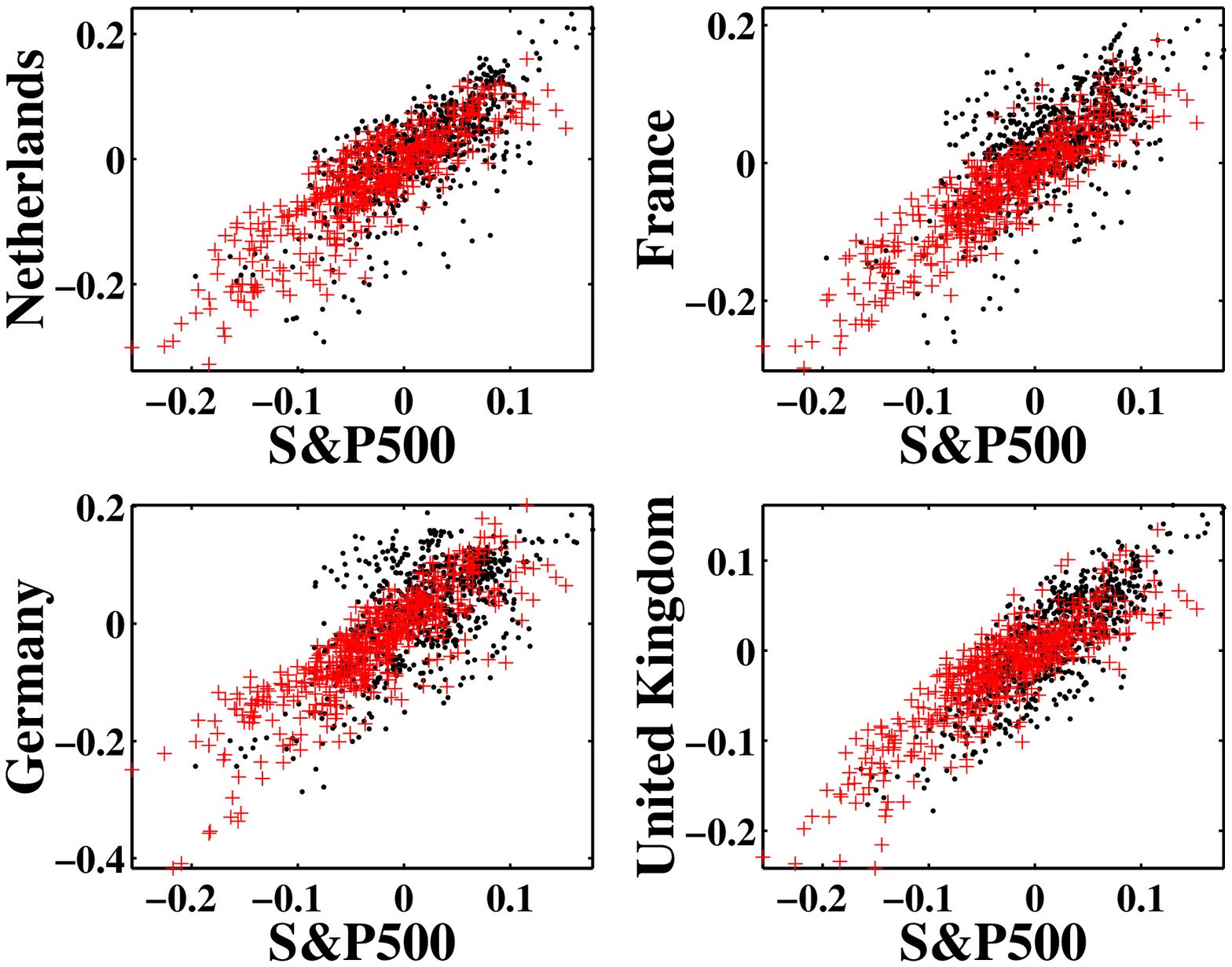,width=15cm, height=12cm}
\end{center}
\caption{Monthly returns of four European market indices as a function
of the monthly returns of the USA S\&P500 index. The black dots
correspond to period [Jun-04-1997, Aug-09-2000]. The $+$ correspond to
the period [Aug-10-2000, Sep-04-2002].} \label{Fig:LinRegL01}
\end{figure}

\begin{figure}
\begin{center}
\epsfig{file=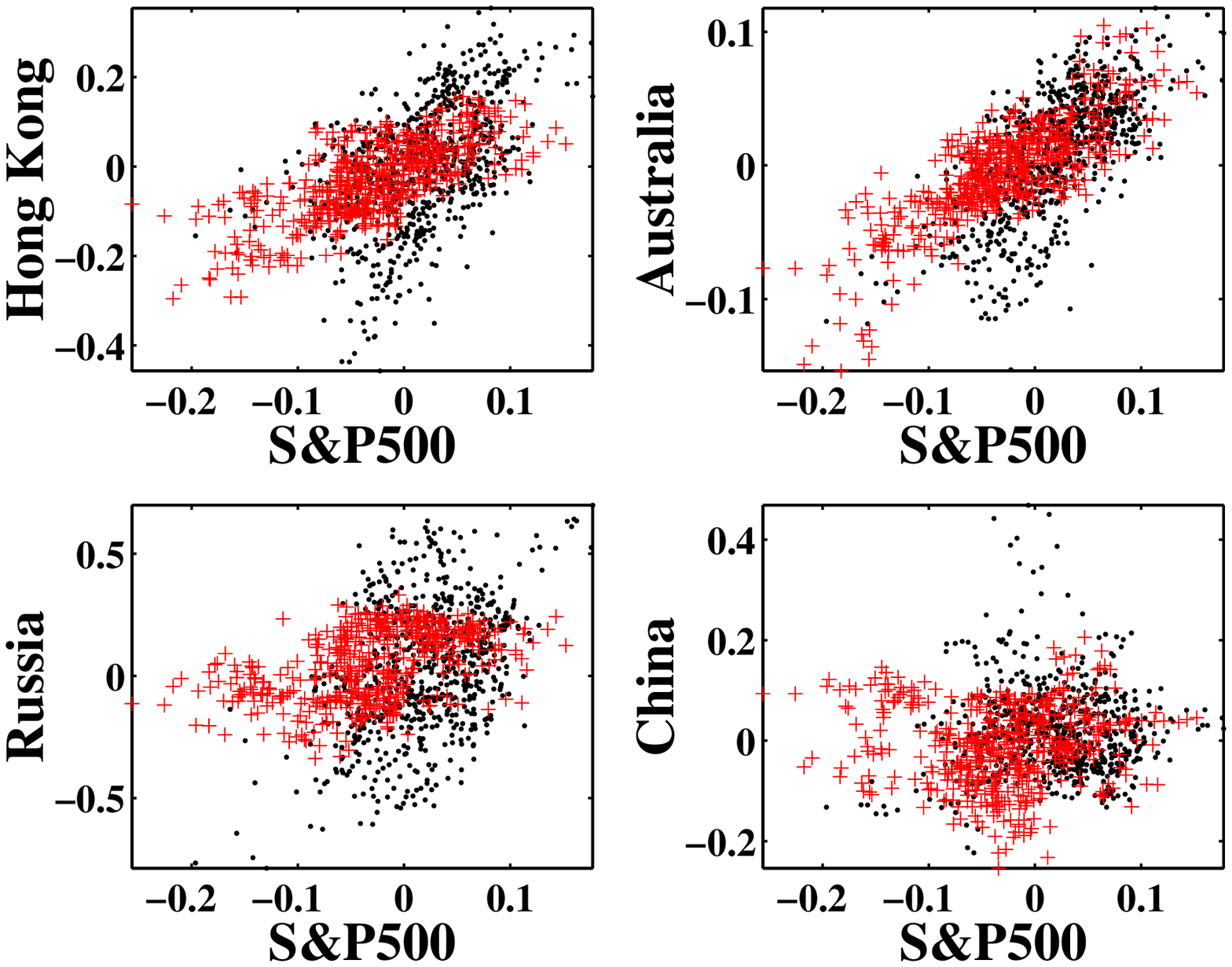,width=15cm, height=12cm}
\end{center}
\caption{Weekly returns of four non-European market indices as a
function of the weekly returns of the USA S\&P500 index.
The black dots
correspond to period [Jun-04-1997, Aug-09-2000]. The $+$ correspond to
the period [Aug-10-2000, Sep-04-2002].} \label{Fig:LinRegL02}
\end{figure}

To refine the evidence for an increase in correlation,
we investigate the correlation between
the USA S\&P500 index and nine other indices
(Netherlands (HL),
France (FR), Japan (JP), Germany (DE), United Kingdom (UK), Hong
Kong (HK), Australia (AU), Russia (RU), and China (CN)) in two periods,
[Jun-04-1997, Aug-09-2000] and [Aug-10-2000, Sep-04-2002].
Table~\ref{TbLinReg} shows the $\beta$ coefficients and corresponding
correlation coefficients $\gamma$ of the weekly returns in these two periods.
The coefficients $\beta_1$ and $\beta_2$ for the two periods of each index are
obtained by using the well-known linear regression of the time series
of returns of each index against the time series of returns of the
S\&P500 index.
Such an approach led Roll \cite{Roll88} to conclude on the existence
of a particularly strong synchronization during and after the crash
of Oct. 1987
seen by the fact that the beta's of the
different indices against a world market index were anomalously large.
The two correlation coefficients $\gamma_1$ and $\gamma_2$ are directly
evaluated. The values of the slope $\beta$
\cite{Roll88} and the linear correlation coefficient $\gamma$ are
listed in Table~\ref{TbLinReg}. Fig.~\ref{Fig:LinRegL01} plots
the returns of four European indices as a function of the returns of
the S\&P500 for
each of the two periods. This figure and Table~\ref{TbLinReg} confirm
a significant increase of the correlations from the period
[Jun-04-1997, Aug-09-2000] to the period [Aug-10-2000, Sep-04-2002].
Fig.~\ref{Fig:LinRegL02} plots
the returns of the indices of Hong Kong,
Australia, Russia and China as a function of the returns of the S\&P500 for
each of the two periods. Table~\ref{TbLinReg} and Fig.~\ref{Fig:LinRegL02}
show a significant increase in correlation from the period
[Jun-04-1997, Aug-09-2000] to the period [Aug-10-2000, Sep-04-2002]
only for Hong Kong and Australia. Russia gives a marginal signal and
China none.

Table~\ref{TbLinReg} and Fig.~\ref{Fig:LinRegL01} clearly confirm
a strong increase in the correlation between the USA stock market
and the European indices and
some non-European indices from the period
[Jun-04-1997, Aug-09-2000] to the period [Aug-10-2000, Sep-04-2002],
in agreement with the evidence of the log-periodic synchronization
documented above.

\subsection{Synchronization of weekly returns}

We now discuss another intuitive measure for the characterization of the
synchronization of weekly returns between different world stock markets.
We use a moving window,
whose size is 65 trading days, corresponding to 13 weeks.
In this moving window, we define
the synchronization factor $R(t)$ as the fraction of weeks among
the 13 weeks for which a
given index return has the same sign as that of the
S\&P500 index. By
definition, $0 \le R(t)\le 1$. $R(t) = 1$ (respectively
$R(t)=0$) corresponds to full synchronization (respectively perfect
anti-synchronization).
$R(t)=0.5$ corresponds to independent time series whose weekly
returns have mutually random signs.

\begin{figure}
\begin{center}
\epsfig{file=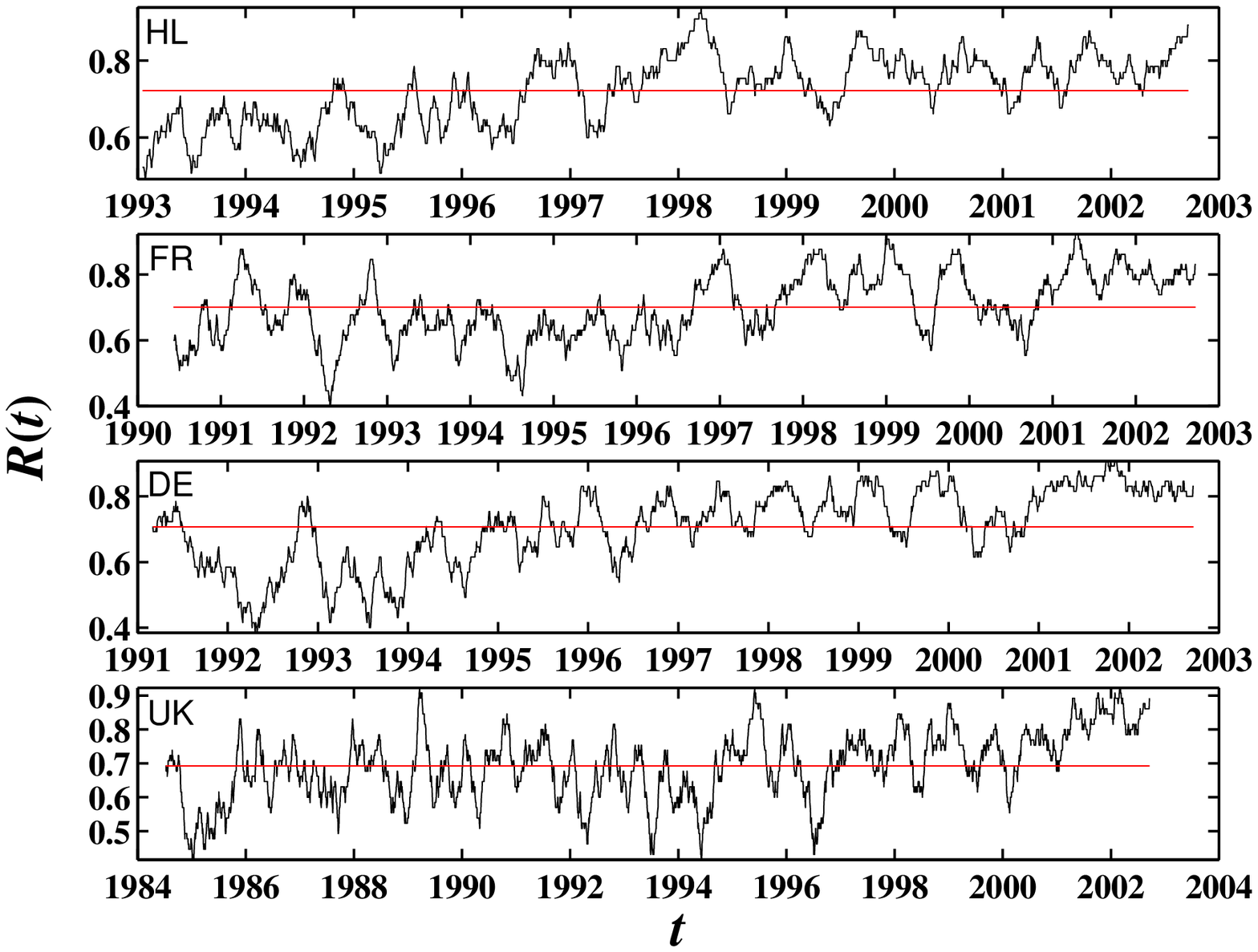,width=15cm, height=12cm}
\end{center}
\caption{The synchronization factor $R(t)$ between the USA S\&P500
index and four European indices (HL, FR, DE and UK), defined as
the fraction of weeks with the same return signs over a moving
13-week window. The horizontal lines show the average of $R(t)$.
The variations are estimated to be $0.06$ by bootstrapping
simulations.} \label{Fig:Rs01}
\end{figure}

\begin{figure}
\begin{center}
\epsfig{file=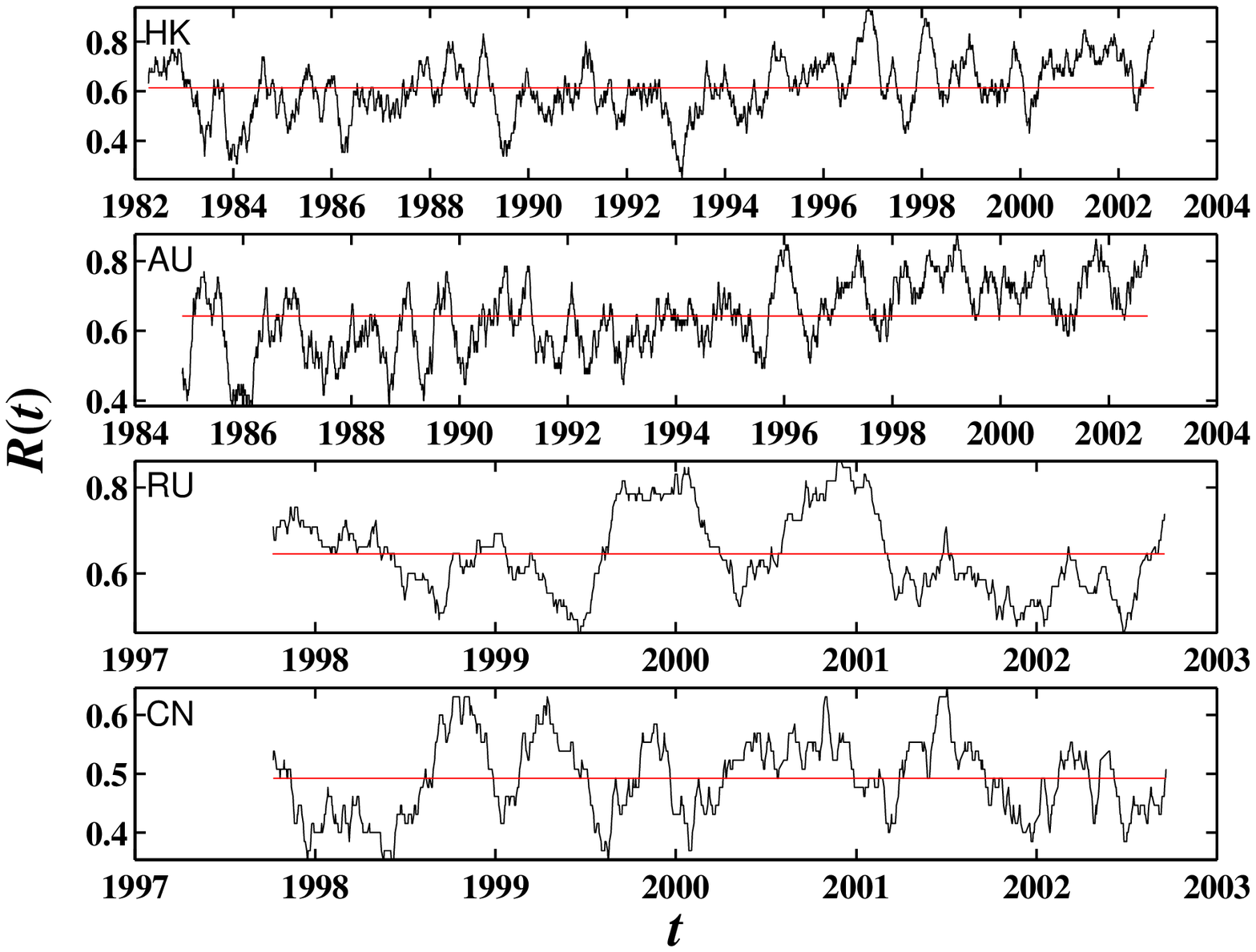,width=15cm, height=12cm}
\end{center}
\caption{The synchronization factor $R(t)$ between the USA S\&P500
index and four non-European indexes (HK, AU, RU and CN), defined
as the fraction of weeks with the same return signs over a moving
13-week window. The horizontal lines show the average of $R(t)$.
The variations are estimated to be $0.06$ by bootstrapping
simulations.} \label{Fig:Rs02}
\end{figure}

We calculate the synchronization factor $R(t)$ between the USA
S\&P500 index and the indices of the Netherlands, France, Japan,
Germany, UK, Hong Kong, Australia, Russia and China. As shown in
Figs.~\ref{Fig:Rs01} and \ref{Fig:Rs02}, all considered indices
have $R(t)$ significantly larger than $0.5$ except for China for
which $R(t)=0.49$. The uncertainties and fluctuations of the
variables $R(t)$ are determined by a bootstrap simulation of 1000
series of reshuffled returns which give a standard deviation
$\sigma_R = 0.06$. Again, the European markets have relatively
higher synchronization factors and their $R(t)$ increase clearly
with time. Not only $R(t)$ is consistently at its highest
long-term average level in the last few years for all markets,
except for Russia and China, we can also note a very strong and
significant increase of $R(t)$ over the last year with much less
fluctuations. Only Russia and China among the eight indices escape
from this world-wide synchronization.

\subsection{Time resolved event synchronization of the index time
series}

In view of the importance of characterizing the dependence between
different markets, we present yet another measure of
the synchronization of weekly returns across different world
stock markets.
The two previous analyses were based on the time series of weekly
returns. The present analysis measures the synchronization between
different index time series by quantifying
the relative timings of specific events in the
time series, following the algorithm
initially introduced in  \cite{EventSyn}.

Given two index time series $p^{(1)}(t)$ and $p^{(2)}(t)$,
we define ``events'' as large market velocities.
We define the market velocity as a coarse-grained measure of the
slope of the price as a function of time. To obtain this coarse-grained
measure, we apply on the prices $p(t)$ a causal Savitzky-Golay
fourth-order polynomial filter with eight points on
the left and no point on the right. The velocity $v(t)$ at time $t$ is
define as the analytical time derivative of the coarse-grained $p(t)$.
``Large velocities'' are defined by the condition $|v(t)| > d$, where
$d$ is a threshold chosen here equal to $d=0.001$. The
times when the velocities $v^{(1)}(t)$ and $v^{(2)}(t)$ of
the two index time series $p^{(1)}(t)$ and $p^{(2)}(t)$ obey the
condition $|v(t)| > d$ are denoted respectively
$t^{(1)}_i$ ($i=1,\ldots,m_1$) and $t^{(2)}_j$ ($j=1,\ldots,m_2$).
The degree of synchronization is then quantified by counting the
number of times an event ($|v(t)| > d$) appears in time series
$p^{(1)}(t)$ shortly after
it appears in time series $p^{(2)}(t)$. This number is estimated by the
following formula
\begin{equation}
c(1|2) = \sum_{i=1}^{m_1} \sum_{j=1}^{m_2} J_{ij} \label{Eq:c12}
\end{equation}
with
\begin{equation}
J_{ij} = \left\{
\begin{array}{ll}
1&{\mathtt{if~~}} 0<t^{(1)}_i-t^{(2)}_j\le \tau_{ij} \\
1/2&{\mathtt{if~~}} t^{(1)}_i=t^{(2)}_j\\
0&{\mathtt{otherwise}}
\end{array} \right., \label{Eq:Jij}
\end{equation}
where
\begin{equation}
\tau_{ij} = \min_{i,j}\left\{t^{(1)}_{i+1}-t^{(1)}_{i},~
t^{(1)}_{i}-t^{(1)}_{i-1},~t^{(2)}_{j+1}-t^{(2)}_{j},~
t^{(2)}_{j}-t^{(2)}_{j-1}\right\}~. \label{Eq:tauij}
\end{equation}
Likewise, $c(2|1)$ is calculated in a similar manner. The
symmetrical combination
\begin{equation}
Q = {{c(1|2)+c(2|1)} \over m_1m_2}~, \label{Eq:Q}
\end{equation}
called the synchronization index,
measures the synchronization of the
events and thus of the two time series.
By construction, $0\le Q \le 1$. The cases of $Q=1$ and $Q=0$
correspond respectively to
full synchronization and absence of synchronization of events
of the two index time series.

\begin{figure}
\begin{center}
\epsfig{file=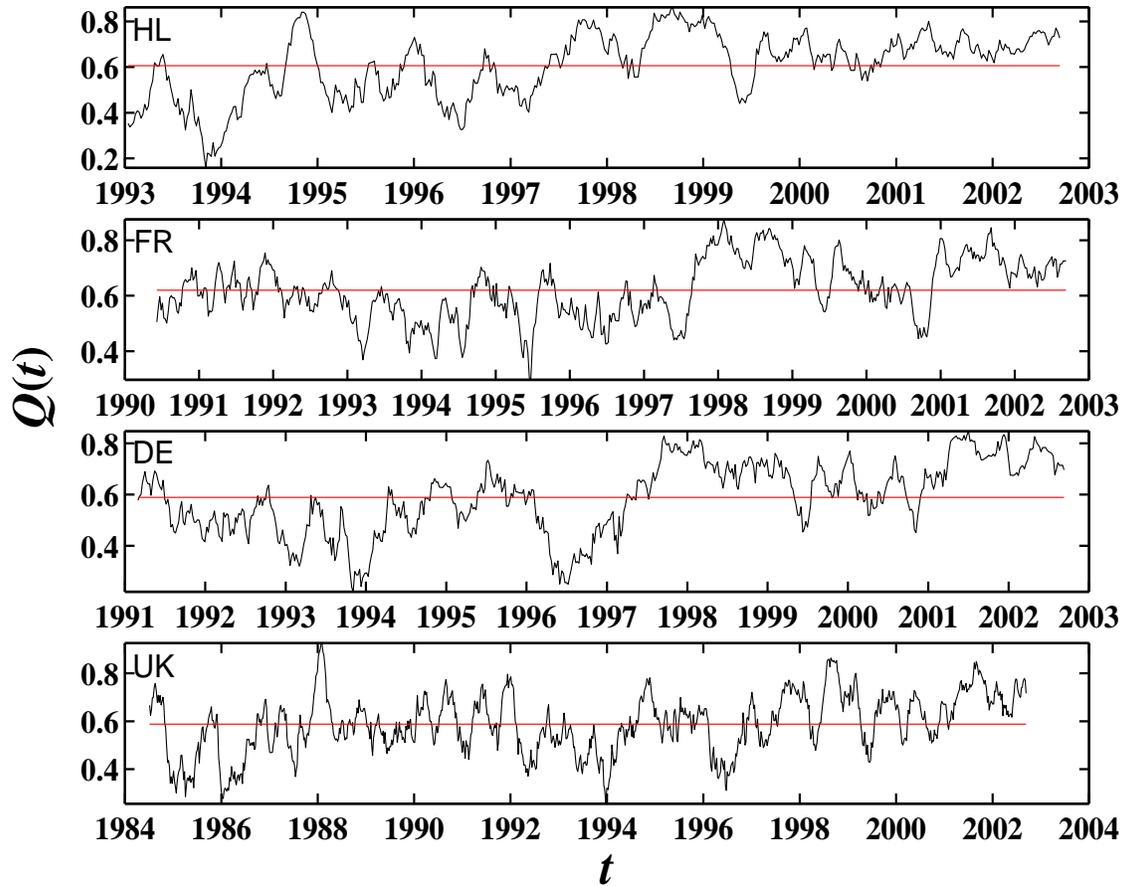,width=15cm, height=12cm}
\end{center}
\caption{Synchronization index $Q(t)$ (see text and
expression (\ref{Eq:Q}) for its definition) between
the USA S\&P500 index and four
European stock indexes (HL, FR, DE and UK), as a
function of time in a running window of 65 trading days.
The horizontal lines show the average of $Q(t)$ over the shown
time interval.} \label{Fig:ES01}
\end{figure}

\begin{figure}
\begin{center}
\epsfig{file=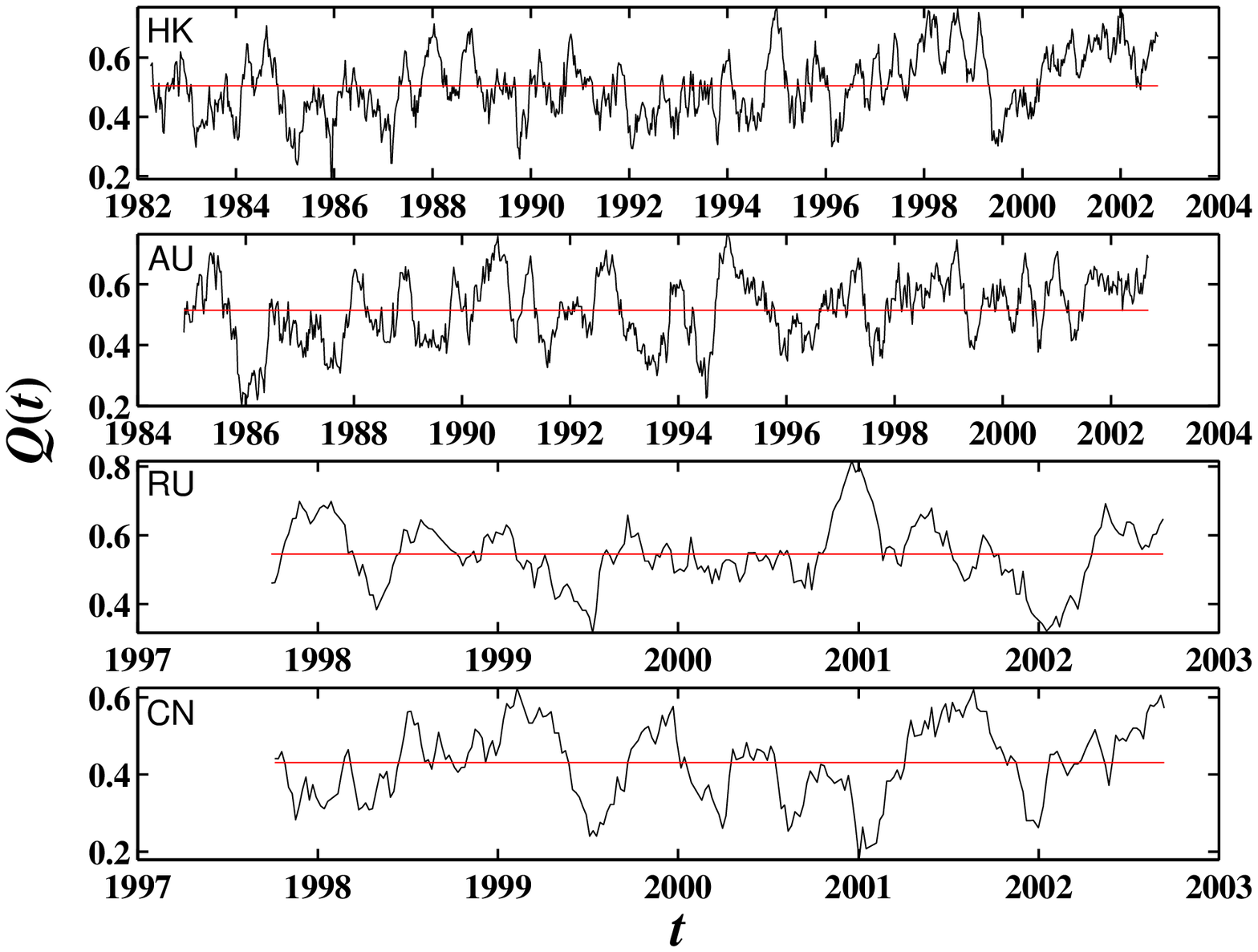,width=15cm, height=12cm}
\end{center}
\caption{Same as Fig.~\ref{Fig:ES01} for
four non-European indexes (HK, AU, RU and CN).} \label{Fig:ES02}
\end{figure}

We calculate $Q(t)$ in a moving window of 65 trading days as
before between the USA
S\&P500 index on the one hand
and the stock markets of the Netherlands, France, Japan, Germany, UK,
Hong Kong,
Australia, Russia and China on the other hand. Fig.~\ref{Fig:ES01} and
Fig.~\ref{Fig:ES02} shown that
all stock markets have $Q(t)$ significantly larger
than $0.5$ except for China. Again, an increasing trend appears
clearly for the European markets. Furthermore, the period
since the winter of 2000 has significantly larger $Q(t)$ compared
with the earlier time for the European markets, Hong Kong and
Australia. $Q(t)$ is especially large and regular
for The Netherland (HL) since mid-1999 and all other
markets also have a very high synchronization index $Q(t)$ since 2001.

\section{Discussion \label{s1:conc}}

Following our previous investigation of the USA Standard and Poor
index anti-bubble that started in August 2000 \cite{SPpredict}, we
have analyzed the major stock market indices worldwide and found
that a vast majority of European and Western countries as well as
many other indices exhibit practically the same log-periodic power
law anti-bubble structure as found for the USA S\&P500 index. In
addition, these anti-bubbles are found to start approximately at
the same time, August 2000, in all these markets. This shows a
remarkable degree of synchronization worldwide which, to our
knowledge, has never been seen at any other time other than during
and in the (short) aftermath of the October 1987 crash.

To test further this synchronization, we have also used several standard and
less standard
measures of correlation, dependence and synchronization between the
USA S\&P500 index and other world markets. These measures confirm
the existence of significant increase of dependence in the last decade
and still a larger increase in the last one-two years. However, these
measures come nowhere close to the clarity of the signal of
the extraordinary strong synchronization found using the log-periodic
power law analysis. This is due to the fact that the log-periodic
power law analysis is not sensitive to detailed phases in the oscillations
(translated in slightly shifted effective time units in different markets)
and detects only the robust universal unit-independent discrete scale
invariant features of the price trajectories.

What triggered the worldwide anti-bubble in August 2000? The
international descent of many of the worldwide stock markets since
2000 suggests the strengthening of globalization and the leading
impact of the USA. In this respect, if history is any guide, the
historical record on financial crises shows that they are often
accompanying surges of globalization in the past, including events
as far back as in the 19th century such as during the gold
standard period of 1880-1913 \cite{global1}. Bordo and Murshid
\cite{Global2} compared various characteristics of the
cross-country transmission of shocks in the financial markets of
both advanced and emerging countries during two periods of
globalization - the pre-World War I classical gold standard era,
1880-1914, and the post-Bretton Woods era, 1975-2000. They found
that financial market shocks were more globalized before 1914
compared to the present and interpret this result by the growing
financial maturity of advanced countries and the widening of the
center to include a more diverse group of countries spanning
several regions. Our findings temper Bordo and Murshid's results
and suggest a possible transition to a stronger integration and
globalization fostered by several factors, including corporate and
financial globalization, and the rapid development, adoption and
use of information and communications technology. Our results also
confirm those of Goetzmann et al. \cite{Goetz} who find that the
correlation structure of the major world equity markets over 150
years vary considerably through time and are highest during
periods of economic and financial integration such as the late
19th and 20th centuries. Goetzmann et al. \cite{Goetz} also stress
that such increase of correlation implies that diversification
benefits to global investing  relies increasingly on investment in
emerging markets, in agreement with our results on the weaker
synchronization of emerging markets.
   Our results can also be seen
to add to the literature on contagion,
usually defined as correlation between markets in
excess of what would be implied by economic fundamentals,
by providing a new technical tool.

\bigskip
{\bf Acknowledgments:}
This work was supported in part by the James S. Mc Donnell Foundation
21st century scientist award/studying complex system.

%\pagebreak

\pagebreak

%table 1
\begin{table}
\begin{center}
\caption{\label{TbBearAB} Parameters of the fits of the indices
indicated in the first column using the first-order formula
(\ref{Eq:fit1}) from $t_{\rm{start}}$ to September, 30, 2002. All
these indices are in the so-called bearish anti-bubble regime,
qualified by the fact that the coefficient $B$ is negative. The
exponents $\alpha$ of the leading power law in formula
(\ref{Eq:fit1}) are found either larger or smaller than $1$,
corresponding to an accelerating (respectively decelerating)
decrease of the prices as a function of time (see text). $\chi$
denotes the root-mean-square (r.m.s.).}
\medskip
\begin{tabular}{ccccccccccc}
\hline\hline
Stock&$t_{\rm{start}}$&$t_c$&$\alpha$&$\omega$&$\phi$&$A$&$10^3B$&$10^
3C$&$10^2\chi$\\\hline
Netherlands&00/09/04&00/08/28&1.05&9.16&1.63&6.53&-0.55&-0.18& 4.30\\
France&00/09/04&00/08/30&0.92&8.88&3.54&8.79&-1.39&-0.33& 3.99\\
USA Dow Jones&00/09/06&00/08/15&1.05&9.76&4.01&9.30&-0.17&-0.11& 3.37\\
USA NASDAQ&00/08/20&00/09/02&0.26&10.00&3.37&8.74&-251&-23.6& 6.36\\
Japan&00/08/28&00/08/06&0.79&7.74&3.34&9.74&-3.40&-0.86& 3.96\\
Belgium&00/11/06&00/06/25&1.52&12.20&2.85&8.02&-0.01& 0.00& 3.36\\
Denmark&00/10/24&00/05/03&0.78&13.37&0.62&5.98&-2.73& 0.47& 3.16\\
Germany&00/09/04&00/08/31&1.05&9.02&5.66&8.87&-0.57& 0.19& 4.39\\
Norway&00/09/05&00/10/02&1.02&8.21&4.77&6.75&-0.60& 0.22& 3.96\\
Spain&00/09/14&00/10/04&0.93&7.52&3.15&6.90&-0.82& 0.31& 3.98\\
Switzerland&00/08/23&00/11/18&1.00&6.76&5.16&9.01&-0.70&-0.26& 3.67\\
UK&00/09/04&00/10/23&1.00&7.58&0.00&8.77&-0.55&-0.17& 3.17\\
Israel&00/08/28&00/09/09&0.18&11.45&4.02&6.61&-205&17.7& 4.51\\
Brazil&00/08/14&00/08/12&0.87&10.15&4.74&9.74&-1.36& 0.56& 6.20\\
Hong Kong&00/07/21&00/02/26&0.90&7.65&1.95&9.90&-1.69&-0.29& 5.27\\
India&00/07/12&00/05/26&0.78&6.34&5.61&8.47&-2.92&-0.85& 4.89\\
Peru&00/09/11&99/10/28&0.15&19.80&1.16&7.45&-118&-20.9& 3.22\\
Taiwan&00/07/17&00/01/24&0.38&9.22&1.01&9.26&-63.0&17.7& 8.50\\
Czech&00/07/28&00/08/13&0.53&4.61&1.19&6.32&-12.5& 5.20& 4.80\\
Argentina&00/07/14&00/05/02&1.54&7.25&5.28&6.23&-0.03&-0.02&12.2\\
Turkey&00/07/10&00/08/28&0.13&9.56&2.88&9.62&-160&63.7&11.8\\
\hline\hline
\end{tabular}
\end{center}
\end{table}

\pagebreak

%table 2
\begin{table}
\begin{center}
\caption{\label{TbBearAB4} Values of the parameters of the fits
to the 21 indices of table \ref{TbBearAB} given in the first column
using formula (\ref{Eq:fit4}) from $t_{\rm{start}}$
to September, 30, 2002. As in table \ref{TbBearAB},
the bearish nature of the markets is identified by
the fact that $B<0$. Note that the amplitude $D$ of the
harmonics is often very significant compared with the amplitude $C$
of the mean angular log-frequency.}
\medskip
\begin{tabular}{ccccccccccccc}
\hline\hline
Stock&$t_{\rm{start}}$&$t_c$&$\alpha$&$\omega$&$\phi_1$&$\phi_2$&$A$&$
10^3B$&$10^3C$&$10^3D$&$10^2\chi$\\\hline\hline
Netherlands&00/09/04&00/07/20&0.80&11.07&4.75&0.91&6.60&-2.82& 0.64&
0.32& 3.72\\
France&00/09/04&00/07/04&0.71&11.24&0.28&4.25&8.90&-5.76&-0.91& 0.48& 3.47\\
USA
DowJones&00/09/06&00/06/18&0.72&12.09&0.83&2.16&9.33&-1.45&-0.65&-0.34
& 3.00\\
USA
NASDAQ&00/08/20&00/09/02&0.26&10.00&3.37&0.00&8.77&-269.2&-24.73&-9.18
& 5.70\\
Japan&00/08/28&00/08/06&0.77&7.74&3.35&0.00&9.74&-3.72&-0.89& 0.21& 3.69\\
Belgium&00/11/06&00/06/21&1.14&12.24&2.67&0.49&8.05&-0.12& 0.05&-0.02& 2.83\\
Denmark&00/10/24&00/05/04&0.64&13.35&0.87&0.00&6.04&-7.51& 1.05& 0.36& 2.86\\
Germany&00/09/04&00/10/06&0.94&8.47&3.61&4.58&8.86&-1.20& 0.41& 0.12& 3.96\\
Norway&00/09/05&00/07/13&0.87&10.92&2.03&1.78&6.82&-1.61&-0.45& 0.13& 3.58\\
Spain&00/09/14&99/07/03&0.88&15.70&1.75&1.28&7.14&-1.14&-0.18& 0.08& 3.46\\
Switzerland&00/08/23&00/11/23&0.95&6.72&5.54&0.00&9.01&-0.98&-0.35&-0.
07& 3.52\\
UK&00/09/04&00/07/21&0.84&10.78&3.56&1.52&8.83&-1.53&-0.30&-0.16& 2.82\\
Israel&00/08/28&00/09/09&0.19&11.41&4.26&0.00&6.60&-194.&17.28&-1.73& 4.49\\
Brazil&00/08/14&00/02/23&1.08&7.99&6.24&4.69&9.83&-0.34&-0.07&-0.08& 5.41\\
HongKong&00/07/21&00/01/30&0.41&7.52&2.21&0.41&10.39&-71.68&-4.88& 3.03& 4.63\\
India&00/07/12&99/12/04&0.03&8.84&5.96&0.73&17.34&-7544&-63&39.5& 4.12\\
Peru&00/09/11&00/02/26&0.54&7.69&2.98&4.99&7.19&-1.67& 1.20& 1.92& 2.76\\
Taiwan&00/07/17&99/05/07&0.31&7.34&1.93&4.44&9.95&-181.0&-14.51&21.50& 5.41\\
Czech&00/07/28&00/08/11&0.40&4.59&1.27&0.77&6.39&-33.90& 9.69&-4.46& 3.94\\
Argentina&00/07/14&99/08/23&2.00&11.23&0.67&4.40&6.25&-0.00&-0.00& 0.00& 8.20\\
Turkey&00/07/10&00/08/28&0.20&9.55&2.94&1.32&9.52&-78.16&47.12&18.47&11.0\\
\hline\hline
\end{tabular}
\end{center}
\end{table}

\clearpage

%table 3

\begin{table}
\begin{center}
\caption{\label{TbBullAB} Values of the parameters of the fits of
the indices indicated in the first column using the first-order
formula (\ref{Eq:fit1}) from $t_{\rm{start}}$ to September, 30,
2002. The bullish nature of the markets refers to the the fact
that $B>0$. The exponents $\alpha$ of the leading power law take
values either greater or less than 1 indicating either an upward
or a downward overall large scale curvature of the price
trajectory.}
\medskip
\begin{tabular}{ccccccccccc}
\hline\hline
Stock&$t_{\rm{start}}$&$t_c$&$\alpha$&$\omega$&$\phi$&$A$&$10^3B$&$10^
3C$&$10^2\chi$\\\hline
Australia&00/09/04&00/08/08&0.77&10.85&3.55&8.08& 0.07&-0.37& 1.76\\
Mexico&00/09/04&00/08/08&0.80&10.11&4.42&8.70& 0.32& 0.86& 4.02\\
Indonesia&00/09/16&00/08/09&1.06&10.30&5.83&5.99& 0.09&-0.21& 3.90\\
Korea&00/10/12&00/11/15&1.37&6.69&5.67&6.29& 0.05&-0.05& 5.00\\
Thailand&00/09/20&00/08/16&0.93&9.45&5.16&5.61& 0.58&-0.39& 4.37\\
Russia&00/12/01&00/10/08&0.92&7.93&3.33&7.24& 2.76&-0.68&4.85\\
\hline\hline
\end{tabular}
\end{center}
\end{table}

%table 4

\begin{table}
\begin{center}
\caption{\label{TbBullAB4} Values of the parameters of the fits of
the indices indicated in the first column using formula
(\ref{Eq:fit4}) from $t_{\rm{start}}$ to September, 30, 2002.}
\medskip
\begin{tabular}{ccccccccccccc}
\hline\hline
Stock&$t_{\rm{start}}$&$t_c$&$\alpha$&$\omega$&$\phi_1$&$\phi_2$&$A$&$
10^3B$&$10^3C$&$10^3D$&$10^2\chi$\\\hline\hline
Australia&00/09/04&00/08/10&0.64&10.89&3.42&0.51&8.08& 0.26&-0.80&-0.26& 1.53\\
Mexico&00/09/04&00/06/27&0.73&5.97&4.23&1.65&8.69& 0.72& 0.39&-1.15& 3.91\\
Indonesia&00/09/16&00/11/16&0.96&3.47&1.33&3.10&6.04&-0.25&-0.50&-0.62& 2.71\\
Korea&00/10/12&00/08/10&1.00&5.04&1.90&1.90&6.24& 0.46&-0.23&-0.25& 4.28\\
Thailand&00/09/20&00/07/27&0.43&5.18&3.62&2.22&5.48&20.83& 3.90& 6.62& 3.54\\
Russia&00/12/01&00/10/12&0.91&7.85&3.90&0.17&7.23& 3.07&-0.75& 0.10& 4.57\\
\hline\hline
\end{tabular}
\end{center}
\end{table}

\pagebreak

%table 5
\begin{table}
\begin{center}
\caption{\label{TbWilk} Likelihood-ratio (Wilk) test of hypothesis
$H_1$ against $H_0$ and of hypothesis $H_2$ against $H_1$, where
$H_0$ is the hypothesis that $C=0$ in (\ref{Eq:fit1}) (pure power
law fit), $H_1$ is the hypothesis that $D=0$ in (\ref{Eq:fit4})
(corresponding to the log-periodic function (\ref{Eq:fit1})
without any harmonics) and $H_2$ is the hypothesis that $D\ne0$,
corresponding to the log-periodic function (\ref{Eq:fit4}) which
includes an harmonic at $2\omega$. Each stock index time series whose
country/area is given in the first column starts at
$t_{\rm{start}}$ and ends at September 30, 2002. The column $n$
gives the number of the data points for each fit. The $\sigma_j$'s
are the standard deviations for hypothesis $H_j$ with $j=0,1,2$ of
the fits to the data associated with formula (\ref{Eq:fit1}) and
(\ref{Eq:fit4}). The $T_{j,j+1}$ are the log-likelihood-ratios
defined in expression (\ref{Eq:T}). The $P_{j,j+1}$ are the
probability for exceeding $T_{j,j+1}$ under the assumption of that
hypothesis $H_j$ holds. }
\medskip
\begin{tabular}{ccccccccccccc}
\hline\hline
Stock&$t_{\rm{start}}$&$n$&$\sigma_0$&$\sigma_1$&$\sigma_2$&$T_{0,1}$
&$T_{1,2}$&${\rm{P}}_{0,1}(\%)$&${\rm{P}}_{1,2}(\%)$\\\hline
Netherlands&00/09/04&508&0.078&0.043&0.037&609.2&148.5&$<10^{-4}$&$<10^{-4}$\\
France&00/09/04&520&0.070&0.040&0.035&589.0&144.2&$<10^{-4}$&$<10^{-4}$\\
USA DowJones&00/09/06&500&0.053&0.034&0.030&459.8&118.1&$<10^{-4}$&$<10^{-4}$\\
USA NASDAQ&00/08/20&511&0.099&0.064&0.057&456.2&111.5&$<10^{-4}$&$<10^{-4}$\\
Japan&00/08/28&498&0.069&0.040&0.037&551.7&70.7&$<10^{-4}$&$<10^{-4}$\\
Belgium&00/11/06&459&0.056&0.034&0.028&466.7&158.6&$<10^{-4}$&$<10^{-4}$\\
Denmark&00/10/24&456&0.052&0.032&0.029&457.1&92.2&$<10^{-4}$&$<10^{-4}$\\
Germany&00/09/04&504&0.083&0.044&0.040&637.9&105.7&$<10^{-4}$&$<10^{-4}$\\
Norway&00/09/05&543&0.081&0.040&0.036&772.0&110.8&$<10^{-4}$&$<10^{-4}$\\
Spain&00/09/14&483&0.062&0.040&0.035&421.9&133.0&$<10^{-4}$&$<10^{-4}$\\
Switzerland&00/08/23&505&0.062&0.037&0.035&526.9&40.9&$<10^{-4}$&$<10^{-4}$\\
UK&00/09/04&506&0.050&0.032&0.028&464.5&120.4&$<10^{-4}$&$<10^{-4}$\\
Israel&00/08/28&390&0.058&0.045&0.045&194.0& 2.7&$<10^{-4}$&$10.03$\\
Brazil&00/08/14&505&0.094&0.062&0.054&418.9&138.0&$<10^{-4}$&$<10^{-4}$\\
HongKong&00/07/21&523&0.070&0.053&0.046&298.3&135.6&$<10^{-4}$&$<10^{-4}$\\
India&00/07/12&528&0.073&0.049&0.041&428.7&180.2&$<10^{-4}$&$<10^{-4}$\\
Peru&00/09/11&488&0.051&0.032&0.028&438.3&151.9&$<10^{-4}$&$<10^{-4}$\\
Taiwan&00/07/17&518&0.114&0.085&0.054&307.3&466.7&$<10^{-4}$&$<10^{-4}$\\
Czech&00/07/28&507&0.052&0.048&0.039&72.4&199.0&$<10^{-4}$&$<10^{-4}$\\
Argentina&00/07/14&502&0.148&0.122&0.082&194.0&401.1&$<10^{-4}$&$<10^{-4}$\\
Turkey&00/07/10&534&0.146&0.118&0.110&233.1&67.8&$<10^{-4}$&$<10^{-4}$\\
Australia&00/09/04&509&0.031&0.018&0.015&588.4&144.1&$<10^{-4}$&$<10^{-4}$\\
Mexico&00/09/04&493&0.078&0.040&0.039&655.4&25.9&$<10^{-4}$&$\approx10^{-4}$\\
Indonesia&00/09/16&454&0.080&0.039&0.027&658.6&329.1&$<10^{-4}$&$<10^{-4}$\\
Korea&00/10/12&456&0.108&0.050&0.043&705.2&142.3&$<10^{-4}$&$<10^{-4}$\\
Thailand&00/09/20&481&0.076&0.044&0.035&535.1&203.0&$<10^{-4}$&$<10^{-4}$\\
Russia&00/12/01&413&0.114&0.048&0.046&704.7&48.1&$<10^{-4}$&$<10^{-4}$\\
\hline\hline
\end{tabular}
\end{center}
\end{table}

\pagebreak

%table 6
\begin{table}
\begin{center}
\caption{\label{TbtcFr} Fitting parameters with
equation (\ref{Eq:fit1}) for the French market stock index with
different $t_{\rm{start}}$ indicated in the first column.}
\medskip
\begin{tabular}{ccccccccccccc}
\hline\hline
$t_{\rm{start}}$&$t_c$&$\alpha$&$\omega$&$\phi$&$A$&$B$&$C$&$\chi$\\\hline
00/06/01&00/08/20&0.91&9.23&4.26&8.81&-0.00154& 0.00035& 0.0387\\
00/07/01&00/08/15&0.89&9.53&5.48&8.81&-0.00168&-0.00038& 0.0389\\
00/08/01&00/08/19&0.89&9.36&3.45&8.81&-0.00165& 0.00038& 0.0393\\
00/09/01&00/08/30&0.91&8.90&3.37&8.79&-0.00145&-0.00035& 0.0399\\
00/10/01&00/10/23&0.97&7.56&3.19&8.74&-0.00101& 0.00029& 0.0400\\
00/11/01&00/10/25&0.98&7.46&0.70&8.74&-0.00096&-0.00027& 0.0407\\
00/12/01&00/10/23&1.05&7.48&0.50&8.72&-0.00061&-0.00018& 0.0413\\
\hline\hline
\end{tabular}
\end{center}
\end{table}

%table 7
\begin{table}
\begin{center}
\caption{\label{TbtcAu} Fitting parameters with
equation (\ref{Eq:fit1}) for the Australian market stock index with
different $t_{\rm{start}}$ indicated in the first column.}
\medskip
\begin{tabular}{ccccccccccccc}
\hline\hline
$t_{\rm{start}}$&$t_c$&$\alpha$&$\omega$&$\phi$&$A$&$B$&$C$&$\chi$\\\hline
00/06/01&00/09/23&0.76&9.41&0.88&8.08& 0.00002&-0.00044& 0.0173\\
00/07/01&00/08/21&0.80&10.25&4.42&8.09& 0.00001& 0.00032& 0.0173\\
00/08/01&00/08/20&0.78&10.29&4.17&8.09& 0.00002& 0.00036& 0.0176\\
00/09/01&00/08/14&0.77&10.57&2.29&8.08& 0.00006& 0.00037& 0.0176\\
00/10/01&00/08/11&0.78&10.68&4.67&8.08& 0.00007&-0.00035& 0.0177\\
00/11/01&00/09/26&0.72&9.35&4.47&8.08& 0.00003& 0.00055& 0.0180\\
00/12/01&00/09/11&0.67&9.78&4.66&8.08& 0.00007&-0.00073& 0.0181\\
\hline\hline
\end{tabular}
\end{center}
\end{table}

%table 8
\begin{table}
\begin{center}
\caption{\label{TbLinReg} $\beta$'s and correlation coefficients
$\gamma$ between the USA S\&P500 index and nine indices
(Netherlands (HL), France (FR), Japan (JP), Germany (DE), United
Kingdom (UK), Hong Kong (HK), Australia (AU), Russia (RU), and
China (CN)) in two periods. Period $1$ is [Jun-04-1997,
Aug-09-2000] and period $2$ is [Aug-10-2000, Sep-04-2002]. The
return is the logarithm of the ratio between two successive prices
with time lag of 30 trading days. } \bigskip
\begin{tabular}{ccccccccccccc}
\hline\hline
  INDEX&HL&FR&JP&DE&UK&HK&AU&RU&CN\\\hline
  $\beta_1$&1.07&0.93&0.98&0.92&0.68&1.34&0.60&2.21&-0.05\\
  $\gamma_1$&0.73&0.67&0.62&0.83&0.49&0.51&0.68&0.42&0.03\\
  $\beta_2$&1.06&1.03&1.17&0.76&0.66&0.92&0.45&0.99&0.02\\
  $\gamma_2$&0.87&0.90&0.86&0.87&0.59&0.72&0.80&0.46&0.02\\
\hline\hline
\end{tabular}
\end{center}
\end{table}

\pagebreak
%FIGURE 1
\clearpage

\clearpage
%FIGURE 2

\clearpage
%FIGURE 3

\clearpage
%FIGURE 4

\clearpage
%FIGURE 5

\clearpage
%FIGURE 6

\clearpage
%FIGURE 7

\clearpage
%FIGURE 8

\clearpage
%FIGURE 9

\clearpage
%FIGURE 10

\clearpage
%FIGURE 11

\clearpage
%FIGURE 12

\clearpage
%FIGURE 13

\clearpage
%FIGURE 14

\clearpage
%FIGURE 15

\clearpage
%FIGURE 16

\clearpage
%FIGURE 17

\clearpage
%FIGURE 18

\clearpage
%FIGURE 19

\clearpage
%FIGURE 20

\clearpage
%FIGURE 21

%FIGURE 22
\clearpage

\clearpage
%FIGURE 23

\clearpage
%FIGURE 24

\clearpage
%FIGURE 25

\clearpage
%FIGURE 26

\clearpage
%FIGURE 27

\clearpage
%FIGURE 28

\clearpage
%FIGURE 29

\clearpage
%FIGURE 30

\clearpage
%FIGURE 31

\clearpage
%FIGURE 32

\clearpage
%FIGURE 33

\clearpage
%FIGURE 34

\clearpage
%FIGURE 35

\clearpage
%FIGURE 36

\clearpage
%FIGURE 37

\clearpage
%FIGURE 38

\clearpage
%FIGURE 39


\begin{thebibliography}{}

\bibitem{Barro}
Barro R.J.,  E.F. Fama, D.R. Fischel, A.H  Meltzer, R. Roll and L.G. Telser,
Black monday and the future of financial markets, in: R.W. Kamphuis, Jr.,
R.C. Kormendi and J.W.H. Watson, eds., Mid American Institute for Public Policy
Research, Inc. and Dow Jones-Irwin, Inc. (1989).

\bibitem{Blanchard} Blanchard, O.J., Speculative Bubbles, Crashes and
Rational Expectations,
Economics Letters 3, 387-389 (1979).

\bibitem{Blanchardwat} Blanchard, O.J. and M.W. Watson, Bubbles,
Rational Expectations and
Speculative Markets, in: Wachtel, P. ,eds., Crisis in Economic and Financial
Structure: Bubbles, Bursts, and Shocks. Lexington Books: Lexington (1982).

\bibitem{Global2} Bordo, M.D. and A.P. Murshid,
Globalization and Changing Patterns in the International
Transmission of Shocks in Financial Markets,
NBER Working Paper No. W9019 (2002)
$http://papers.ssrn.com/paper.taf?abstract\_id=316798$

\bibitem{sunspots} Cass, D., Sunspots and incomplete financial
markets: The General case,
Philadelphia, University of Pennsylvania (1991).

\bibitem{Taboo} Cvijovic, D. and J. Klinowski, Taboo Search: An
Approach to the Multiple Minima Problem, Science 267, 664-666 (1995).

\bibitem{Drozdz} Drozdz, S., Ruf, F., Speth, J. and Wojcik, M.,
Imprints of log-periodic self-similarity in the stock market,
European Physical Journal 10, 589-593 (1999).

\bibitem{Weir} Gluzman, S. and D. Sornette,
Log-periodic route to fractal functions, Phys. Rev. E 65, 036142
(2002).

\bibitem{Goetz} Goetzmann, W.N., L. Li and K.G. Rouwenhorst,
Long-Term Global Market Correlations, NBER Working Paper No. W8612
(2001)// $http://papers.ssrn.com/paper.taf?abstract\_id=291287$

\bibitem{Holden} Holden K, Peel D.A. and Thompson J.L., Economic
Forecasting: An Introduction (Cambridge University Press, Cambridge,
1990) pp.59.

\bibitem{Idesor1} Ide, K. and D. Sornette,
Oscillatory Finite-Time Singularities in Finance, Population and Rupture,
Physica A  307 (1-2), 63-106 (2002).

\bibitem{crashcom} Johansen, A., Comment on ``Are financial crashes
predictable?'' Europhys. Lett. 60, 809-810 (2002). see also
Remarks on reply to Johansen's comment, at cond-mat/0206479

\bibitem{CriCrash00} Johansen, A., O. Ledoit and D. Sornette,
Crashes as critical points, International Journal of Theoretical
and Applied Finance 3, 219-255 (2000).

\bibitem{outl1} Johansen, A. and D. Sornette, Stock market crashes
are outliers,
European Physical Journal B 1, 141-143 (1998).

\bibitem{Nikkei99} Johansen, A. and D. Sornette, Financial
``anti-bubbles'': Log-periodicity in Gold and Nikkei collapses,
Int. J. Mod. Phys. C 10, 563-575 (1999).

\bibitem{JS1999}
Johansen, A. and D. Sornette, Critical Crashes, RISK 12, 91-94
(1999).

\bibitem{emergent} Johansen, A. and D. Sornette, Bubbles and
anti-bubbles in Latin-American, Asian and Western Stock markets:
An emprical study, Int. J. Theor. Appl. Fin. 4, 853-920 (2000).

\bibitem{evalNikkei} Johansen, A. and D. Sornette,
Evaluation of the quantitative prediction of a trend reversal on
the Japanese stock market in 1999, Int. J. Mod. Phys. C 11,
359-364 (2000).

\bibitem{JS2000} Johansen, A. and D. Sornette,
The Nasdaq crash of April 2000: Yet another example of
log-periodicity in a speculative bubble ending in a crash, Eur.
Phys J. B 17, 319-328 (2000).

\bibitem{outl2} Johansen, A. and D. Sornette,
Large Stock Market Price Drawdowns Are Outliers, Journal of Risk
4, 69-110 (2002).

\bibitem{epsilondd} Johansen, A. and D. Sornette,
Endogenous versus Exogenous Crashes in Financial Markets, preprint
at cond-mat/0210509.

\bibitem{JohSorLed99} Johansen, A., D. Sornette and O. Ledoit, Predicting
financial crashes using discrete scale invariance, Journal of Risk 1,
5-32 (1999).

\bibitem{Luxsor} Lux, T. and D. Sornette,
On Rational Bubbles and Fat Tails, Journal
of Money, Credit and Banking part 1, vol. 34, No. 3, 589-610 (2002).

\bibitem{global1} Neal, L.D. and M. Weidenmier,
Crises in the Global Economy from Tulips to Today: Contagion and
Consequences, NBER Working Paper No. W9147 (2002)
preprint at $http://papers.ssrn.com/paper.taf?abstract\_id=328697$

\bibitem{Malsorbub} Malevergne, Y. and D. Sornette,
Multi-dimensional Rational Bubbles and fat tails, Quantitative
Finance 1, 533-541 (2001).

\bibitem{Press} Press, W., S. Teukolsky, W. Vetterling and B.
Flannery, Numerical Recipes in FORTRAN: The Art of Scientific
Computing (Cambridge University, Cambridge, 1996).

\bibitem{EventSyn} Quiroga, R., T. Kreuz and P. Grassberger,
Event synchronization: a simple and fast method to measure
synchronicity and time delay patterns, Phys. Rev. E 66, 041904
(2002).

\bibitem{Rao} Rao, C., Linear statistical Inference and Its
Applications (New York, Wiley, 1965) chap. 6, section 6e.3.

\bibitem{Roehnersor} Roehner, B.M. and D. Sornette,
``Thermometers'' of Speculative Frenzy, European Physical Journal B
16, 729-739 (2000).

\bibitem{Roll88} Roll, R., The International Crash of October 1987,
in {\it{Black Monday and The Future of Financial Markets}}, edited
by A. Meltzer, Dow-Jones-Irwin (1988).

\bibitem{Shefrin} Shefrin, H.,
Beyond greed and fear: understanding behavioral finance and the
psychology of investing (Boston, Mass.: Harvard Business School Press, 2000).

\bibitem{Shillervol} Shiller, R.J.,
{\it Market volatility} (Cambridge, Mass.: MIT Press, 1989).

\bibitem{Shillerexu} Shiller, R.J.,
{\it Irrational exuberance} (Princeton University Press, Princeton, NJ., 2000).

\bibitem{Shleifer} Shleifer, A.,
Inefficient markets: an introduction to behavioral finance
(New York: Oxford University Press, 2000).

\bibitem{SorDSI} Sornette, D., Discrete scale invariance and complex
dimensions, Physics Reports 297, 239-270 (1998).

\bibitem{bookcrash} Sornette, D., Why Stock Markets Crash
(Critical Events in Complex Financial Systems), Princeton
University Press, Princeton, NJ, 2003.

\bibitem{Sorandersen} Sornette, D. and J.V. Andersen,
A Nonlinear Super-Exponential Rational Model of Speculative
Financial Bubbles, Int. J. Mod. Phys. C 13, 171-188 (2002).

\bibitem{Idesor2} Sornette, D. and K. Ide,
Theory of self-similar oscillatory finite-time singularities in
Finance, Population and Rupture, Int. J. Mod. Phys. C 14, 267-275
(2002).

\bibitem{SJ1998} Sornette, D. and A. Johansen,
A Hierarchical Model of Financial Crashes, Physica A 261, 581-598 (1998).

\bibitem{SorJoh01} Sornette, D. and A. Johansen, Significance of
log-periodic precursors to financial crashes, Quantitative Finance
1, 452-471 (2001).

\bibitem{SJB} Sornette, D., A. Johansen and  J.-P. Bouchaud,
Stock market crashes, Precursors and Replicas, J.Phys.I France 6,
167-175 (1996).

\bibitem{sormalbubble} Sornette, D. and Y. Malevergne,
  From Rational Bubbles to Crashes, Physica A 299, 40-59 (2001).

\bibitem{SPpredict} D. Sornette and W.-X. Zhou, The US 2000-2002
Market Descent: How Much Longer and Deeper?
Quantitative Finance 2, 468-481 (2002).

\bibitem{Zhoustatsi} Zhou, W.-X. and Didier Sornette,
Statistical Significance of Periodicity and Log-Periodicity with
Heavy-Tailed Correlated Noise, Int. J. Mod. Phys. C 13, 137-170
(2002).

\end{thebibliography}
\end{document}